\documentclass[fleqn,usenatbib]{mnras}

\usepackage{newtxtext,newtxmath}
\usepackage{subcaption}

\usepackage[T1]{fontenc}
\DeclareRobustCommand{\VAN}[3]{#2}
\let\VANthebibliography\thebibliography
\def\thebibliography{\DeclareRobustCommand{\VAN}[3]{##3}\VANthebibliography}

\usepackage{graphicx}	
\usepackage{amsmath}	
\usepackage{multicol}

\title[Nebulae around Wolf-Rayet stars in M33]{ Ring nebulae around Wolf-Rayet stars in M33 as seen by SITELLE}

\author[S. Tuquet et al.]{Selin Tuquet,$^{1,3}$
Nicole St-Louis,$^{1,3}$\thanks{E-mail: nicole.st-louis@umontreal.ca}
Laurent Drissen,$^{2,3}$
Sylvain Raaijmakers,$^{1,3}$
\newauthor
Laurie Rousseau-Nepton,$^{4}$
René Pierre Martin,$^{5}$
Carmelle Robert,$^{2,3}$
Philippe Amram$^{6}$
\\
$^{1}$Département de Physique,  Université de Montréal, Campus MIL, 1375 Ave.Thérèse-Lavoie-Roux, Montréal (QC), H2V 0B3, Canada\\
$^{2}$Département de physique, de génie physique et d’optique, Université Laval, Québec (QC), G1V 0A6, Canada\\
$^{3}$Centre de recherche en astrophysique du Québec (CRAQ)\\
$^{4}$Dunlap Institute of Astronomy and Astrophysics, University of Toronto, 50 St George St, Toronto (On), M5S 3H4, Canada\\
$^{5}$Department of Physics and Astronomy, University of Hawaii at
Hilo, Hilo, HI, United States\\
$^{6}$Aix Marseille Univ, CNRS, CNES, LAM, Marseille, France\\
}

\date{Accepted XXX. Received YYY; in original form ZZZ}

\pubyear{2021}

\begin{document}
\label{firstpage}
\pagerange{\pageref{firstpage}--\pageref{lastpage}}
\maketitle

\begin{abstract}
    We have conducted an analysis of nebulae around Wolf-Rayet (WR) stars in M33 using data collected by the imaging Fourier transform spectrometer SITELLE at the Canada-France-Hawaii telescope as part of the SIGNALS Large Program. Of the 211 known Wolf-Rayet stars in M33, 178 are located in the fields observed in this study. We present the results of this analysis in the form of a comprehensive summary of all nebulae found around the observed WR stars. Based on three criteria we find to be the most effective for their detection, we detect a clear association with a circumstellar bubble around 33 of them (19\%). Our results show that the presence of bubbles does not correlate with the spectral type of the central star. The mean diameter of the WR nebulae we have found is 21 parsecs.
\end{abstract}




\section{Introduction}

Classical Wolf-Rayet (cWR) stars are an advanced stage of evolution of massive stars and are known for their spectra that include broad, strong emission lines \citep[e.g.][]{Crowther_2007}. According to the Conti scenario \citep{1975MSRSL...9..193C}, these stars represent the last stage in the evolution of massive stars before they explode as supernovae. In this same scenario, the WR phase is the result of an O-type star that has lost its outer layers due to strong winds, thus revealing products of nuclear reactions occurring in its core. The first elements that are revealed are helium and nitrogen, products of the main sequence H-burning phase, through the CNO cycle \citep[]{beals1933classification}. More massive stars will reveal the products of the subsequent He-burning phase, that is carbon (WC stage) and some hotter stars can reach the WO stage where the emission from oxygen is also increased.  The broad emission lines are due to the high mass-loss rates of these stars and these create dense, bubble-like nebulae around them \citep[e.g.][]{Chu_2016}.

Several formation mechanisms can lead to the presence of a nebulosity around a WR star. \citet{Chu1981GalacticRN} classified WR nebulae into three categories :
\begin{itemize}
    \item Radiatively excited H{\sc ii} regions (R-type), where the WR star only contributes to the ionization of its surrounding interstellar medium (ISM), such as RCW 78 \citep{1983ApJS...53..937C}. Chemical abundances of these nebulae do not differ from those of their surroundings.
    \item Stellar ejecta (E-type), which are short-lived, thus only found if recently ejected or protected inside a bubble, such as M1-67 around WR 124 studied for example by \citet{S_vigny_2020}. Strong helium and nitrogen enhancement is common in these nebulae.
    \item Wind-blown bubbles (W-type), which consist of ISM material swept-up by the strong winds of the WR star and preceding evolutionary phases : NGC 3199, surrounding WR 18, is an example of a W-type nebula, where chemically-enriched filaments coexist with zones of solar abundances, indicative of its mixed origin \citep{2017ApJ...846...76T}.
\end{itemize}

 \begin{figure*}
\centering
\includegraphics[scale=0.6]
{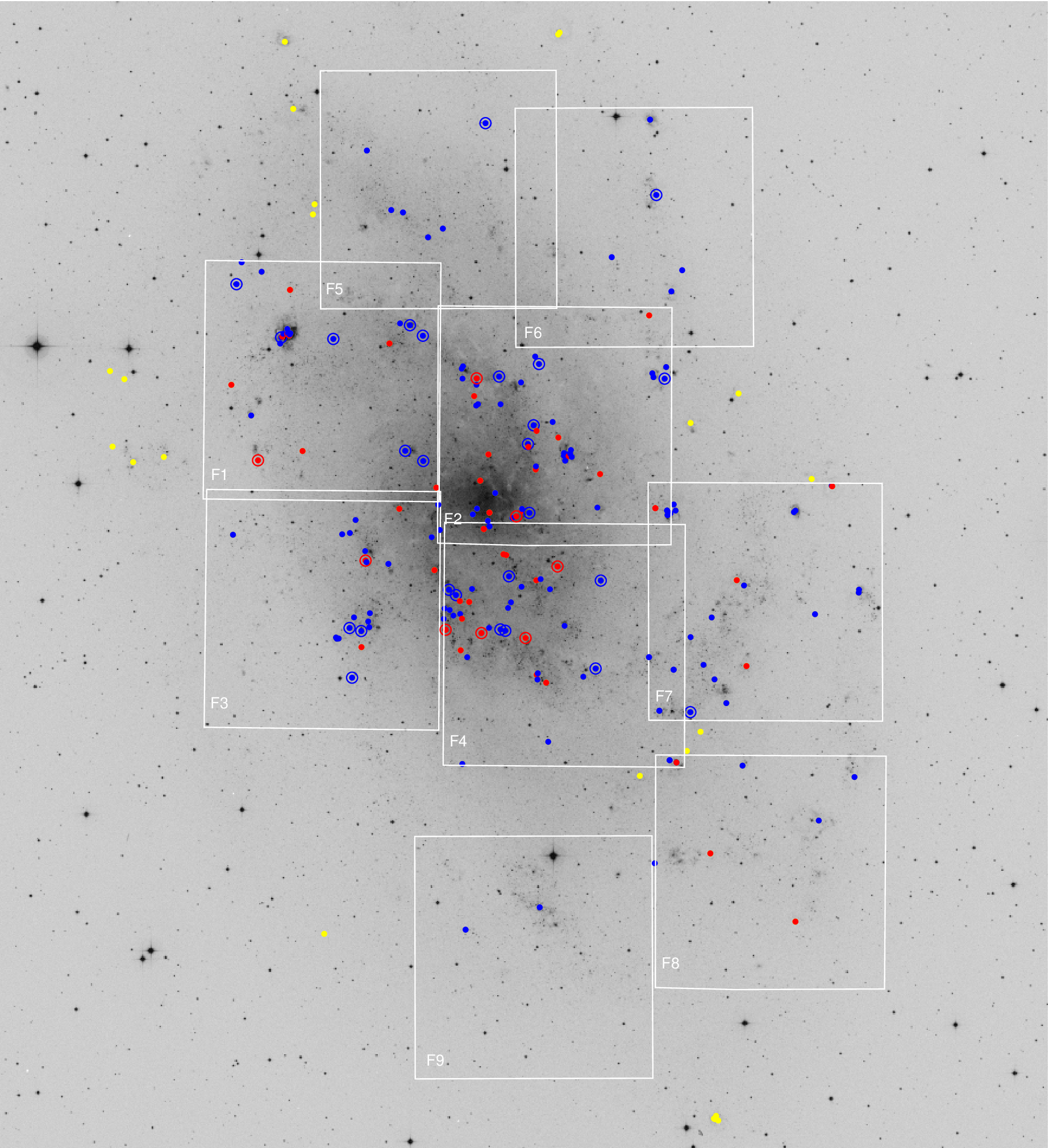}
\caption{Footprints of the nine $11'\times 11'$  fields studied with SITELLE superimposed on a Digital Sky Survey of M33 (Courtesy Palomar Observatory - Space Telescope Science Institute). North is up and East to the left. We indicate WN stars in blue and WC stars in red and the remaining WR stars located outside the SITELLE fields in yellow. Confirmed bubbles are indicated by larger circles around the corresponding WR stars.}
\label{fig:m33}
\end{figure*}

\begin{table*}
        \centering
        \caption{Characteristics of the datacubes}
        \label{tab:fields}
        \begin{tabular}{lccccccl}
                \hline
                Field &  RA (J2000)& Dec (J2000)& Filter & Resolving Power& Exposure/step & Num. Steps& Date of Observation\\
                \hline
                Field 1 (F1) & 01:34:24.1 & +30:44:56 &SN3 & 2900 & 18s & 505 & October 12, 2017 \\
                 &  & &SN2 & 1020 & 38s & 219 & September 28, 2017 \\
                 &  &  &SN1 & 1020 & 49s & 172 & September 28, 2017 \\ 
                Field 2 (F2) & 01:33:36.2 & +30:42:57 &SN3 & 2900 & 18s & 505 & October 13, 2017 \\
                 &  & &SN2 & 1020 & 38s & 219 & September 28, 2017 \\
                 &  &  &SN1 & 1020 & 49s & 172 & September 28, 2017 \\
                Field 3 (F3) & 01:34:24.1 & +30:34:33 &SN3 & 2900 & 18s & 505 & October 17, 2017 \\
                 &  & &SN2 & 1020 & 38s & 219 & September 28, 2017 \\
                 &  &  &SN1 & 1020 & 49s & 172 & September 28, 2017 \\
                Field 4 (F4) & 01:33:33 & +30:32:55 &SN3 & 2200 & 18s & 403 & October 13, 2017 \\
                 &  & &SN2 & 1020 & 38s & 219 & September 28, 2017 \\
                 &  &  &SN1 & 1020 & 49s & 172 & September 28, 2017 \\
                Field 5 (F5) & 01:34:00.1 & +30:53:35 &SN3 & 5000 & 13s & 842 & October 2, 2021 \\
                 &  & &SN2 & 1020 & 46s & 219 & October 4, 2021 \\
                 &  &  &SN1 & 1020 & 59s & 171 & October 6, 2021 \\
                Field 6 (F6) & 01:33:19.1 & +30:51:51 &SN3 & 5000 & 13s & 842 & September 30, 2021 \\
                 &  & &SN2 & 1020 & 46s& 219 & November 5, 2021 \\
                 &  &  &SN1 & 1020 & 59s & 171 & November 10, 2021 \\
                Field 7 (F7) & 01:32:50.7 & +30:34:50 &SN3 & 5000 & 13s & 842 & October 4, 2018 \\
                 &  & &SN2 & 1020 & 46s & 220 & October 11, 2018 \\
                 &  &  &SN1 & 1000 & 59s & 171 & October 12, 2018 \\   
                Field 8 (F8) & 01:32:51.1 & +30:22:34 &SN3 & 5000 & 13s & 842 & October 15, 2020 \\
                 &  & &SN2 & 1020 & 46s & 219 & December 8, 2020 \\
                 &01:32:47.1  &   +30:22:44&SN1 & 1000 & 59s & 171 & 17 November 2020 \\  
                Field 9 (F9) & 01:33:41.6 & +30:18:47 &SN3 & 5000 & 13s & 842 & September 29, 2019 \\   
                 & 01:33:42.6 & +30:20:57&SN2 & 1020 & 46s & 220 & December 11, 2020 \\
                 & 01:33:38.6 & +30:21:08  &SN1 & 1000 & 59s & 171 & November 14, 2020 \\                            
                \hline
        \end{tabular}
\end{table*}

The shape of a nebula can also carry information on the physical mechanism that led to its ejection. We know that most massive stars evolve in binaries. \citet{2012Sci...337..444S} concluded that more than two thirds of all massive stars will exchange mass with a companion during their lifetime and in one third of the cases, this will lead to a merger. Although not well studied yet, this can very likely influence the distribution of the ejected gas found around massive stars. Although the morphology of a WR nebula can be indicative of its origin, a detailed quantitative chemical and kinematical analysis is necessary to assess the impact of the central star and its stellar wind as well as its formation history.
Superbubbles \citep[][]{2008IAUS..250..341C} can also be found around groups of massive stars and share a similar structure to bubbles blown by individual stars.

Studying the morphology, abundance pattern and dynamics of these circumstellar bubbles enlightens us on their formation mechanisms, the nature and the mass-loss history of the star that generated them and the interactions between stellar winds and the ISM. 

The goal in carrying out a study of the population of bubbles in another galaxy is not only its convenience compared to the Milky Way, because of our unfavorable viewpoint within the disc leading to higher extinction. It also enlightens us on the impact of the abundance of metals in the galaxy on the presence, size and structure of the bubbles. The general properties of the bubbles could, for example, lead to new insight on the impact of metallicity on the strength of the wind of the stars or on the evolutionary paths they have followed.

The nearby spiral galaxy M33 is particularly interesting for this type of work because its WR population has already been well documented. \citet{Neugent_2011} and later \citet[]{2014} list 211 spectroscopically confirmed WR stars, and estimate their survey complete to 95\%. Moreover, its proximity and nearly face-on orientation provide a global view of the galaxy and facilitates the study of the WR nebulae. This galaxy also has a well-known metalicity gradient with values for oxygen that are nearly solar in the center and decrease with radius with a slope of -0.037$\pm$0.007 dex\ kpc$^{-1}$  \citep[e.g.][]{2022ApJ...939...44R}. Through this paper, we use the most reliable distance of M33 determined by \cite{cluster}, namely 860 kpc: at that distance, $1'' = 4.17$ pc.  

Using at KPNO's 2.1-m telescope one of the first generations of astronomical CCDs and an H$\alpha$ filter, \citet{article} discovered a dozen nebulae highly likely to be physically associated with WR stars in M33, along with eight more potential candidates. The completeness of the surveys for WR stars was then not as high as it is now, in particular for the weak-lines WN stars.

Our current work aims at considerably improving on this earlier study with the analysis of hyperspectral imagery and thus present an extensive and up-to-date comparative study of nebulae around the vast majority of known WR stars in M33. We used data collected by the imaging Fourier transform spectrometer (iFTS) SITELLE \citep[]{drissen2019sitelle} at the Canada-France-Hawaii Telescope as part of the SIGNALS Large Program which aims at studying a large sample of H{\sc ii} regions in nearby galaxies \citep[][]{rousseau2019signals}. 
We present the data used in this work in Section 2 and a general overview of our approach to identify WR bubbles in Section 3. A brief description of each identified WR nebula is given in Section 4. General statistics and a comparison of our results with those found in the Galaxy and the LMC are presented and discussed in Section 5. We summarize our findings and conclude in Section 6.

\section{Observations}

\subsection{SITELLE}

M33 was a prime target for the SIGNALS program, being very rich in H{\sc ii} regions and supernova remnants. Nine SITELLE fields (Figure \ref{fig:m33} and Table\,\ref{tab:fields}) were therefore prioritized and observed, the four central ones being part of SIGNALS' proof-of-concept study shortly after the commissioning of SITELLE. 

\begin{figure*}
	\includegraphics[width=\columnwidth]{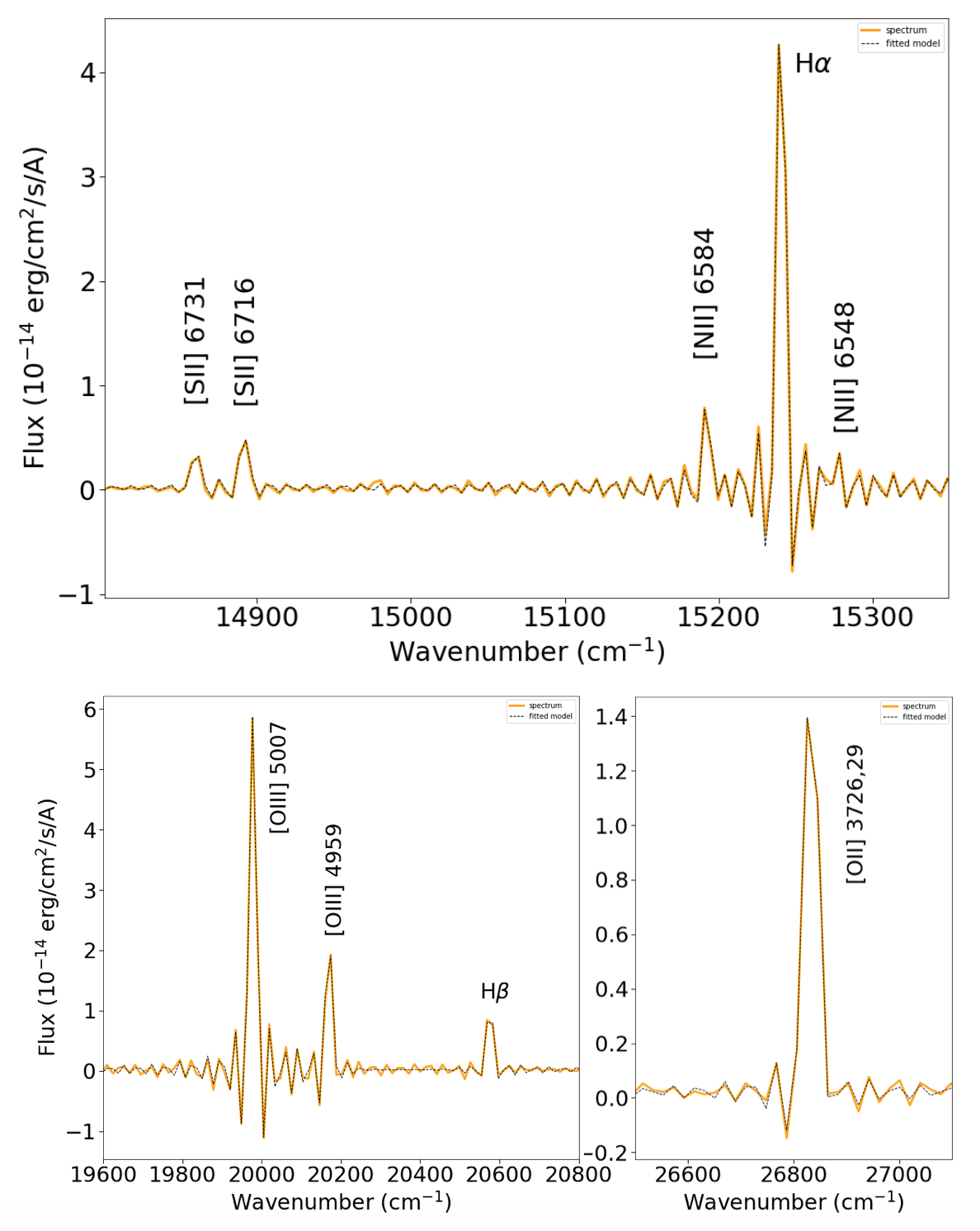}
    \caption{Integrated spectrum of the nebula around the WN star J013327.76+303150.9 in a 32-pixel diameter region using datacubes obtained by SITELLE with the SN3 (upper panel), SN2 (lower left panel) and SN1 (lower right panel) filters. The orange line is the observed spectrum while the dotted black line is the fit with ORCS. Note that the oscillations on the sides of each line are not noise, but rather the sidelobes from the sinc function, and are thus properly fitted by the data analysis software ORCS.}
    \label{fig:spectrum}
\end{figure*}

 SITELLE acquires interferograms of the field with two CCD detectors (2048 $\times$ 2064 pixels each) attached to the complementary output ports of a Michelson interferometer. Data reduction involving Fourier transforms then converts these two interferometric cubes into a single spectroscopic data cube, providing spatially resolved spectra of the sources in an $11' \times 11'$ field of view with a sampling of $0.32''$/pixel, in selected bandpasses of the visible range. The spectral resolution can be chosen at will (up to R$\sim 10^4$), before the beginning of the data acquisition, to be adapted to the needs of the observer. Three bandpasses, centered on the H{\sc ii} regions' bright lines, are used in the SIGNALS survey: SN3 (647 - 685 nm, R = 5000), SN2 (482 - 513 nm, R = 1000) and SN1 (363 - 386 nm, R = 1000). The SN3 datacubes (that include H$\alpha$, the [N{\sc ii}] $\lambda\lambda$ 6548,84 doublet and the [S{\sc ii}] $\lambda\lambda$ 6717,31 doublet) with their higher spectral resolution provide the required kinematics information, with a precision reaching the sub-km/s level in the brightest regions. Note that because the four central fields (Fields 1 to 4) were acquired before the official beginning of SIGNALS, their spectral resolution and total exposure times did not quite meet SIGNALS' standards, but are nevertheless invaluable to the present work because they include the vast majority of WR stars in this galaxy (Table \ref{tab:nbwrfield}).
 An example of a spectrum obtained with the data from SITELLE is shown in Fig. \ref{fig:spectrum}.
 
\begin{table*}
	\centering
	\caption{Number of WR stars found in SITELLE fields of M33}
	\label{tab:nbwrfield}
\begin{tabular}{ |c|c| }
   \hline
 Field &  Number of WR stars \\
   \hline
    Field 1 (F1) & 26 \\
    Field 2 (F2) & 58  \\
    Field 3 (F3) & 21 \\
    Field 4 (F4) & 45 \\
    Field 5 (F5) & 6 \\
    Field 6 (F6) & 5 \\
    Field 7 (F7) & 24 \\       
    Field 8 (F8) & 7 \\
    Field 9 (F9) & 2 \\       
\hline
\end{tabular}
\end{table*}

Data reduction was performed using ORBS and ORCS, SITELLE's dedicated data reduction and analysis pipelines. Details of the automated procedure as well as the sources of uncertainties can be found in \citet{Martin2021}. The wavelength calibration is performed using a high resolution laser source but may show distortions due to aberrations and deformations in the optical structure. To refine the wavelength calibration, we measured the centroid positions of the night sky OH emission lines (in the SN3 filter) in spectra extracted from circular regions sampling the entire FOV following the procedure described in \citet{Martin2018}. A velocity correction map is obtained by interpolating values on the night sky velocity grid, reducing the uncertainty on the velocity calibration across the entire field to less than 2 km/s. We also applied a barycentric correction to our observed spectra. Photometric calibrations were performed using images and datacubes of spectrophotometric standard stars. We note that the natural units of iFTS data are wavenumbers (cm$^{-1}$); ORCS always extracts the information on fluxes, velocity and velocity dispersion using wavenumbers and produces maps of these quantities along with maps of their uncertainties.

While the Fourier transform of the interferograms produces a spectral cube, their sum creates a deep image corresponding to the integral flux within the filter bandpass. We thus examined the stellar content of each region to identify the WR star and its neighbors and roughly determine their color by comparing deep SN1 (375 nm) and SN3 (665 nm) images. These were complemented by
Hubble Space Telescope images of our fields which were available for all stars in our fields except one.

\subsection{Hubble Space Telescope}

Images from the Hubble Space Telescope's Advanced Camera for Surveys (ACS; filters F475W and F814W) in the visible and Wide Field Camera 3 (WFC3; filters F275W and F336W) in the near UV, are used in this study to get a holistic view of the stellar population inside the fields of interest (see \citet[]{2021ApJS..253...53W} for a detailed analysis of the stellar population of M33 based on these images). The color information enabled the identification of possible ionization sources inside each candidate WR nebula.

\subsection{Maps alignment and seeing adjustments}

In order to study and compare spectral line maps from different datacubes, their misalignment and resolution mismatch have been adjusted using Python scripts, mainly from the Numpy, Scipy and Astropy libraries. Reference stars (unsaturated, isolated and visible in all datacubes) were chosen and modeled as a 2D gaussians and their parameters were used to align all three datacubes (by aligning the centroid of the gaussians). When line ratio maps were required, the maps with the best seeing were artificially degraded by convolving them with a kernel that results in the same gaussian width as the reference star in the map with the worst seeing. 

\section{Analysis}

\subsection{Identification of the nebulae}

Out of the 211 WR stars known in M33, 178 (84\%) are located in the fields observed by SITELLE, which are illustrated in Figure \ref{fig:m33} (see also Table \ref{tab:nbwrfield}). Eleven of those are common to two different fields.

We first extracted and inspected a 100x100 pixel (32$'' \times 32''$) frame in the H$\alpha$ and [O{\sc iii}]$\lambda$5007 line maps centered on each of the 178 WR stars, using their coordinates from \citet{Neugent_2011}. This provided general information about their immediate surroundings and allowed us to identify a first group of circumstellar nebulae, seemingly centered or almost centered on the star, in a circle-like shape, either complete or not.

We then inspected the HST images to determine if the WR star was the only (or at least the brightest) and most centrally located blue star within the bubble to be its associated star.
\\

Based on this first overview of the WR stars environment, we have adopted three criteria to decide if a WR star has an associated bubble or not :

\begin{enumerate}
    \item[1.] There is a nebula surrounding the WR star, that is reminiscent of a bubble (complete / symmetrical or not).
    
     \item[2.] The WR star is the brightest blue star located near the center of the bubble.
     
     \item[3.] The WR star has a clear impact on the ionization of the bubble or surrounding gas.
    
\end{enumerate}

The third criterion comes from the fact that in most cases, the [O{\sc iii}]$\lambda$5007 line is the best indicator of the presence of a bubble since, most of the time, it highlights the closest parts of the nebula to the WR star, which renders the bubble-like shape most discernible. Moreover, we found the [O{\sc iii}]$\lambda$5007 / [O{\sc ii}]$\lambda$3726+3729 and [O{\sc iii}]$\lambda$5007 / H$\beta$ ratios to also be amongst the clearest diagnostics for the presence of circumstellar bubbles. 

In numerous cases, a bright and large nebula (30 - 80 pc) was first identified but a closer inspection allowed the detection of a much smaller and fainter nebula immediately surrounding the WR star, much more likely to be physically associated with the current evolutionary phase.

When at least two of the three criteria are satisfied, we consider the WR star to be associated with a bubble. This leads us to identify 33 WR bubbles in M33. We chose to label these bubbles according to the identification already provided to their associated WR star, linked to their J2000 coordinates. 
The information on all confirmed WR nebulae we identified in this work is summarized in Table \ref{tab:bubblesv2},  while Figure \ref{fig:stamps1} presents $32'' \times 32''$ H$\alpha$ images of each of them. The diameter of each nebula indicated in Table \ref{tab:bubblesv2} was determined by overlaying an ellipse over the brightest part of the ring (shown in Figure \ref{fig:stamps1}); in the case of nebulae identified as clumps, we provide the characteristic dimensions. A brief description of each object is then presented in Section 4.1. Observed spectral line and line ratio maps for the identified bubbles are presented in Appendix B. 

\begin{table*}
	\centering
	\caption{Properties of the nebulae around WR stars in M33 identified in this study. For stars that have nested bubbles, we provide the diameter of the larger bubble in parenthesis. Other identifications in the last column are from \citet{1983ApJ...273..576M} (MC), \citet{1987AJ.....94.1538M} (MCA) and \citet{1991AJ....102..927A} (AM).}
	\label{tab:example_table}
\begin{tabular}{llrrll}
               \hline
  Name (Star)  & Field &    Sp. Type &  Diameter (pc)  &  Shape &                    Notes \\
               \hline 
J013307.80+302951.1 & F7 &      WNE &  16 &     ring &        \\
J013312.95+304459.4 & F2 &       WN &       30 &       ring &         \\
J013314.56+305319.6 & F6 &     WN3+neb &          25 &     arc &              ... \\
J013326.60+303550.3 &  F4 &     WN6 &          $\le$ 4 &      clump &                      \\
J013327.76+303150.9 & F4 &      WN &          15 (40) &     ring & MC27 (Nested) \\
J013334.04+304117.2 &  F2 &     WN &          15 (26) &      clump &                     Nested \\
J013335.73+303629.1 &  F4 &     WC &          20 &     ring &                     MC34 \\
J013339.52+304540.5 &  F2 &     B0.5Ia+WNE &          15$\times$30 &     bipolar   &           \\
J013340.69+304253.7 & F2 &      WN &          15 &      arc &             AM 11 \\
J013341.65+303855.2 &  F2 &     WN &          26 &     ring &                     MCA9 \\
J013341.91+304202.7 &  F2 &     WN &          23$\times$12 &    clump &                MCA7 \\
J013342.53+303314.7 & F4 &   WC4/5 &          22 (37) &       ring &                     MC44 (Nested)\\
J013344.40+303845.9 & F2 &     WC4 &           11 &     ring &                      ... \\
J013345.99+303602.7 & F4 &    WN4b &           13 (40) &       ring & MC46 (Nested) \\
J013346.80+303334.5 & F4 &  WN6/C4 &          12 &       clump &                     MC48 \\
J013347.83+303338.1 &  F4 &    WN4 &           24 &       arc/clump &                      \\
J013347.96+304506.6 &  F2 &     WN &           11 &     ring &                     MC51 \\
J013350.71+305636.7 &  F5 &    WN3+neb &           16 (40) &       arc &                    Nested \\
J013351.84+303328.4 & F4 &     WC6 &           35 &       ring &                     MC55 \\
J013352.71+304502.0 & F2 &     WC4 &           26 &       arc &                     MC57 \\
J013357.20+303512.0 & F4 &     WNL &           36 &       arc &                    MCA14 \\
J013358.69+303526.5 & F4 &     WN &           43 &       ring &                     ... \\
J013359.39+303337.5 &  F4 &    WC6 &           24 &       arc &                     MC65 \\
J013404.07+304658.3 & F1 &     WN6 &           13 (45) &       ring &                      Nested \\
J013406.80+304727.0 & F1 & WN7+neb &           15 (65) &       arc &                      Nested \\
J013407.85+304145.1 &  F1 &     WN &           28 &       ring &                     MC67 \\
J013416.28+303646.4 & F3 & WC6+neb &           13 (28) &      ring &                     MC70 (Nested) \\
J013417.21+303334.7 & F3 & WN3+neb &           13 &    arc &                      ... \\
J013419.16+303127.7 & F3 &     WN3 &           16 &       arc &                      ... \\
J013419.68+303343.0 & F3 & WN3-4+neb &           11&      ring &                      ... \\
J013423.02+304650.0 & F1 &     WN4 &           23 &       arc &        ... \\
J013438.98+304119.8 & F1 &     WC4 &           15 &       arc &                      ... \\
J013443.51+304919.4 & F1 &    WN4 &           9 (25) &      ring &    Nested \\

\hline
\label{tab:bubblesv2}
\end{tabular}
\end{table*}
\vfill\eject

\begin{figure*}
\includegraphics[width=0.9\linewidth]{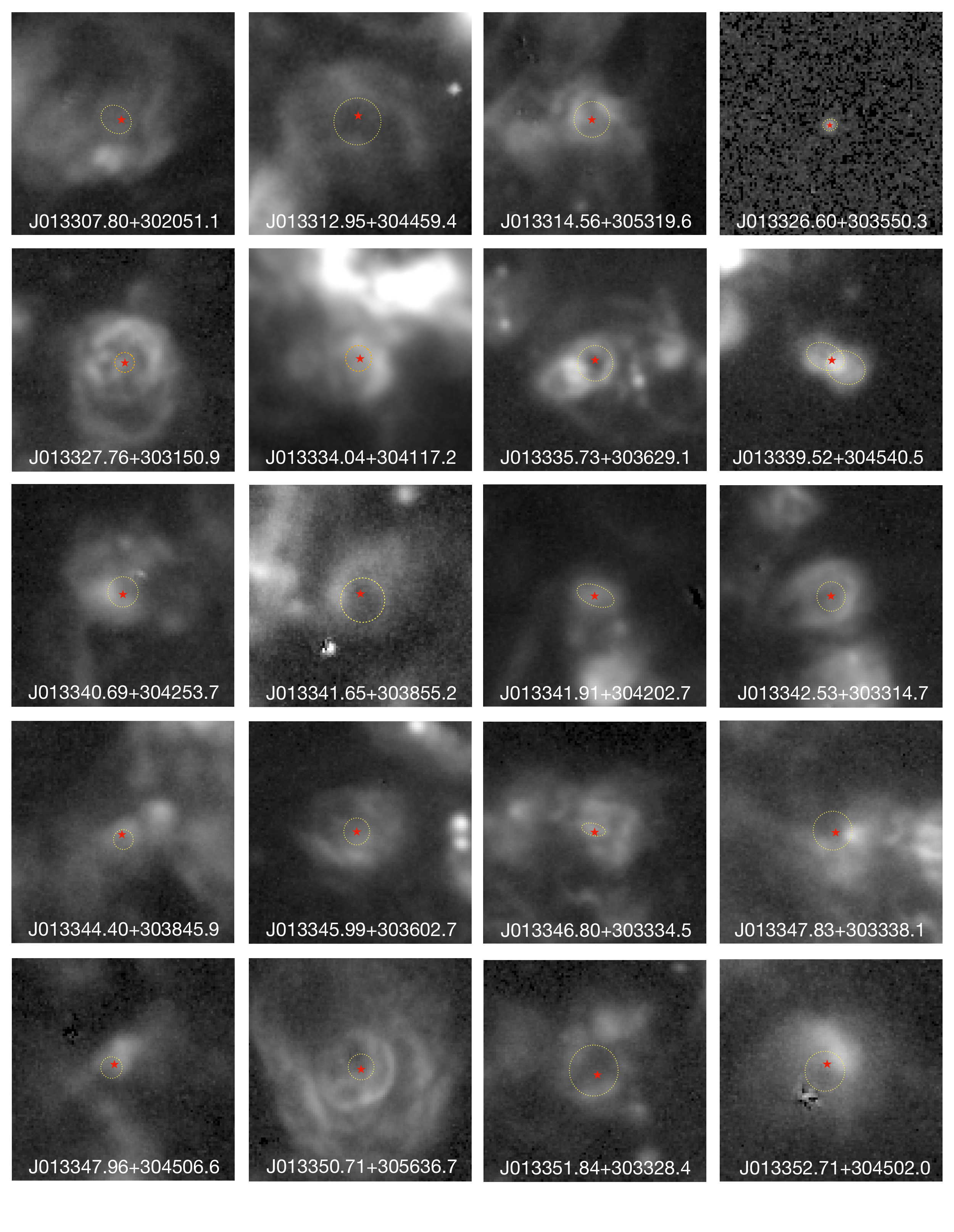}
\caption{H$\alpha$ images of the nebulae associated with WR stars in M33 (32$'' \times 32''$, with North at the top and East to the left). The location of the WR star is indicated with a red star, while the yellow ellipses delineate the structure that we interpret as the WR nebula.}
\label{fig:stamps1}

\end{figure*}

\begin{figure*}
\setcounter{figure}{2}
\includegraphics[width=\textwidth]{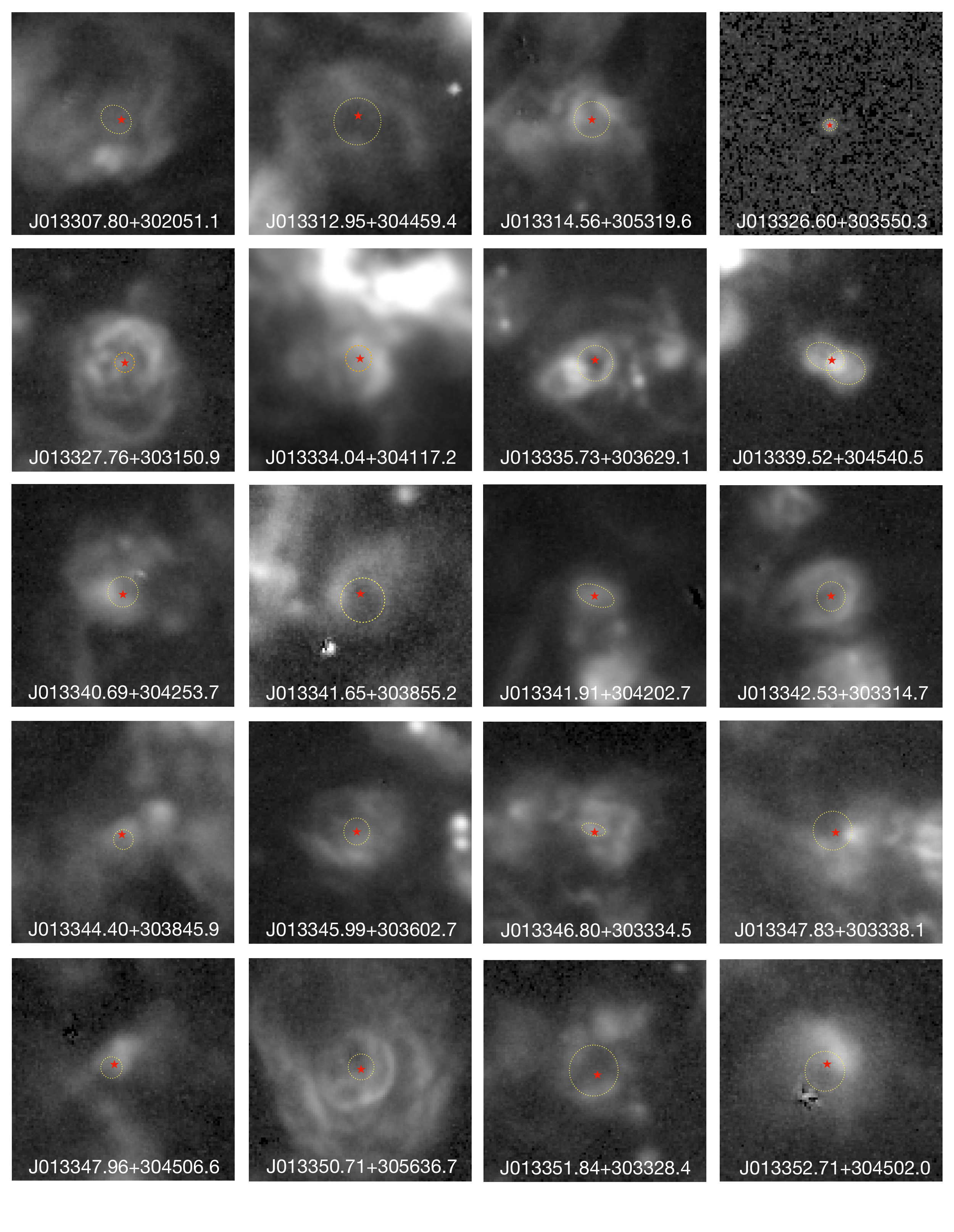}
\caption{(Continued)}
\label{fig:stamps2}
\end{figure*}

\subsection{Spectral lines and line ratio maps}

For WR stars that seemed to be associated with a bubble-like nebula, 
we carried out a more detailed investigation by looking at the same 100x100 pixel images in the lines covered by the SITELLE filters: H$\beta$ (SN2), [N{\sc ii}]$\lambda$6583 (SN3) , [S{\sc ii}]$\lambda$6731+6717 (SN3) , [O{\sc ii}]$\lambda$3726+3729 (SN1), and [O{\sc iii}]$\lambda$5007 (SN2). We also inspected the  H$\alpha$ velocity dispersion map in this same region, along with the velocity maps and some line-ratio maps that we have chosen to diagnose different types of phenomena: 

\begin{itemize}
    \item The H$\alpha$ / H$\beta$ ratio: provides a reliable estimate of the extinction along the line of sight.
    \item The [N{\sc ii}]$\lambda$6583/H$\alpha$ ratio depends on the ionization parameter.
    \item The  [S{\sc ii}]$\lambda$6731+6717/H$\alpha$ ratio also depends on the ionization parameter. Indeed, according to \cite{Kewley_2019}, [S{\sc ii}] lines are produced in a partially ionized zone at the edge of H{\sc ii} regions, which depends on the ionization parameter.
    \item The [S{\sc ii}]$\lambda$6731/[S{\sc ii}]$\lambda$6717 ratio is sensitive to the electron density.
    \item The [O{\sc iii}]$\lambda$5007/[O{\sc ii}]$\lambda$3726+3729 ratio is a clear ionization diagnostic.
    \item The [O{\sc iii}]$\lambda$5007/H$\beta$ ratio also depends on the ionization parameter.
\end{itemize}

These maps provide us with a quantifiable appraisal of the surroundings of each WR star, either in regards to the morphology/shape and composition of the nebulae (with spectral line maps), its dynamics (velocity and velocity dispersion maps), or the ionization parameter, electron density, and excitation sources. A detailed analysis of these maps is beyond the scope of this paper and will be presented elsewhere, but Appendix A presents graphs of the ionization diagnostic line ratios as a function of the distance from the WR star.

\begin{figure*}
    \includegraphics[width=0.9\textwidth]{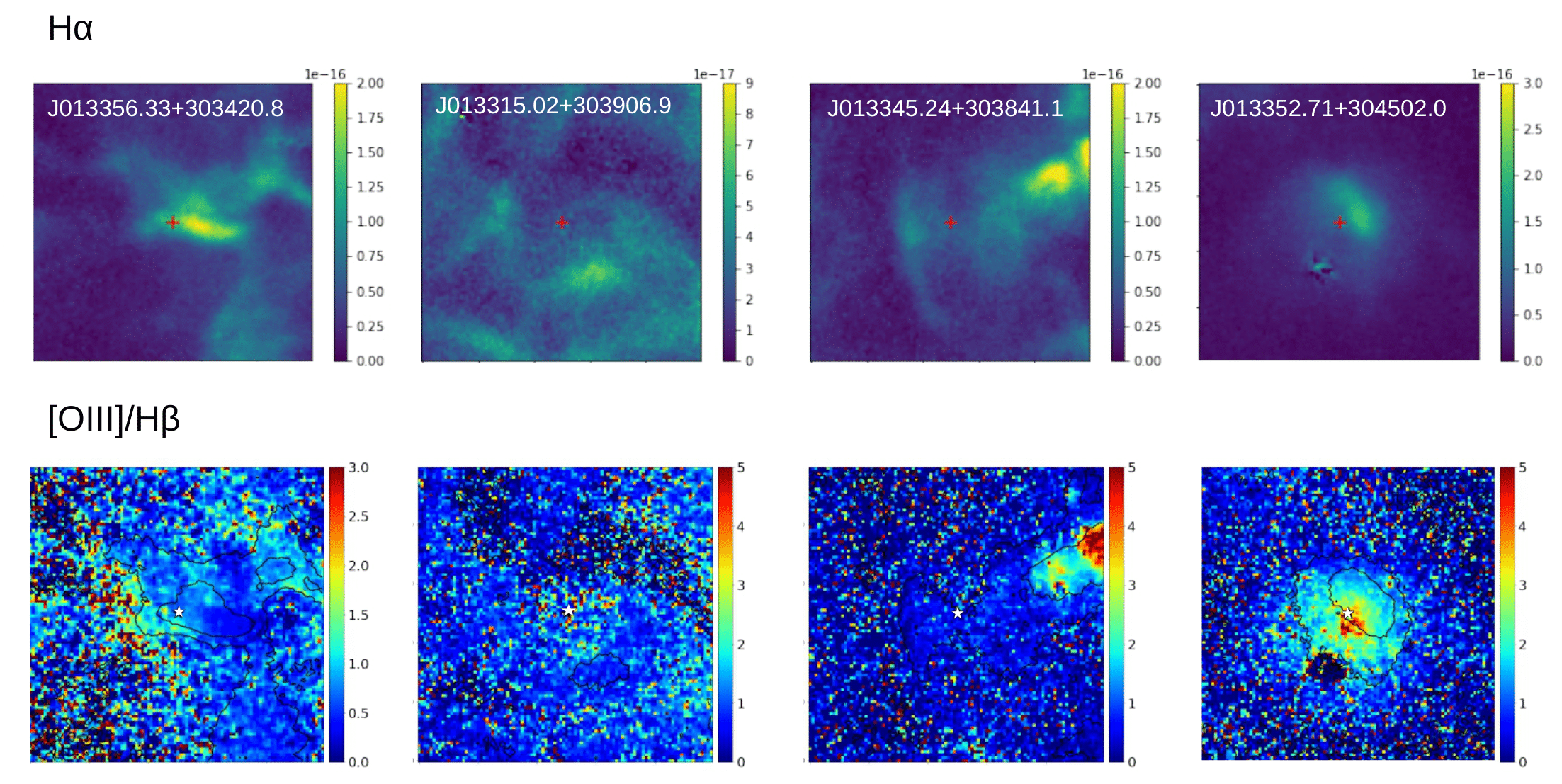}
\caption[Example of a WR bubble vs. other WR stars]{An example of four WR stars, with only one surrounded by a nebula confirmed as a WR bubble in this study (J013352.71+304502.0, see Section 4.1). The red cross (for H$\alpha$ images) and white star ([O{\sc iii}]/H$\beta$ images) in the middle indicates the location of the WR star. The fields represent a 32$''\times$ 32$''$ (130 pc on a side) region centered on the WR star. The flux is in erg/cm$^2$/s, as in all subsequent figures.}
\label{fig:exbubble}
\end{figure*}

As mentioned above, the [O{\sc iii}]$\lambda$5007/[O{\sc ii}]$\lambda$3726+3729 and [O{\sc iii}]$\lambda$5007/H$\beta$ ratios were particularly useful in identifying the WR nebulae. When comparing the surroundings of the 178 WR stars in these line ratios, we found, in most cases, a clear contrast between isolated stars with no specific associated nebula, or foreground/background nebulosity that does not seem to have a correlation with the WR star, and those that are associated with a circumstellar bubble that is partially impacted by or entirely blown by them. Indeed, in the former case, the star or its surroundings do not stand out from the background ionized gas while in the second case, we see a clear, usually circular structure perfectly centered on the WR location with a definite increase in the value of these line ratios, which most likely indicates that the gas has been ionized by the WR star. A clear example of such a case is illustrated in Figure \ref{fig:exbubble} for the nebula  J013352.71+304502.0,  together with other examples for which this is not the case.  \\

In this paper, all maps represent a field of 32$'' \times 32''$  ($\sim$ 130 pc $\times$ 130 pc) centered on a WR star, unless otherwise specified.\\

\section{Individual objects}

In this section, we briefly describe the characteristics of the nebulae we have classified as WR bubbles. 

\vskip 0.2truecm
\noindent  J013307.80+302951.1 -  An elongated nebula with the WNE star slightly off-center, not unlike NGC 6888 in the Milky Way. The influence of the WR star is particularly obvious in the [O{\sc iii}]/[O{\sc ii}] and velocity dispersion maps. This nebula is itself located in a much larger nebular complex with luminous OB stars, but the WR star is fairly isolated and is the brightest within the ring nebula.

\vskip 0.2truecm
\noindent J013312.95+304459.4 - An almost complete circular structure around the WN star, surrounded by a much larger and complex nebula.

\vskip 0.2truecm
\noindent  J013314.56+305319.6 - Patchy, arc-shaped nebula visible in all line maps.  The WR star is the sole bright blue star within the region in the HST images. The nebula is very prominent in [O{\sc iii}]/[O{\sc ii}] and [O{\sc iii}]/$H\beta$ ratio maps.

\vskip 0.2truecm
\noindent  J013326.60+303550.3 - By far the smallest nebula associated with a WR star in M33: coincident with the WN6 star, it is barely resolved ($\sim$ 4 pc in diameter), but slightly ellipsoidal, in all nebular images. The WR star is the sole bright star in the HST images. The nebula is very bright in [O{\sc iii}], but not detectable in [O{\sc ii}].

\begin{figure*}
\includegraphics[width=\textwidth]{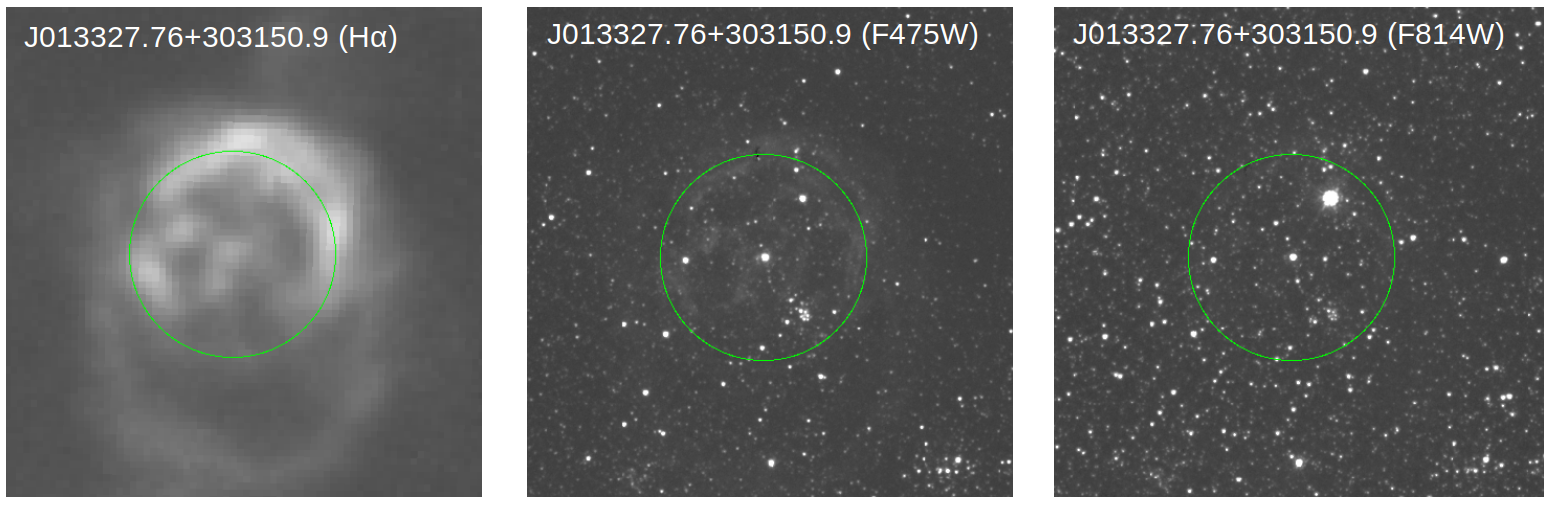}\par
\caption{H$\alpha$ image of the region surrounding J013327.76+303150.9 (left) obtained with SITELLE. Center and right : the same region as seen by the HST (with the F475W and F814W filters, respectively). The green circle is centered on the WR star and has a radius of 5" (20 pc). We note the presence of a small, dense cluster to the south-west and red a foreground star to the north-west of the WR star (see text). }
\label{fig:hstsw6}
\end{figure*}

\begin{figure*}
\includegraphics[width=0.7\textwidth]{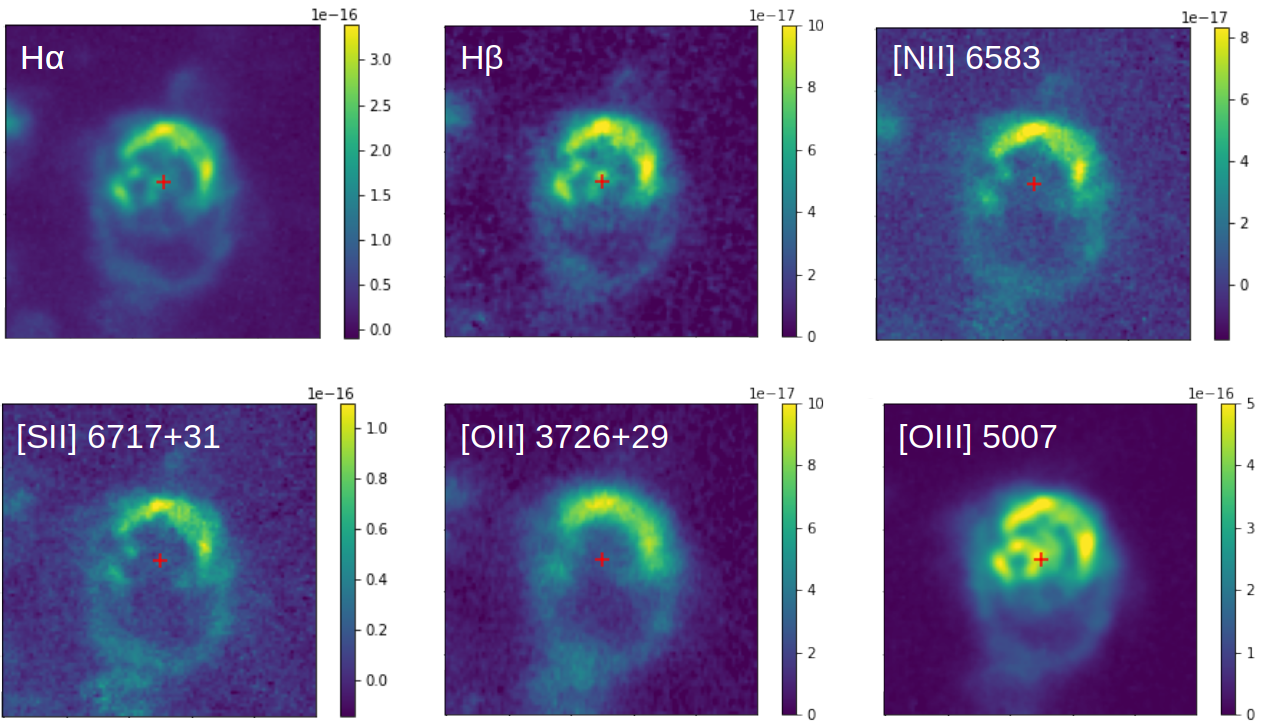}
\caption[Some spectral line maps of F4-1]{Some spectral line maps of the WR bubble J013327.76+303150.9. Top, from left to right : H$\alpha$,  H$\beta$, [N{\sc ii}] $\lambda$ 6583. Bottom, from left to right : [S{\sc ii}] $\lambda$ 6731 + 6717 (SN3), [O{\sc ii}] $\lambda$ 3726 + 3729 and [O{\sc iii}] $\lambda$ 5007. The red cross at the middle is where the central WR star of J013327.76+303150.9 is located. The images show a 32$''\times$ 32$''$ (130 pc on a side) region centered on the WR star. Flux units are erg cm$^{-2}$ s$^{-1}$.}
\label{fig:SW6}
\end{figure*}

\begin{figure*}
\includegraphics[width=\textwidth]{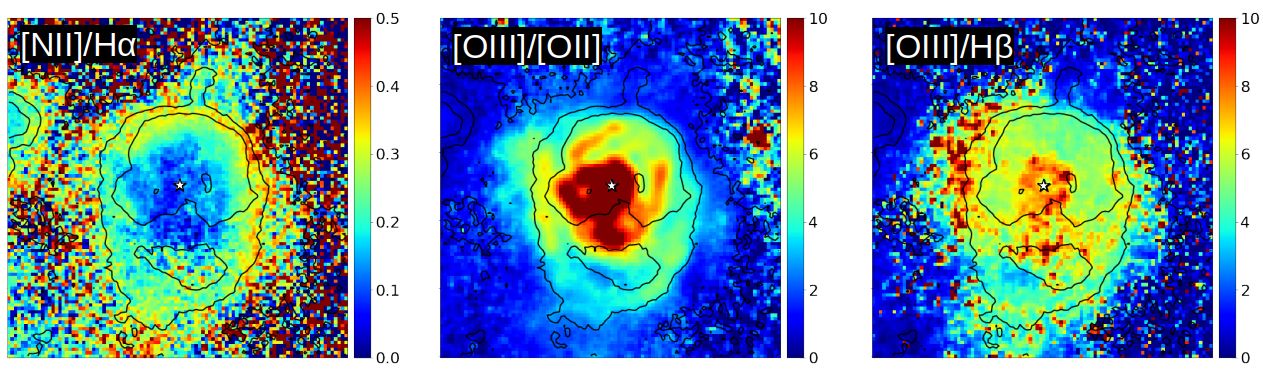}
\caption{Line ratio maps of J013327.76+303150.9 showing the same region as Figure \ref{fig:SW6}. From left to right : [N{\sc ii}] / $H_{\alpha}$, [O{\sc iii}]/[O{\sc ii}], [O{\sc iii}] / $H_{\beta}$.The overlaid contours are of the H$\alpha$ flux.}
\label{fig:ratioSW6}
\end{figure*}

\begin{figure}
\includegraphics[width=\columnwidth]{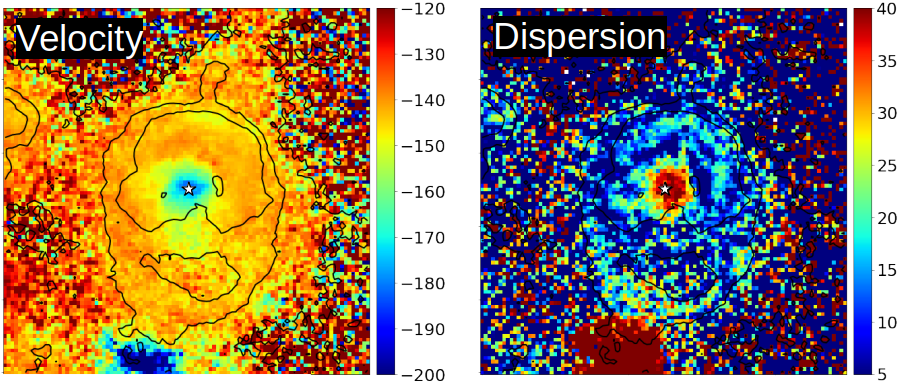}
\caption{Heliocentric velocity (left) and velocity dispersion (right) maps, in units of km/s, for J013327.76+303150.9. We note the presence of a SNR to the south-west of the star, with very large values of velocity dispersion.}
\label{fig:velo_f41}
\end{figure}

\begin{figure}
\includegraphics[width=\columnwidth]{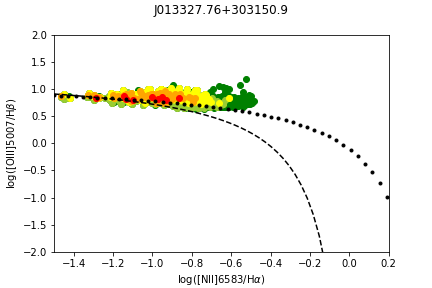}
\includegraphics[width=\columnwidth]{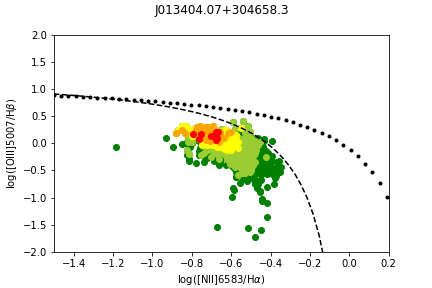}
\caption{BPT diagram of J013327.76+303150.9 (up) and J013404.07+304658.3 (down) with the borders between purely photoionized regions (below the lines) and shocked regions (above the lines) indicated by the predictions from \citet[]{2001} (dotted line) and \citet{10.1111/j.1365-2966.2003.07154.x} (dashed line). The red points are pixels in a 2-pixel radius from the WR star, orange is 5-pixel, yellow is 10-pixel, light green is 15-pixel and green is 20-pixel radius.}
\label{fig:bpt}
\end{figure}

\vskip 0.2truecm
\noindent J013327.76+303150.9 - This nebula surrounding a WN star (see Figures \ref{fig:hstsw6}, \ref{fig:SW6}, \ref{fig:ratioSW6}, \ref{fig:velo_f41},  \ref{fig:bpt}), discovered by \citet[]{article}, is morphologically very interesting and complex; we describe it in more detail than other nebulae to illustrate our process of selecting the most likely genuine WR nebulae. Its structure can be divided into three distinct parts. Immediately surrounding the WR star, a patchy circular nebula is found with a diameter of $\sim 3'' = 12.5$ pc in the Balmer lines.  It is very bright in [O{\sc iii}] but hardly visible in the low ionisation lines (Fig. \ref{fig:SW6}). It is seen more clearly in the [O{\sc iii}]/H$\beta$ map (right panel of Fig. \ref{fig:ratioSW6}). The second, larger circular nebula  (diameter of $\sim 10'' = 40$ pc) is much brighter in all the lines (except at its southern section), and is also centerered on the WR star. Figure \ref{fig:spectrum} shows an integrated spectrum of this nebula up to the outer edge of this second bubble. Finally, a fainter oval structure extends to the south, up to the supernova remnant L10-034 \citep{2018ApJ...855..140L}.

HST images of this nebula (Fig. \ref{fig:hstsw6}) are quite revealing. Because the F475W filter includes the strong [O{\sc iii}] $\lambda$ 5007 emission line, the nebula is visible, however weakly. The clumpy structure of the first two shells is visible, but the outer, oval nebula is too faint. These images also reveal the underlying stellar population, physically associated or simply in the line-of-sight. The WR star is clearly the brightest blue star inside both shells, and it is at the very center of both; it is with great confidence the only contributor to both the formation of the inner shell and its ionisation.   A very red, bright star is located at the edge of the outer shell. A verification of the Gaia catalogue\footnote{https://gea.esac.esa.int/archive/} reveals that this is actually a foreground objects located at a distance of 430 pc.   A few fainter blue stars are located inside the large shell, mostly close to its edge, and even a small cluster of a half-dozen faint blue stars is visible to the south-west of the WR star. If they are physically within this shell and not merely along the line-of-sight, they could collectively contribute to a small fraction of its ionisation but cannot have played a role in its formation through their winds.

    The line ratio maps (see Figure \ref{fig:ratioSW6}) indicate a weak [N{\sc ii}]/H$\alpha$ ratio in the center but the intensity increases from 0 to 0.35 around the edges of the bigger bubble with increasing distance from the star. 
    As for the [O{\sc iii}] / [O{\sc ii}] ratio map, it unambiguously demonstrates the
    predominant role of the WN star in the ionisation of the nebula.
    
    In Figure \ref{fig:velo_f41}, we present the velocity and velocity dispersion maps of J013327.76+303150.9. We clearly detect a difference in velocity of the small, inner nebula compared to the surrounding gas; its velocity dispersion is significantly higher than that of the larger ones, suggesting a direct, recent impact of the WR wind. \\

Figure \ref{fig:bpt} (left panel) presents a Baldwin, Phillips \&\ Terlevich diagram (BPT, \cite{1981PASP...93....5B}) of the region within a 20 pixel (27 pc) radius of the WR star, relating the [N{\sc ii}]/$H_{\alpha}$ and [O{\sc iii}]/$H_{\beta}$ line ratios. The BPT diagram was originally meant to study galaxies, in order to differentiate star-forming galaxies (where the emission mostly originates from classical H{\sc ii} regions) from shock-driven environments such as AGNs. As explained by \cite{2006MNRAS.372..961K}, \citet{2001} used a combination of stellar population synthesis model and photoionization models to create a theoretical prediction of a "maximum starbust line" (dashed line in Figure \ref{fig:bpt}). \citet{10.1111/j.1365-2966.2003.07154.x} added to that prediction a second line (dotted line in Figure \ref{fig:bpt}) dividing pure star-forming galaxies from objects with both AGN and star formation contributions in their spectra. 
Interpreting a BPT diagram of a small nebula in which each point represents an individual pixel is not trivial, but it can nevertheless be used to identify shock regions.

Along with the BPT diagram of J013327.76+303150.9, we show for comparison that of J013404.07+304658.3, another nebula ionised by a WN star, in the lower panel of Figure \ref{fig:bpt}. The latter shows the expected pattern : the dots are all below the theoretical lines, which indicates a purely photoionized region, and they shift to the right and bottom side as they get further from the WR star. As for J013327.76+303150.9, we also observe a shift to the right as one moves away from the WR star (see also the left panel of Figure \ref{fig:ratioSW6}), but the [O{\sc iii}]/$H_{\beta}$ line ratio does not dramatically decrease, on average. We note that the unusual aspect of the BPT diagram of J013327.76+303150.9 is not so much the increase of the [N{\sc ii}]/H$\alpha$ ratio as one moves away from the star (this is very common in WR nebulae, as well as PNe), but the presence of high [O{\sc iii}]/$H_{\beta}$ ratio regions outside of the two inner, nested ring nebulae: to their upper left and at the bottom. There is, at the moment, no clear explanation for this behaviour. \\

J013327.76+303150.9 is a clear case of intricate bubbles resulting from multiple phases of wind-ISM interactions, and it is tempting to sketch an evolutionary scenario based on our data. The outer, elliptical nebula with a lower surface brightness is approximately 15.5$'' \times 13.5''$, corresponding to a physical size of about 65 pc $\times$ 52 pc. Assuming an expansion velocity of 15 km/s, its lifetime is thus approximately 4 million years, quite compatible with the lifetime of an O star on the main sequence. Therefore, we conclude that this component of the nebula can most likely be attributed to the action of the current WR star when still on the main sequence.  The brighter, spherical nebula could then be the result of the interaction between the stellar wind of the now-WR star with gas ejected in a previous evolutionary stage (Luminous Blue Variable or Red Supergiant, for example). The fast WR wind has penetrated and even overtaken the slower gas ejected in the previous phase, as can be clearly seen in the [O{\sc iii}]/H$\beta$ maps. We note an obvious asymmetry in the flux of its intensity profile: it is very bright in the northern part, where it almost overlaps (interacts?) with the fainter elliptical structure, but much fainter albeit obviously present in the south, where the edge of the elliptical bubble is much further away. Although we are unable to verify this, it could possibly be explained if the star was moving towards the north. Finally, in this scenario, the patchy circular nebula near the WR star could correspond ejecta from the current WR wind that has been excited by the passage of the inward-facing reverse shock after the interaction of the WR wind with the surrounding medium, leading to a bright nebula. If this is the case, it would be more likely to show peculiar chemical abundances, including helium and nitrogen enhancements and oxygen depletion. Our data are not capable of confirming this, but long-slit or 3D spectra including He lines as well as auroral lines of oxygen and nitrogen ([O{\sc iii}] $\lambda$ 4363, [N{\sc ii}] $\lambda$ 5755) would be. This interpretation could also possibly explain the higher velocity dispersion observed for this gas (see Figure \ref{fig:velo_f41}).

\vskip 0.2truecm
\noindent  J013334.04+304117.2) - This WN star is adjacent to the giant H{\sc ii} region NGC 595. It lies at the northern edge of a clear, almost complete circular nebula with a diameter of 26 pc. The WR star is however in the center of a smaller, clumpy structure that particularly stands out in the [O{\sc iii}] image, obtained with a better seeing, as well as in the [O{\sc iii}]/$H\beta$ and [O{\sc iii}]/[O{\sc ii}] maps. HST images show that there are no other bright stars, red or blue, within the large bubble. One explanation for this nested structure might be that the larger ring is the bubble blown during an earlier phase and the smaller one the more recent ejecta filling the previously hollow bubble. There is, to our knowledge, no specific spectral subtype for this star, but the [O{\sc iii}]/[O{\sc ii}] and [O{\sc iii}]/$H\beta$ maps show the lowest values of all the confirmed WR bubbles, suggesting that this is actually a late-type WN star.

\vskip 0.2truecm
\noindent J013335.73+303629.1 - Complete circular ring, with the WC star centraly located. HST images show two bright stars within 0.5$''$ from each other, so the unique role of the WR star in shaping and ionizing the nebula cannot be assessed with certainty. This nebula is adjacent to a bright supernova remnant located to the south-east that stands out in the velocity dispersion and line ratio maps.

\vskip 0.2truecm
\noindent  J013339.52+304540.5 - This very bright B0.5Ia + WNE star is at the center of a large ($15 \times 30$ pc) hourglass-like nebula, visible only in the low ionization lines: it does not show up at all in the [O{\sc iii}] map. No other bright star is visible in HST images within the nebula. 

\vskip 0.2truecm
\noindent J013340.69+304253.7 - This nebula is located inside a large and complex nebular structure that includes many bright blue stars according to the HST images. A clear arc is seen to the south-east of the WN star; it is particularly prominent in the [O{\sc iii}] image as well as the [O{\sc iii}]/[O{\sc ii}] and [O{\sc iii}]/$H\beta$ maps. Within this arc-like structure, the WR star is by far the brightest and most centrally located.

\vskip 0.2truecm
\noindent J013341.65+303855.2 - This is an elongated ring visible in the Balmer and [O{\sc iii}] lines that is  open to the south. In the low ionisation lines, the ring is more circular and extended. The WN star is centrally located, but HST images show stars of similar magnitude close to the northern and southern edges of the ring that could also contribute to the ionization and excitation of the gas.

\vskip 0.2truecm
\noindent J013341.91+304202.7 - Very elongated (bipolar?) nebula. HST images show 2 bright blue stars of approximately equal brightness, separated by 0.7$''$, at the center of this nebula; the WN star is the eastern member of the pair. Thus we cannot attribute unambiguously the morphology nor the ionization of the nebula solely to the WR star.

\vskip 0.2truecm
\noindent J013342.53+303314.7 - This interesting nebula surrounds the WC4/5 star, also identified as MC44 \citep[for an illustration of its spectrum]{Schild90}, that is clearly the only bright star visible in the HST images within its boundaries.  Its horseshoe shape, with an opening to the east, could be due to an asymmetry in the density of the ISM with which the stellar wind has interacted. The genuine WR nebula is very likely the smaller complete ring within the larger one, particularly visible in H$\alpha$. In view of the early spectral type of its central star, which suggests a high effective temperature of T$_{\rm eff} \sim 80 - 110$ kK \citep[]{Sander2012}, the rather low [O{\sc iii}]/[O{\sc ii}] and [O{\sc iii}]/$H\beta$ line ratios ($\sim 1$) are surprising.

\vskip 0.2truecm
\noindent J013344.40+303845.9 -  This nebula consists of a ring open to the south-east with the WC4 star close to the northern edge where the nebula is brighter. There is a fainter star 0.4$''$ to the east in HST images.

\vskip 0.2truecm
\noindent J013345.99+303602.7 - Surrounding the WN4 star also known as MC46, this nebula has already been studied by \cite{article} who found a bright, smaller loop centered on the WR star, itself located in a bigger, asymmetric bubble open to the east. The latter is visible in all spectral line maps of interest, but the smaller bubble is only clearly visible in the $H\alpha$ and [O{\sc iii}] maps. The outer bubble is particularly well defined in the low ionisation [N{\sc ii}] and [S{\sc ii}] lines. The inner bubble, brighter in the south-west direction, displays a significantly different velocity and a higher velocity dispersion than the outer bubble: it is obviously younger and still influenced by the WR wind. The HST images confirm that the WR star is the only hot star in the vicinity and that it is located right in the center of the bubble. The [O{\sc iii}]/[O{\sc ii}] and [O{\sc iii}]/$H\beta$ ratio maps clearly show the effect of the WR star on the ionization of the bubble : the ratios gradually decrease with increasing distance from the star in a pattern that is perfectly centered and symmetrical around it. The values of these ratios are among the highest observed for all the bubbles we have studied, which is coherent with a WN4 spectral type associated with this WR star.

\vskip 0.2truecm
\noindent J013346.80+303334.5 -  The WN6/WC4 star coincides with the brightest part (length $\sim$ 11 pc) of the edge of a larger structure that extends to the north, itself encompassed within a larger nebula to the west. This section of the nebula stands out in the [O{\sc iii}]/[O{\sc ii}] map indicating that the enhancement of the 11 pc structure local to the WR star seen in H$\alpha$ is most likely due to the present action of the star.

\vskip 0.2truecm
\noindent J013347.83+303338.1 - The WR star is the sole bright blue star within the nebular region. The WR wind has probably carved and ionised this "hole" in the surrounding ISM. The nebula appears as a clumpy arc open to the North.

\vskip 0.2truecm
\noindent J013347.96+304506.6 - In this case, the WN star is close to the northern edge of this ring, which is open to the south-east. The ring is much better defined in the [O{\sc iii}] image, obtained with a better seeing. The velocity dispersion inside the ring is significantly larger ($\sim 35$ km/s) than that of nearby H{\sc ii} regions ($\sim 18$ km/s). The HST images show that the WR star is the brightest star inside the ring.

\vskip 0.2truecm
\noindent  J013350.71+305636.7 - Nested inside a large and bright H{\sc ii} region \citep[number 637 in the catalogue of][]{Courtes87}, this bubble is circular with the WN star perfectly centered within it. Another bright hot star is seen at the northern edge of the bubble, and a group of fainter ones close to its south-east border: the WR star is clearly a member of a young cluster of massive stars linked to the large H{\sc ii} region. But the prominent role of the WR star in the ionisation and kinematics of the central bubble is particularly obvious in the velocity and velocity dispersion maps, as well as in the [O{\sc iii}]/[O{\sc ii}] line ratio maps.

\vskip 0.2truecm
\noindent  J013351.84+303328.4 - Circular ring-shaped nebula seen in Balmer and low-ionization lines but not so clearly in [O{\sc iii}]. Many red stars from the global population of M33 are seen in the HST visible images, but only one, the WC6, in the near UV images, which is centrally located.

\vskip 0.2truecm
\noindent J013352.71+304502.0 - This is a particularly interesting case, as it displays a very bright arc to the north-west of the WC4 star.  A much fainter, narrow arc is also visible to the south; this sets the limits of the WR nebula listed in Table\,\ref{tab:example_table}. The WR star is not perfectly at the center of the bubble,  itself located in a more extended nebula but it is the only bright star in the region, according to HST images.  The most important influence of the WR star becomes clear in the [O{\sc iii}]/[O{\sc ii}] and [O{\sc iii}]/$H\beta$ ratio maps.

\vskip 0.2truecm
\noindent J013357.20+303512.0 - This is a large, incomplete ring open to the north-west and that is much brighter in the east.The most important influence of the WR star becomes clear in the [O{\sc iii}]/[O{\sc ii}] and [O{\sc iii}]/$H\beta$ ratio maps.

\vskip 0.2truecm
\noindent J013358.69+303526.5 - The point source seen in H$\alpha$ at the center of the patchy structure is stellar in origin as it is much wider that expected for a nebula. HST images suggest that the central source could be a tight cluster, so we cannot ascertain whether the WN star is the sole contributor to the formation and ionization of the nebula.

\vskip 0.2truecm
\noindent J013359.39+303337.5 - Close to the edge of the field, the large ring (diameter $\sim 45$ pc) is not completely visible in our data, but the WC6 star is more likely to be linked with the smaller ring which is more prominent on its southern edge, close to the WR star. The [O{\sc iii}]/[O{\sc ii}] and [O{\sc iii}]/$H\beta$ ratio are exceptionally low.

\vskip 0.2truecm
\noindent J013404.07+304658.3 - The ionizing star, of WN6 subtype, is one of the five candidates mentioned by \citet[]{Neugent_2011} that have been confirmed as WR stars later-on by \cite{2014}.  It is by far the brightest star within the large (diameter $\sim$ 45 pc) nebula, and it lies within a smaller circular ring (13 pc in diameter) most clearly visible in [O{\sc iii}] and very likely to be physically associated with the star. This inner ring also clearly stands out in the [O{\sc iii}]/[O{\sc ii}] and [O{\sc iii}]/$H\beta$ ratio maps.
   
\vskip 0.2truecm
\noindent J013406.80+304727.0 - Although the WN7 star is probably not the sole contributor to the formation and ionization of the large arc-like structure, open to the west (HST images show a loose star cluster inside the arc but the WR star is by far the brightest), it is surrounded by a much smaller (diameter $\sim$ 15 pc), incomplete shell that particularly stands out in the line ratio maps. We also note the presence of a bright, dense ionized spot at the southern edge of the large arc, 5$''$ to the south of the WR star; the continuum source is not a single star according to the HST images, but rather a tight cluster.

\vskip 0.2truecm
\noindent J013407.85+304145.1 - Sole bright star within an incomplete circular structure according to the HST images, the WN star is slightly off-center, closer to the much brighter, southern edge of the nebula, which also shows velocity dispersions close to 40 km/s and high values of [O{\sc iii}]/[O{\sc ii}].

\vskip 0.2truecm
\noindent J013416.28+303646.4 - This nebula, with the WC6 star located very close to the center, comprises a small circular ring nebula (13 pc in diameter) encircled into a larger one open to the East. HST images of this field show that the WR  is in a rich stellar field, but is by far the brightest star within the limits of the bubble. The Ofpe/WN9 star J013416.07+303642.1 clearly stands out in the Balmer lines images as well as the velocity dispersion one: it is located just outside the south-west edge of the WR nebula. Our SITELLE spectrum of this star, which is 2 magnitudes brighter than the WC6 star, shows a strong, but relatively narrow, H$\alpha$ line and a weaker HeI$\lambda$6678, characteristic of cool, transition-type Of/WN stars. The maps do not show any obvious impact of this star on either the WR nebula nor the larger bubble to its south.

\vskip 0.2truecm
\noindent J013417.21+303334.7 - Arc-like structure mainly visible (in particular in the [O{\sc iii}]) to the east of the WN3 star, which is the brightest star inside the nebula in HST images . The arc stands out very well in the HST/ACS F475W image (a broad filter which includes the [O{\sc iii}] line).

\vskip 0.2truecm
\noindent J013419.16+303127.7 - Arc-like structure mainly visible to the north of the WN3 star. Although the WR star is at the edge of the arc in SITELLE data, HST/ACS F475W image shows that the brighter part of the arc in [O{\sc iii}] is 3.5 pc to the NW of the star. There is also a brighter, larger arc further away.

\vskip 0.2truecm
\noindent J013419.68+303343.0 - Clear bubble within a large and complex structure. HST images show two stars of equal brightness within 0.5$''$ of each other, so we cannot ascertain the exact role of the WR star.
     
\vskip 0.2truecm
\noindent J013423.02+304650.0 - The ring nebula is part of a larger, complex nebular structure. It is not well defined in the H$\alpha$ image and is almost invisible in the low ionization lines, but stands out in the  [O{\sc iii}] image and even more in the and  [O{\sc iii}]/[O{\sc ii}] ratio map: the WN4 star clearly has a strong ionizing impact on this nebula. HST images show the presence of two much fainter stars close to the WR star, unlikely to have any influence on the ring nebula.
 
\vskip 0.2truecm
\noindent J013438.98+304119.8 - The WC4 is close to the brightest, arc-shaped part of an extended ellipsoidal nebula. The [O{\sc iii}]/[O{\sc ii}] and [O{\sc iii}]/$H\beta$ ratio maps testify to the ionizing power of the WR star. HST images show the presence of another blue star located 0.8$''$ to the north-east of the WR star but with one magnitude fainter; its flux and location make it unlikely to contribute to the shape or the ionization of the nebula.

\vskip 0.2truecm
\noindent J013443.51+304919.4 - The WN4 star is well placed within a very clear ring, open to the south-west. This ring is even more prominent in the [O{\sc iii}] image but completely disappears in the [O{\sc ii}] image. The [O{\sc iii}] image, obtained with a better seeing than the H$\alpha$ one, shows the presence of a smaller (9 pc diameter), complete ring around the WR star, which is likely to be the one blown during the WR phase. Both these inner and outer rings are surrounded by an even larger one ($80 \times 40$ pc) visible more clearly in lower ionization lines. No HST image of this region is available. We note the presence of a bright, foreground star to the East of the nebula.

\section{Discussion}

\subsection{Statistics and correlations}

The information on all confirmed WR nebulae we identified in this work is summarized in Table \ref{tab:example_table}. We find 33 WR stars to be surrounded by nebulae that seem to be formed and/or ionized mainly by them. 
Considering that there is no uncertainty on the total number of studied WR stars (since the survey of \cite{Neugent_2011} is considered to be 95\% complete, and even if there were more WR stars, we considered the current population of 178 observed WR stars only), we find that less than a fifth of all WR stars in the observed fields are associated with a {\it bona fide} WR bubble.
a detection rate of 19 \%. 
\par

\begin{table*}
	\centering
	\caption{Spectral subtypes of observed WR stars}
	\label{tab:stats}
\begin{tabular}{ |l|cc| }
   \hline
     & WR stars in SITELLE fields & WR stars with bubbles \\
   \hline
    WNE & 65 &  13 \\
    WN & 24 &  9 \\
    WNL & 39 & 2 \\
    WN/WC & 3 & 1 \\
    WCE  & 37 &  7\\
    WC & 8 &  1 \\
    WCL & 2 & 0 \\
   \hline
    \textbf{Total} & \textbf{178} & \textbf{33} \\
       
\hline
\end{tabular}
\end{table*}

As shown in Table \ref{tab:stats}, among the 33 WR stars with confirmed bubbles, 24 are classified as WN, eight as WC and the last one, J013346.80+303334.5, is a transition-type WN/WC star. Transition-type stars, as indicated by their names, show strong lines in both carbon and nitrogen and are thought to be in a transition stage between the WN and WC evolutionary phases \citep{1989ApJ...344..870M}. If we adopt WN6 and hotter WN stars as "early", then we also find that among the 24 WN-type stars with bubbles, 13 of them are of WNE type and  two of them are WNL. The other nine stars are only classified as "WN" type with no further details. As for the WC stars, seven of them are classified WCE and the other one is only classified as a WC-type star. To compare with the total population of WR stars in M33, out of the 178 observed stars, 128 of them are classified as WN and 47 of them as WC stars, with three transition-type stars. For the WN stars, 65 of them are further classified as WNE and 39 as WNL. As for the WC stars, there are 37 WCE and two WCL. We find WC/WN = 0.37 for the observed WR population in M33 and WC/WN = 0.33 for the ones with confirmed bubbles around them. Thus, we find no clear correlation between the spectral type and the presence of bubbles around WR stars. \\

Although the bubbles are very different in appearance, we have distinguished three morphological categories based on the global shape of the nebula. The first appear as an "arc", that is an incomplete circular shape for which the brightest parts are usually closest to the WR star. These could be similar to the Galactic nebula MR100 \citep[around the star WR134; ][]{1993ApJS...85..137M} or NGC3199 \citep[around the star WR18; ][]{1994ApJS...95..151M}, for instance.  Some examples of arc-like nebulae are J013417.21+303334.7 and J013340.69+304253.7. We separate those from the complete "ring"-like nebulae such as J013327.76+303150.9 or J013416.28+303646.4 that are more similar to NGC2359 surrounding the star WR7 in the Galaxy \citep{1993ApJS...85..137M}. The last category is the "clumpy" type where the nebula does not have an apparent ring-like or arc-like shape but is seemingly centered on the WR star and the ratio maps indicate that the WR star has an impact in at least the ionisation or excitation of the gas. We find 16 bubbles to be ring-shaped, 12 arc-shaped, four clumpy, and one bipolar. 11 of the  16 ring nebulae, three out of the  four clumpy ones, and  nine of the 12 arc-shaped ones are found around WN-type stars.  Nine of the bubbles we found around WR stars in M33 are nested, with a smaller bubble inside a bigger one, and seven of those are associated with WN-type stars. For the clumpy-looking nebulae around WN-type stars, one is associated with a  WNE star and the other two do not have a more specific classification. One is found around a WN6/C4 star.\\

The dimensions of the WR bubbles we identified have first been determined by fitting, by eye, an ellipse  on the region of the bubble that appears the brightest.
Since the ellipse has been fitted by eye, we estimate our uncertainty on the diameter to be around 5 px $\sim$ 7 pc. The mean diameter for simple bubbles or the smallest one in the case of nested ones is 19 pc. We find 19 pc for WN stars only, and  21 pc for WC stars. We therefore find no real correlation between the bubble size and the spectral type of the central star. \\

\begin{figure*}
    \includegraphics[width=\linewidth]{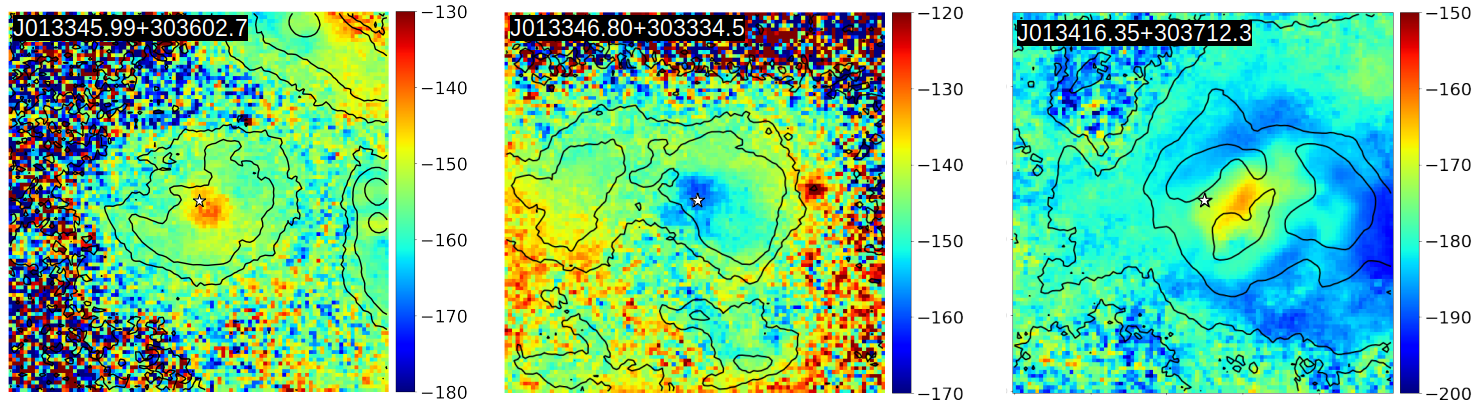}\par 
\caption{Examples of velocity maps for three WR bubbles : J013345.99+303602.7, J013346.80+303334.5 and J013416.35+303712.3. The velocity is heliocentric and in km/s.}
\label{fig:velocity}
\end{figure*}

We expected by examining the kinematics of the bubbles, to be able to reveal their expansion, which should translate by a higher velocity dispersion at their center and decreasing with distance from the star, or by changes in the velocity itself across the nebula. But surprisingly the velocity is globally uniform in most cases, with rare exceptions (Figures \ref{fig:velo_f41} and \ref{fig:velocity}). 

Finally, most WR bubbles are clearly visible in H$\alpha$ but the individual impact of the WR star is usually more distinguishable in [O{\sc iii}] images. Whenever there are entangled bubbles, that is a smaller, circumstellar bubble as defined by \cite{2021} contained within a bigger, so-called interstellar bubble, the contrast between them is also the most noticeable in the [O{\sc iii}] line-map, as is the case for J013327.76+303150.9 and J013404.07+304658.3 for example. The brightest part of the arc is always the one closest to the WR star. Although there are a few exceptions, the [O{\sc iii}]/[O{\sc ii}] and [O{\sc iii}]/H$\beta$ ratio seems to be the strongest for bubbles around WNE type stars (See Figure A1).   \\

We find only a few WR stars to be completely isolated from the presence of any hydrogen emission in its vicinity, although it might also be the case for some others since the gas seen in their surroundings could be located along the line-of-sight but not physically related to the star. The lack of surrounding interstellar medium or the dissipation of the wind-blown bubble could explain these rare cases.

\subsection{Evolutionary scenario}

According to \cite{1981ApJ...249..195C}, and, more recently, \cite{Chu_2016}, the first impact a hot star has on its surroundings is by radiatively exciting it (R-type nebula). Then, the high luminosity leads to the formation of stellar winds, which in turn can lead to the surrounding material being blown as a W-type nebula during the main sequence phase when the star is of O spectral type although the low expansion velocities render these difficult to detect. At this point, the star has a fast but not too dense stellar wind. The next evolutionary phase, either LBV or RSG (depending on the star's initial mass) consists of a slower but denser wind, gradually filling the cavity created by the wind-blown bubble composed of interstellar material pushed by the O-star wind. As the star evolves to its final WR stage, the stellar wind pushes in turn the circumstellar nebula accumulated in the previous phase, mostly composed of products of nuclear reactions that took place at the core of the star during the previous phases. That is when we see nested bubbles as in the case of J013327.76+303150.9. For a more detailed account of the the structure and evolution of (often multiple) shells surrounding WR stars, we refer the reader to \citet{1995AJ....109.1839M}.\\

In line with the presently adopted evolutionary scenarios of Wolf-Rayet stars \citep[e.g.][]{Crowther_2007}, the most massive stars first evolve to the WN stage (late-type, then early) which then evolve into WC stars. The study of Galactic WR nebulae conducted by \cite{Chu_2016} leads to the conclusion that most WR nebulae are found around WN stars and that there is a direct correlation between the spectral type of the central star and the morphology of the surrounding nebula : mostly clumpy, and dominated by stellar ejecta (E-type) for WNL stars, swept-up interstellar material for WNE stars, and usually incomplete circumstellar bubbles (comparable to our 'arc' type morphology) around intermediate WN stars (WN5-6).\\

In our study of WR nebulae in M33, we do not find such a clear correlation for the presence of bubbles around WN compared to WC stars as we find that the WC/WN ratio is the same for WR stars with bubbles than for the global WR population. We were unable to search for a correlation between the spectral type of the central star and the morphology of the bubble, as although most of our clumpy nebulae (three out of four) are found around WN stars, in two cases, we do not know if they are of late or early type based on the present classification.  Nine out of the 12 arc-shaped nebulae are found around WN stars but this ratio is the same as the ratio of WN stars in the global WR population of M33. However,  as can be seen in Figures A1 and A2,we do seem to find that the nebulae around WNE-type stars are more highly ionized than those around WNL and WC stars (with a few exceptions), which would support the fact that they are younger and hotter bubbles that probably cool down as they expand and dissipate.   \\

\subsection{Comparison with previous studies}

\subsubsection{Comparison with the Milky Way}
    
    The surveys conducted by \cite{1994ApJS...93..229M} and \cite{1994ApJS...95..151M} indicate that of the 114 Galactic WR stars they studied, more than a quarter have associated ring nebula. 32\% of the WR stars observed are of WN-type and 20\% of WC-type. Of the 38 possible ring nebulae found, 22 are around WN stars, 13 around WC stars, one around a triplet of Wolf-Rayet stars, one around a WO star and the last one around a transition WN/WC star. Preference toward rings being observed around WR + OB binaries and WN type stars are noted. Indeed, when binaries and groups of WR stars are removed, single WN stars are twice more likely to be surrounded by a ring nebula than single WC stars, according to their study. They have also found several intricated rings like the ones we found in our survey. The same explanation as the one we suggested is given for this phenomenon, that is the possible multiple ejection and wind stages that may have taken place during the evolution of the massive star. Note that seven of the nine nested nebulae we identified are found around WN stars. This could possibly be explained by the merger of the nested bubbles as the star evolves to later stages. \\ 
    
    In the latest paper of this survey \citep{1997ApJ...475..188M}, the authors find a final detection rate of 35\%, with 40 bubbles out of 114 observed WR stars. In comparison, we find  19\% of WR stars observed in M33 to have an associated bubble, which is much smaller. However, we note that our criteria are more strict as we only selected cases where we were convinced that the impact of the star in its current WR phase on the ionization of the bubble or surrounding gas was clearly (with a higher [O{\sc iii}]/[O{\sc ii}], for example). Also, the WC/WN ratio for all observed WR stars in M33 versus only those with a nebula leads us to conclude that we are unable to confirm the same preference observed around WN stars in their case. Note, however that we have little information on the binarity of the WR stars in M33 and therefore we cannot quite generate the same statistics.They also find nebulae associated with WC stars to be generally larger, which is not the case of nebulae we found in M33.  We find that the nebulae in M33 to be of similar size, on average, as those in the Galaxy (see their Table 2). However, if we consider only their W and E type nebulae, their average diameter is $\sim$ 12 pc (see their Figure 12), which is smaller than our average diameter of 19 pc. We have detected only one nebula significantly smaller than 10 pc (around J013326.60+303550.3), analogous to NGC 6888 or NGC 2359 in the Milky Way. This is unlikely to be caused by an observational bias since we detect all planetary nebulae known in M33 with the same data cubes, unless some of these small ejecta-type WR bubbles are hidden by the rim of a much brighter H{\sc ii} region (which is the case of the nebula associated with WR134, for example). \\

     The results in the Galaxy seem to support the idea that the presence of a nebula around a WR star is a sign of youth, as WN stars are thought to be, on average, younger than WC stars. Bearing in mind the caveat that we are unable to correct for binary effects, our results do not support this hypothesis. It remains to be seen if M33 is an isolated case in this regard and if so why that is.

    \subsubsection{Comparison with the LMC}
    
    \cite{2021} have conducted a survey of the WR nebula population in the LMC.  They  found small bubbles (<50 pc diameter) around 12\% of the 154 known WR stars (15\% of the ones outside 30 Dor) , which is not too far from our detection rate of 19\%. For comparison, the mean diameter of bubbles in M33 is 19 pc ($<$ 43 pc).  Approximately 82\% of the WR population in the LMC are WN stars and 88\% of the ones with an associated bubble are of WN-type. In the case of M33, we find that 24 out of 33 WR nebula, (73\%) are found around WN stars, which is slightly lower.  WN stars in our sample represent 72\%\ of the total (128 out of 178) population. Our results are thus quite similar to those of the LMC. \\
    
    Their study also focuses on the differences between interstellar (usually bigger, blown during the MS phase) and circumstellar (smaller, ejected circumstellar matter blown during the WR phase) bubbles. To distinguish which type a nebula detected around a WR star is, one could use abundance analysis, and this was done for three of the bubbles observed by \cite{2021}. Only that found around the star WR 19 was confirmed as a circumstellar bubble using the N/O ratios. For the intertwined bubbles in M33, our observations also seem to support an evolutionary scenario implying multiple fast wind phases sweeping first the ISM, then the circumstellar material ejected by the star and showing the products of reactions that have taken place at its core. \\

\section{Conclusions}

We have studied the surroundings of 178 of the 211 confirmed WR stars in M33 \citep{Neugent_2011,2014} in order to search for the presence of circumstellar nebulae using hyperspectral data cubes obtained with SITELLE at the Canada-France-Hawaii telescope complemented by images from the Hubble Space Telescope's Advanced Camera for Surveys. \\

Using the criteria that we estimate to be the most reliable, we were able to identify 33 nebulae around WR stars that satisfy at least two of our three criteria and that we thus classify as WR bubbles. This results in a  19\% detection rate, which is lower than the Galactic rate but slightly higher than those in the LMC. \\ 
 
Twenty-four of our 33 nebulae are found around WN-type stars, 8 around WC stars and one around a transition-type star. We find no clear correlation between the spectral type of the central star and the presence of a surrounding nebula since the WC/WN ratio is the same for the total known WR population and the WR stars associated with bubbles. However, we are unable with the data in hand to quantify the role of binaries in these statistics. We categorize the bubbles into three morphological categories : clumpy, an incomplete arc, or ring-shaped.\\

According to currently adopted evolutionary scenarios, massive stars first blow a cavity in their surrounding interstellar medium during the main-sequence phase, which then is filled with a slow, dense wind during an intermediate evolutionary phase. This material is then blown during the WR stage, creating two nested bubbles until both bubbles merge. We identified such a structure only around eight WR stars in M33, mainly WN stars, suggesting that the bubbles merge as the star evolves. \\

The absence of visible nebulae clearly associated with a majority of WR stars, in M33 as in other Local Group galaxies, needs to be better understood; this has as much to do with visibility (which is enhanced, for instance, when the WR wind compresses the surrounding ejecta) as with evolution. Now that dozens, if not hundreds, of such cases are known, a dedicated study is certainly warranted, combining observations and numerical simulations. \\

A literature search of studies on Galactic bubbles and the LMC counterparts also helps us to confront our results with that of different environments and confirm some of our hypothesis. Our study completes the previous survey of WR nebulae in M33 conducted by \cite{article} with the addition of 22 confirmed bubbles using our criteria. \\

This work only presents an overview of the WR nebulae population in M33 and of their general characteristics.
Further progress in the study of M33's WR nebulae would require dispersive spectroscopy using larger telescopes. First, to perform a quantitative analysis of the ionising star's spectrum to measure their temperature and precise spectral subtype. Second, to measure the chemical abundances across the nebulae using the faint auroral lines of [O{\sc iii}]$\lambda$4363 and [N{\sc ii}]$\lambda$5755.\\

\section*{Acknowledgements}
This paper is based on data obtained for SIGNALS, a large program conducted at the Canada-France-Hawaii Telescope (CFHT), which is  operated  by  the  National  Research  Council of Canada,  the  Institut  National  des  Sciences  de  l'Univers  of the  Centre  National  de  la  Recherche  Scientifique of France and the University of Hawaii.
The observations  were obtained  with  SITELLE,  a  joint project  of  Universit\'e  Laval,  ABB,  Universit\'e  de  Montr\'eal and  the  CFHT, with support from the Canada Foundation for Innovation, the National Sciences and Engineering Research Council of Canada (NSERC) and the Fonds de Recherche du Qu\'ebec - Nature et Technologies (FRQNT). The authors wish to recognize and acknowledge the very significant cultural role that the summit of Mauna Kea has always had within
the indigenous Hawaiian community. We are most grateful to have
the opportunity to conduct observations from this mountain.
NSL, LD and CR are grateful to NSERC and FRQNT for financial support.

\section*{Data Availability}

Data cubes from the SIGNALS survey are publicly available at the Canadian Astronomy data Centre (https://www.cadc-ccda.hia-iha.nrc-cnrc.gc.ca/en/cfht/). Maps extracted from these cubes and presented in this article will be shared on reasonable request
to the corresponding author.



\bibliographystyle{mnras}
\bibliography{WRM33} 




\appendix

\section{Line ratio variations}

\begin{figure*}
    \includegraphics[width=0.9\linewidth]{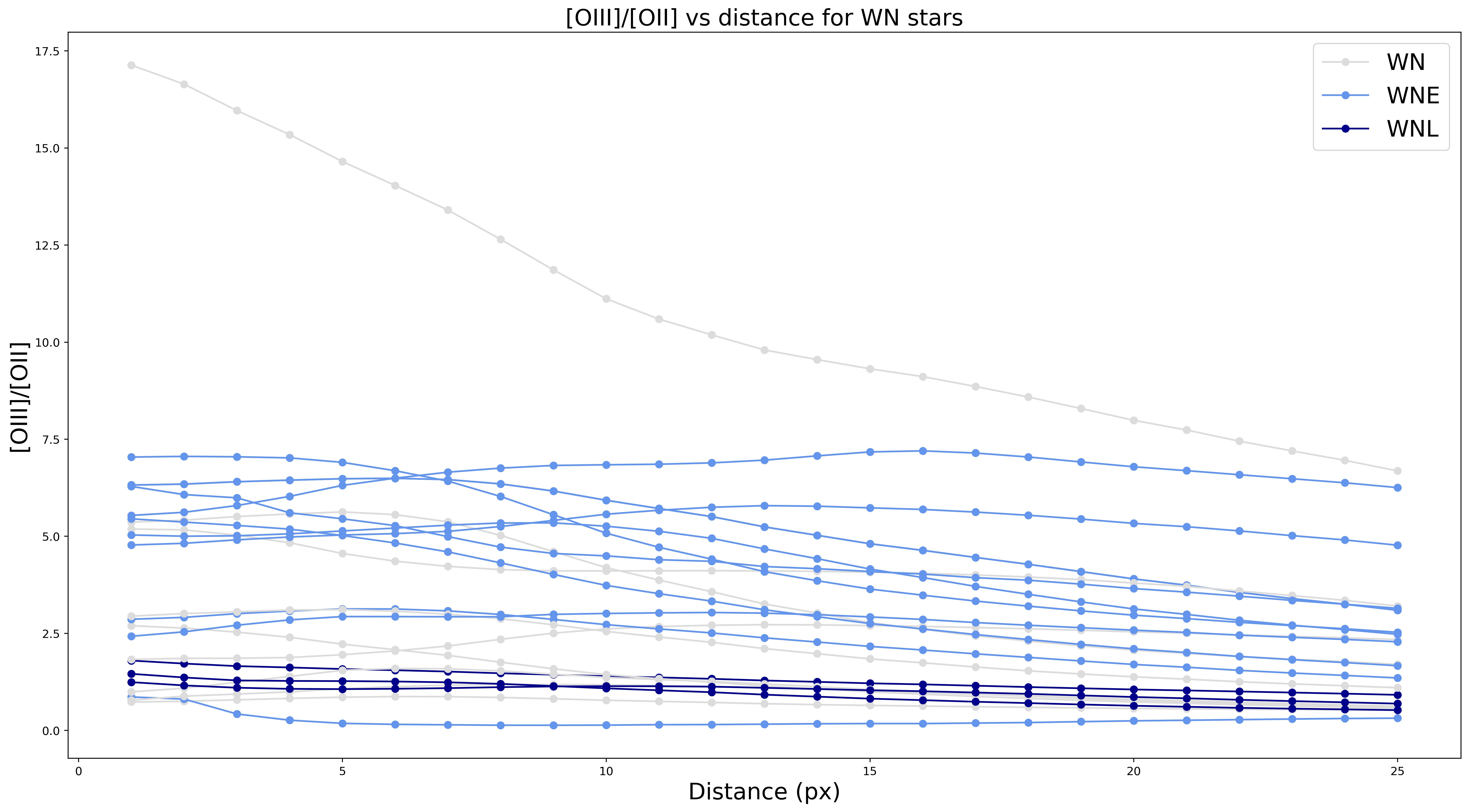}\par 
\caption{The [O{\sc iii}]/[O{\sc ii}] ratio versus the distance from the central star (in pixels) for our WR bubbles around WN-type stars. The color indicates the subtype of the WR star.}
\label{fig:OIII_OII_WN}
\end{figure*}

\begin{figure*}
    \includegraphics[width=0.9\linewidth]{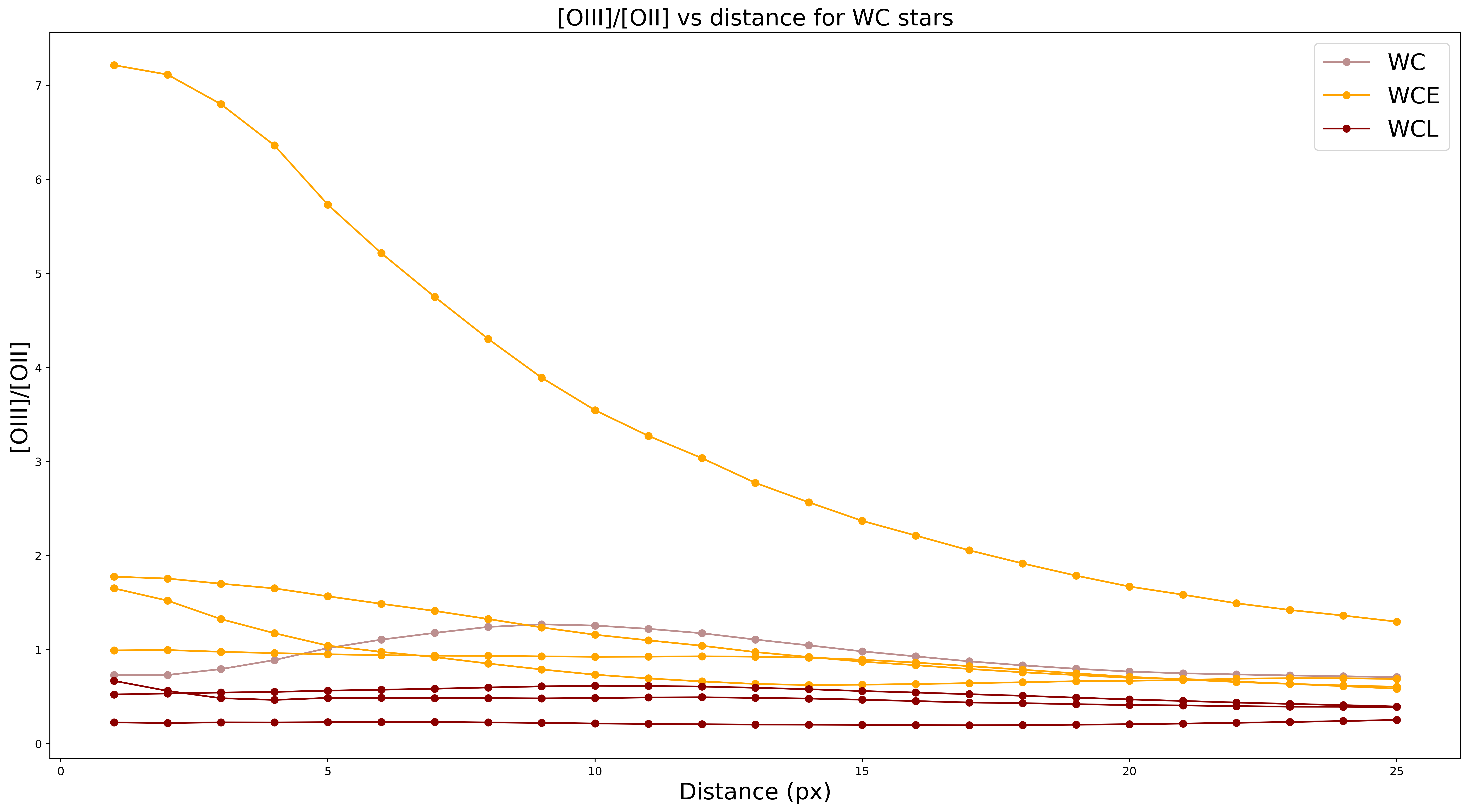}\par 
\caption{The [O{\sc iii}]/[O{\sc ii}] ratio versus the distance from the central star (in pixels) for our WR bubbles around WC-type stars. The color indicates the subtype of the WR star.}
\label{fig:OIII_OII_WC}
\end{figure*}

\begin{figure*}
    \includegraphics[width=0.9\linewidth]{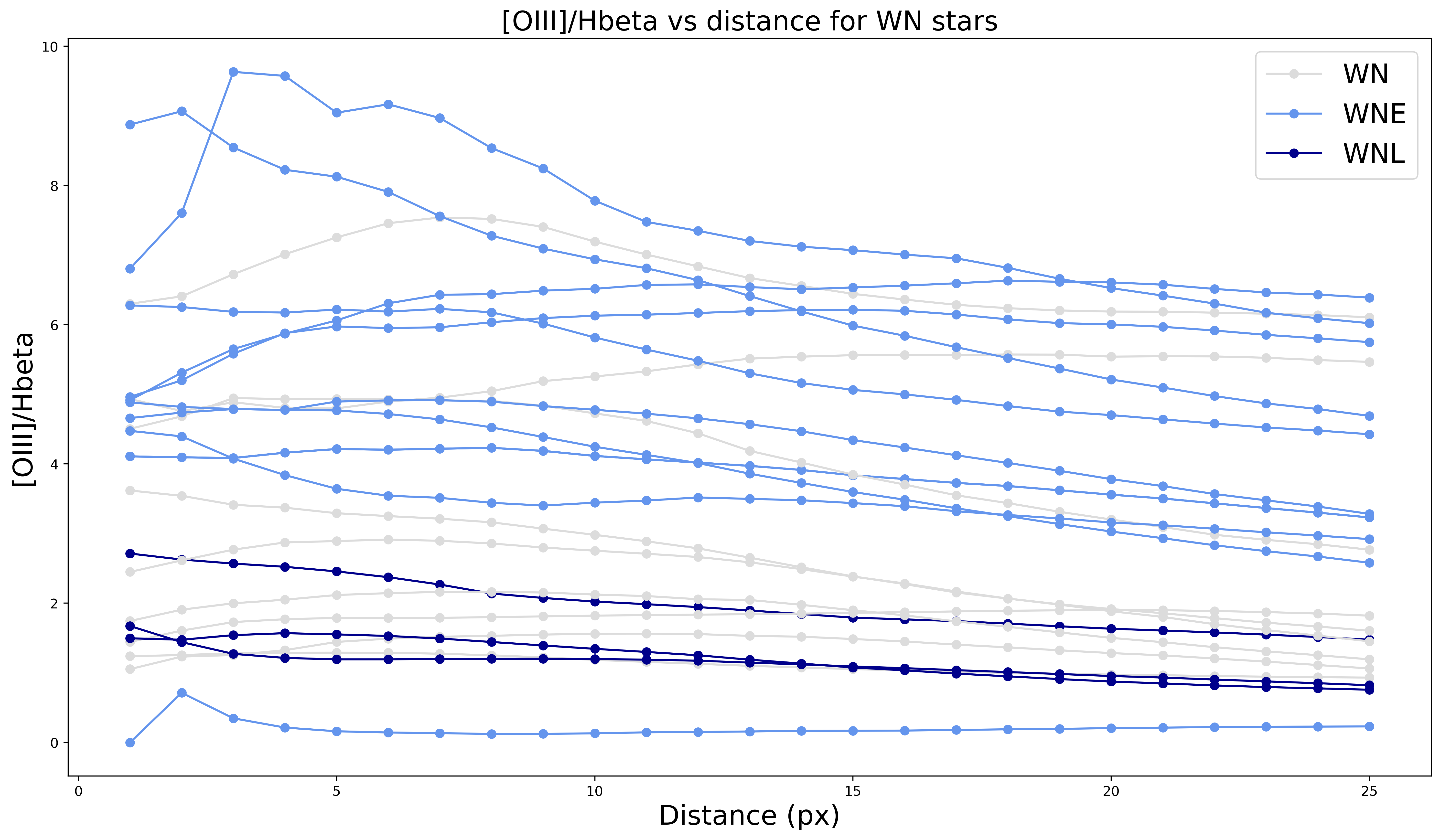}\par 
\caption{The [O{\sc iii}]/H$\beta$ ratio versus the distance from the central star (in pixels) for our WR bubbles around WN-type stars. The color indicates the subtype of the WR star.}
\label{fig:OIII_Hbeta_WN}
\end{figure*}

\begin{figure*}
    \includegraphics[width=0.9\linewidth]{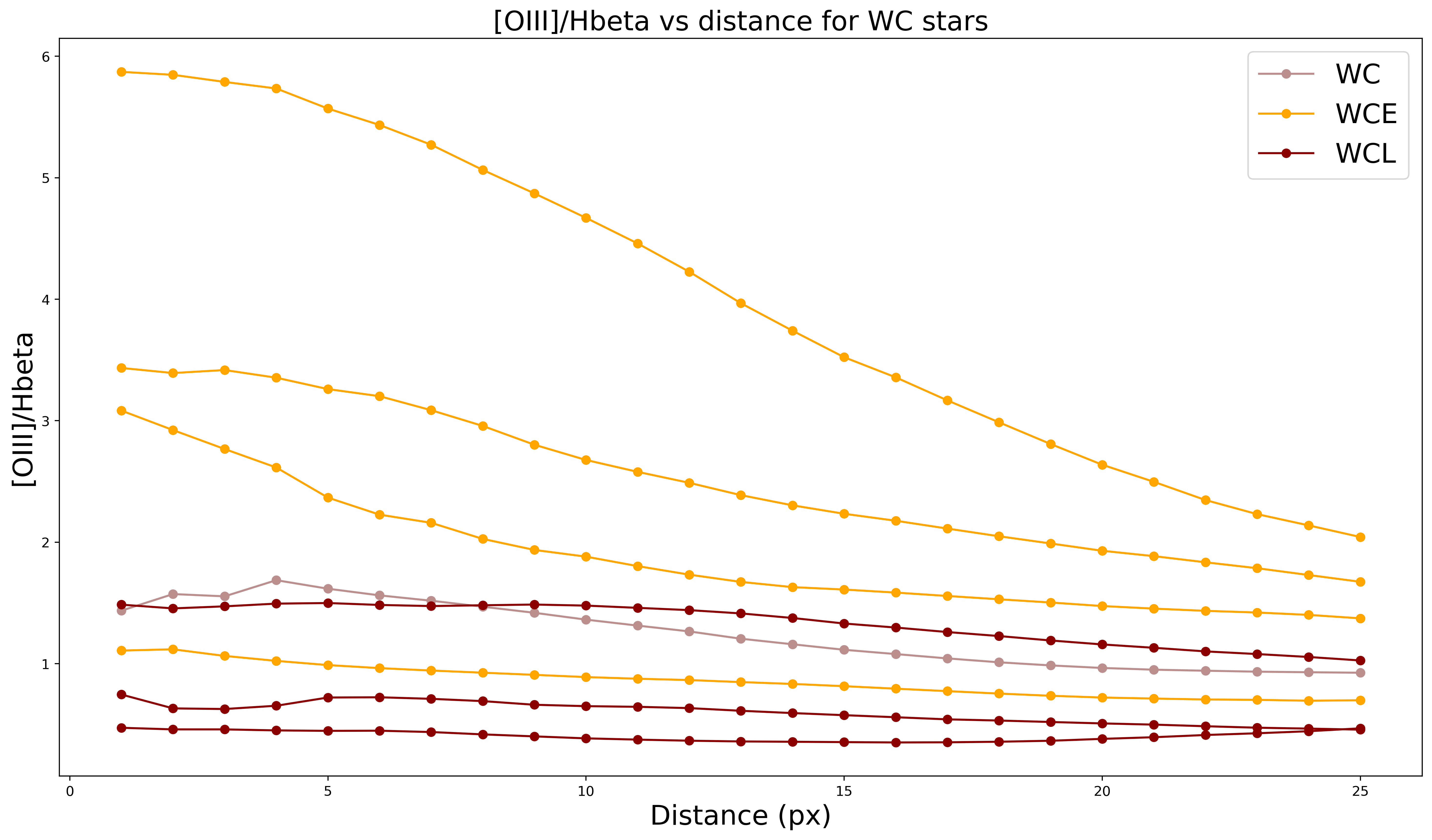}\par 
\caption{The [O{\sc iii}]/H$\beta$ ratio versus the distance from the central star (in pixels) for our WR bubbles around WC-type stars. The color indicates the subtype of the WR star.}
\label{fig:OIII_Hbeta_WC}
\end{figure*}

We show here an example of the analysis that could be done with our data (Figures \ref{fig:OIII_OII_WN},\ref{fig:OIII_OII_WC},\ref{fig:OIII_Hbeta_WN} and \ref{fig:OIII_Hbeta_WC}). We have studied the intensity of the [O{\sc iii}]/[O{\sc ii}] and the [O{\sc iii}]/H$\beta$ line ratios versus the distance from the central star for some of our WR bubbles. We find that the intensity decreases with increasing distance from the star as expected, illustrating the star's role in the ionization of the nebula. The intensities and their evolution are also coherent with the subtype of the stars : we find that late-type stars have a less intense line ratio than early-type stars in most cases. However, the discrepancy between late-type WN stars and early-type WC stars seems too big, especially for the [O{\sc iii}]/[O{\sc ii}] line ratio. We also identify an "outlier" WN star for which the values of the line ratios are much more intense than the other stars : J013327.76+303150.9. A possible interpretation would be that the WR star is actually in a binary system with a hot component.

\section{Confirmed WR bubbles in M33}
We present images of all confirmed bubbles in M33 (which are also listed in Table \ref{tab:bubblesv2}) in some of the mentioned wavelengths and ratios.

\begin{figure*}
    \includegraphics[page=1,width=0.886\linewidth]{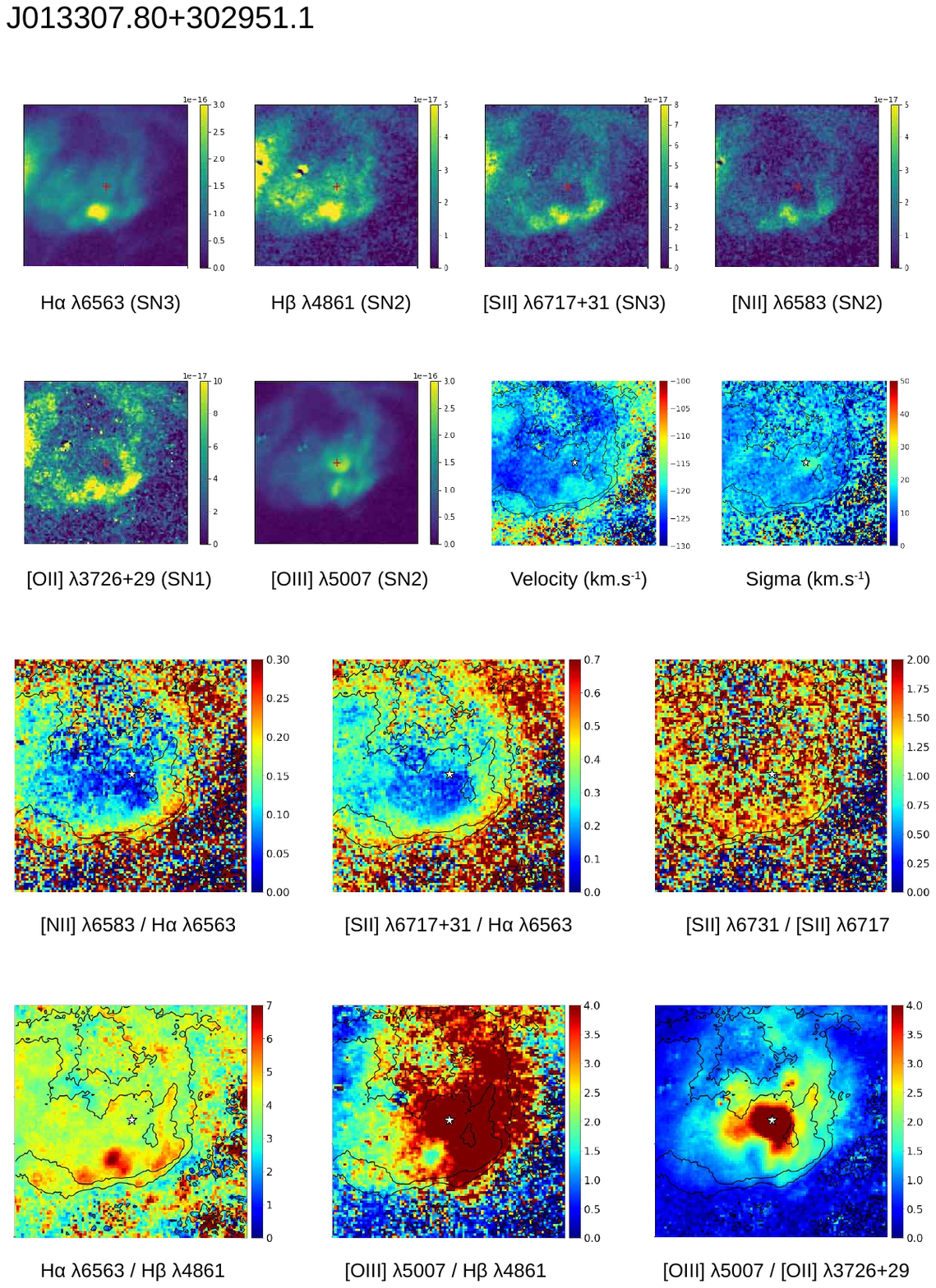}\par 
\caption{Spectral line, line ratio, velocity and  velocity dispersion maps of the bubble around J013307.80+302951.1. The images are 32$''$ $\times$ 32$''$ each and centered on the WR star (indicated by a red cross or white star in the middle). North is up, East to the left. The velocity is heliocentric and sigma is the velocity dispersion. Flux is in erg/cm$^2$/s, velocity and velocity dispersion in km/s.}
\label{fig:bubble1}
\end{figure*}


\begin{figure*}
    \includegraphics[page=2,width=0.9\linewidth]{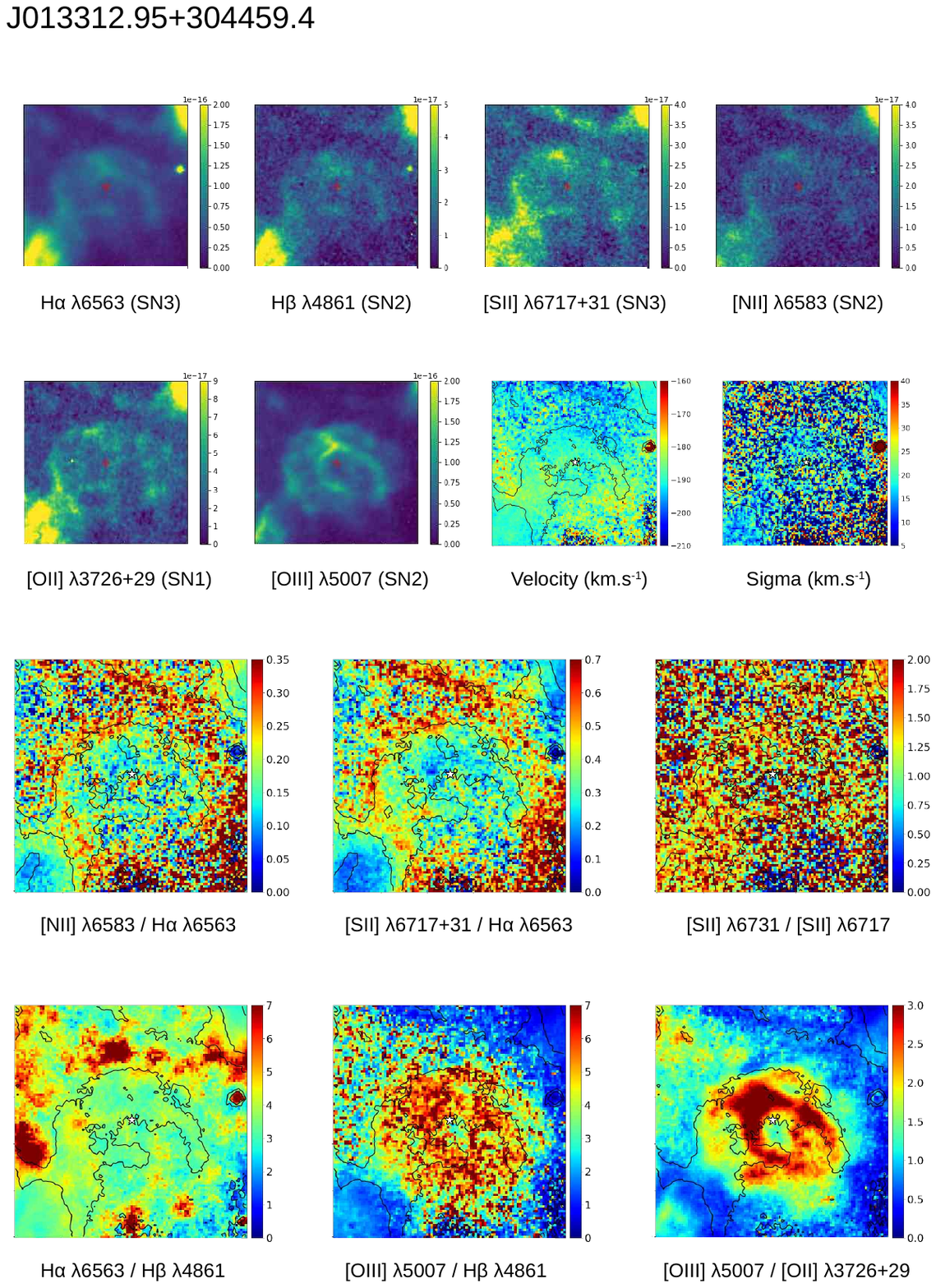}\par 
\caption{Same as Fig. \ref{fig:bubble1} for J013312.95+304459.4}
\label{fig:bubble2}
\end{figure*}

\begin{figure*}
    \includegraphics[page=3,width=0.9\linewidth]{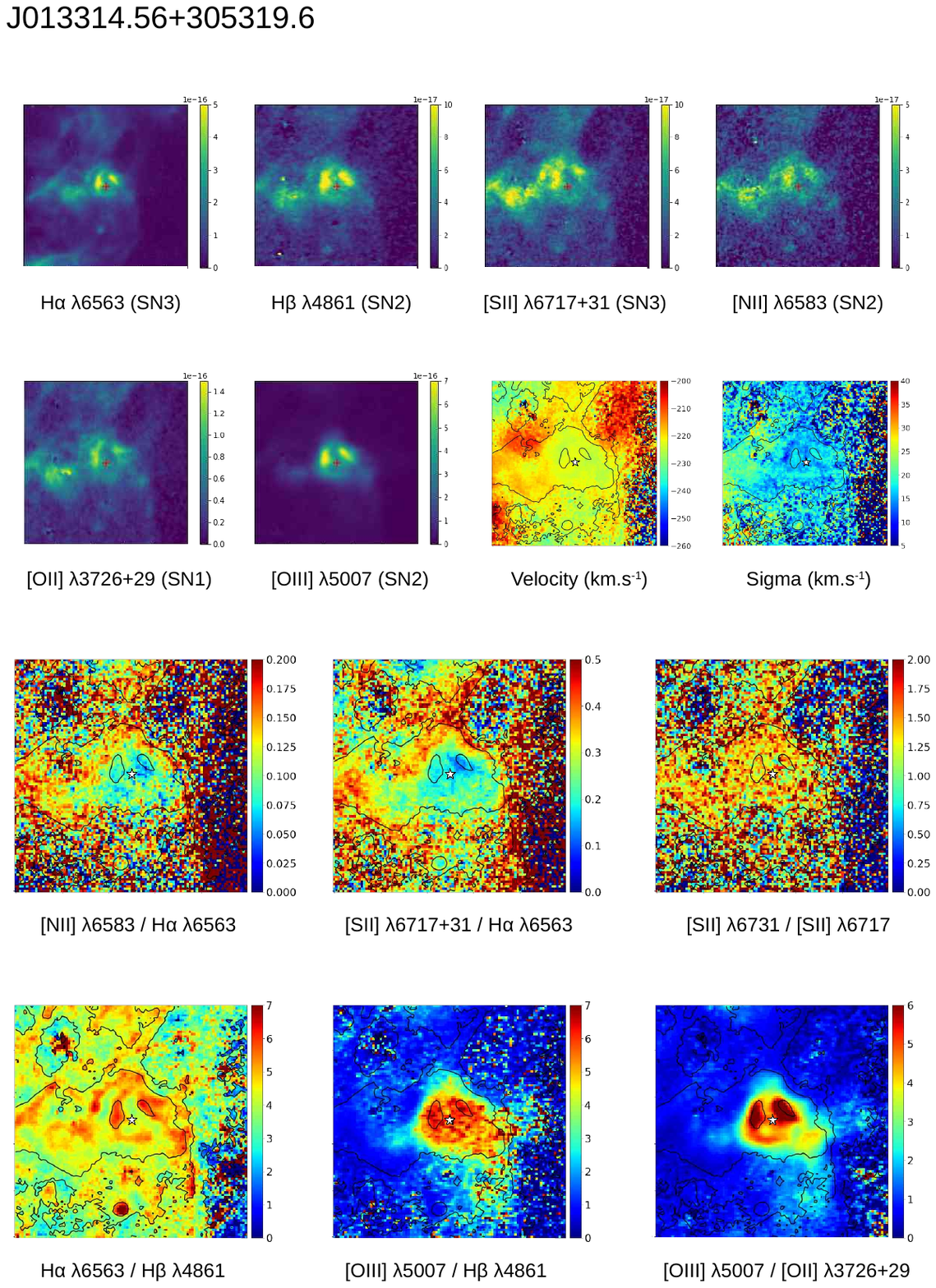}\par 
\label{fig:bubble3}
\caption{Same as Fig. \ref{fig:bubble1} for J013314.56+305319.6}
\end{figure*}

\begin{figure*}
    \includegraphics[page=4,width=0.9\linewidth]{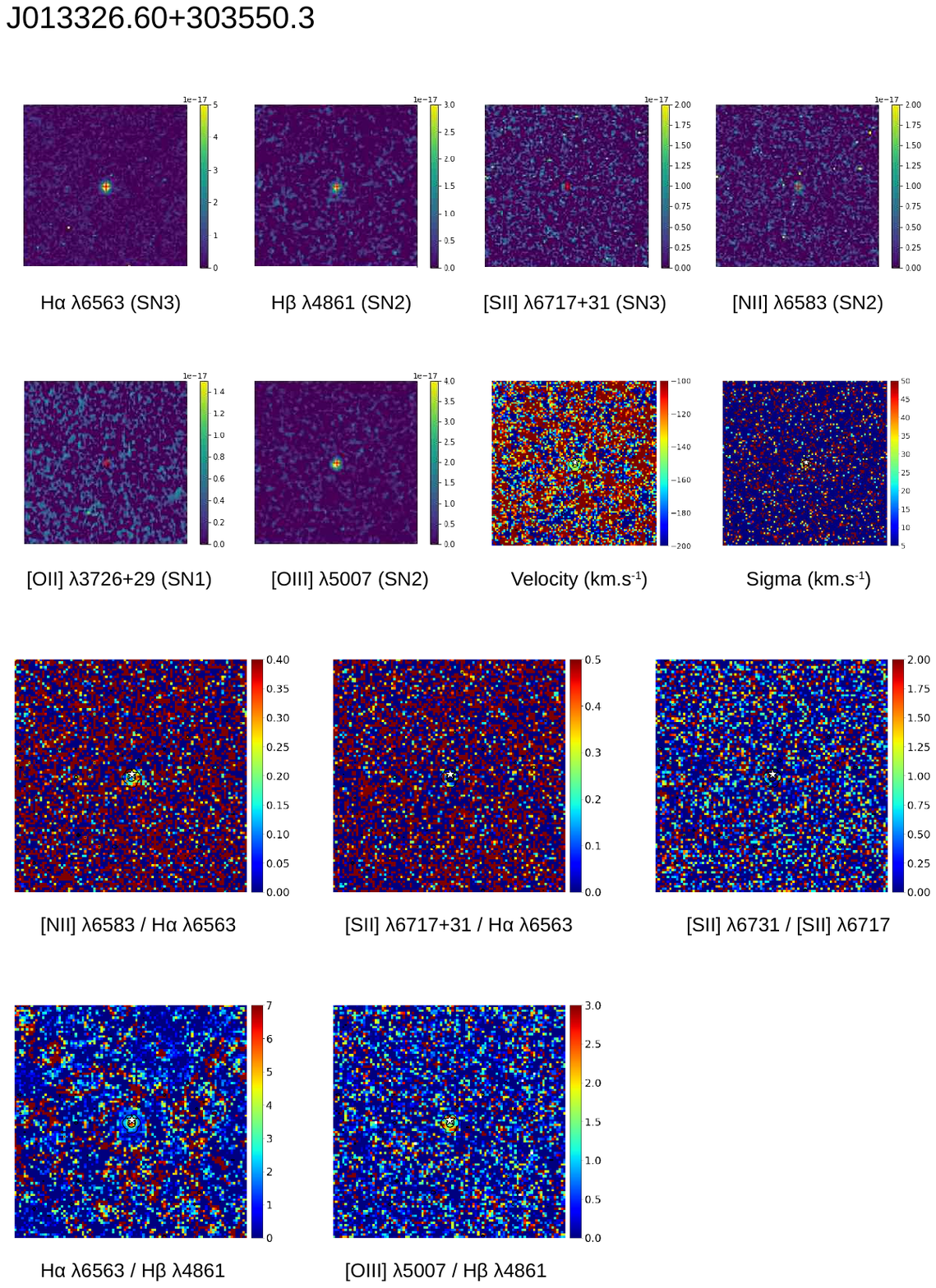}\par 
\label{fig:bubble4}
\caption{Same as Fig. \ref{fig:bubble1} for J013326.60+303550.3. In view of the complete non detection of this nebula in [O{\sc ii}], we have omitted the [O{\sc iii}]/[O{\sc ii}] map.}
\end{figure*}

\begin{figure*}
    \includegraphics[page=5,width=0.9\linewidth]{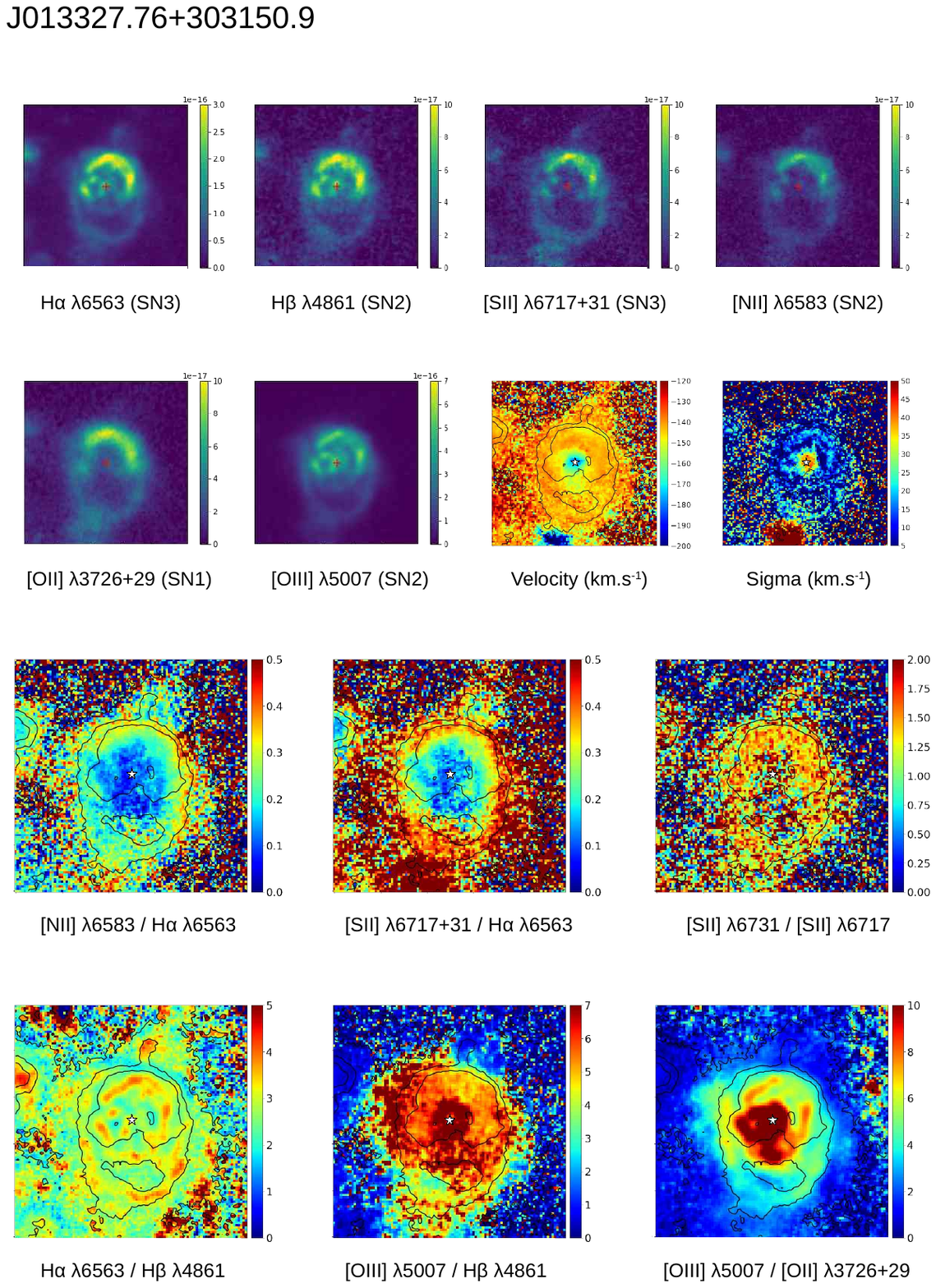}\par 
\label{fig:bubble5}
\caption{Same as Fig. \ref{fig:bubble1} for J013327.76+303150.9}
\end{figure*}

\begin{figure*}
    \includegraphics[page=6,width=0.9\linewidth]{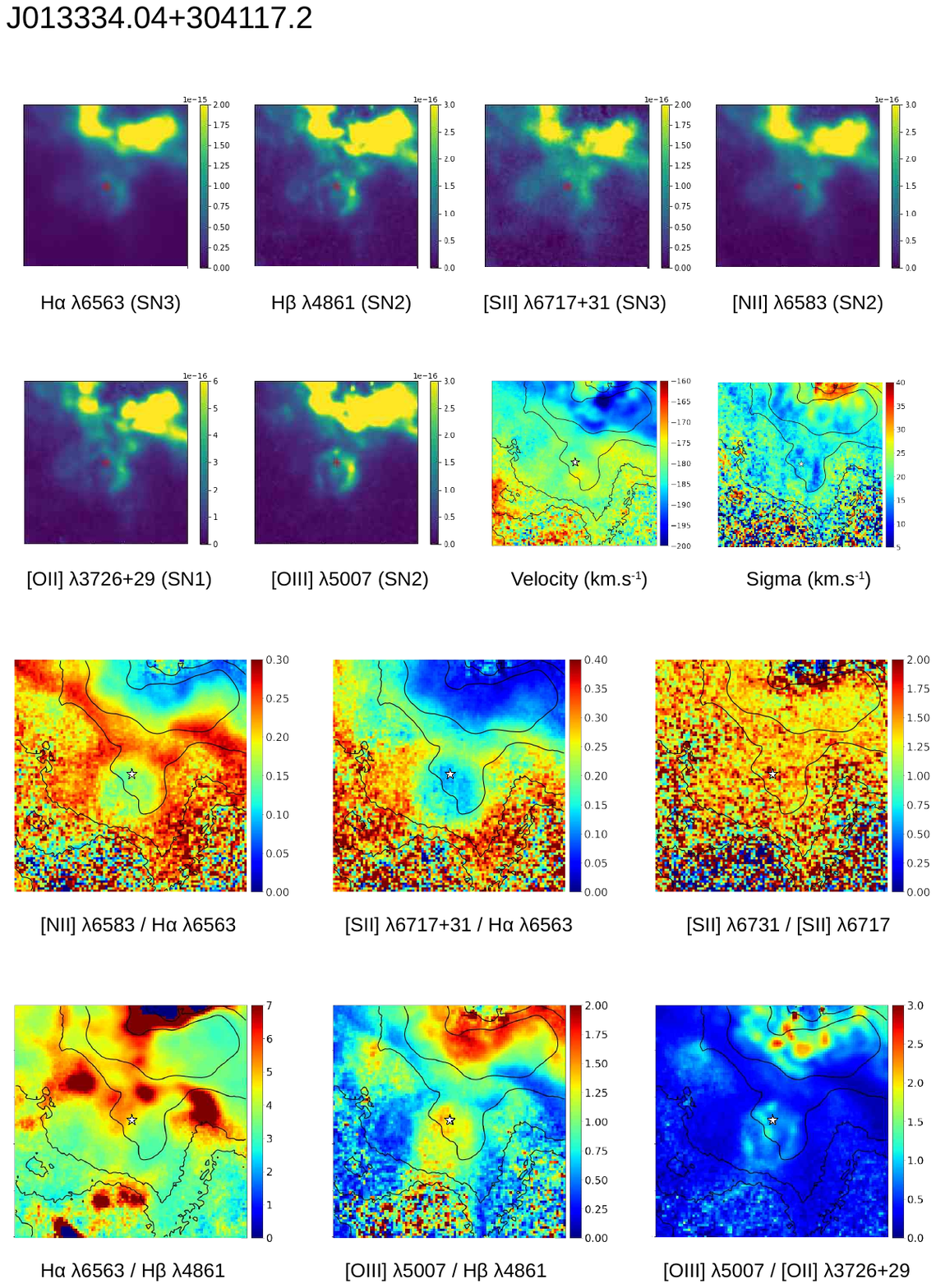}\par 
\label{fig:bubble6}
\caption{Same as Fig. \ref{fig:bubble1} for J013334.04+304117.2}
\end{figure*}

\begin{figure*}
    \includegraphics[page=7,width=0.9\linewidth]{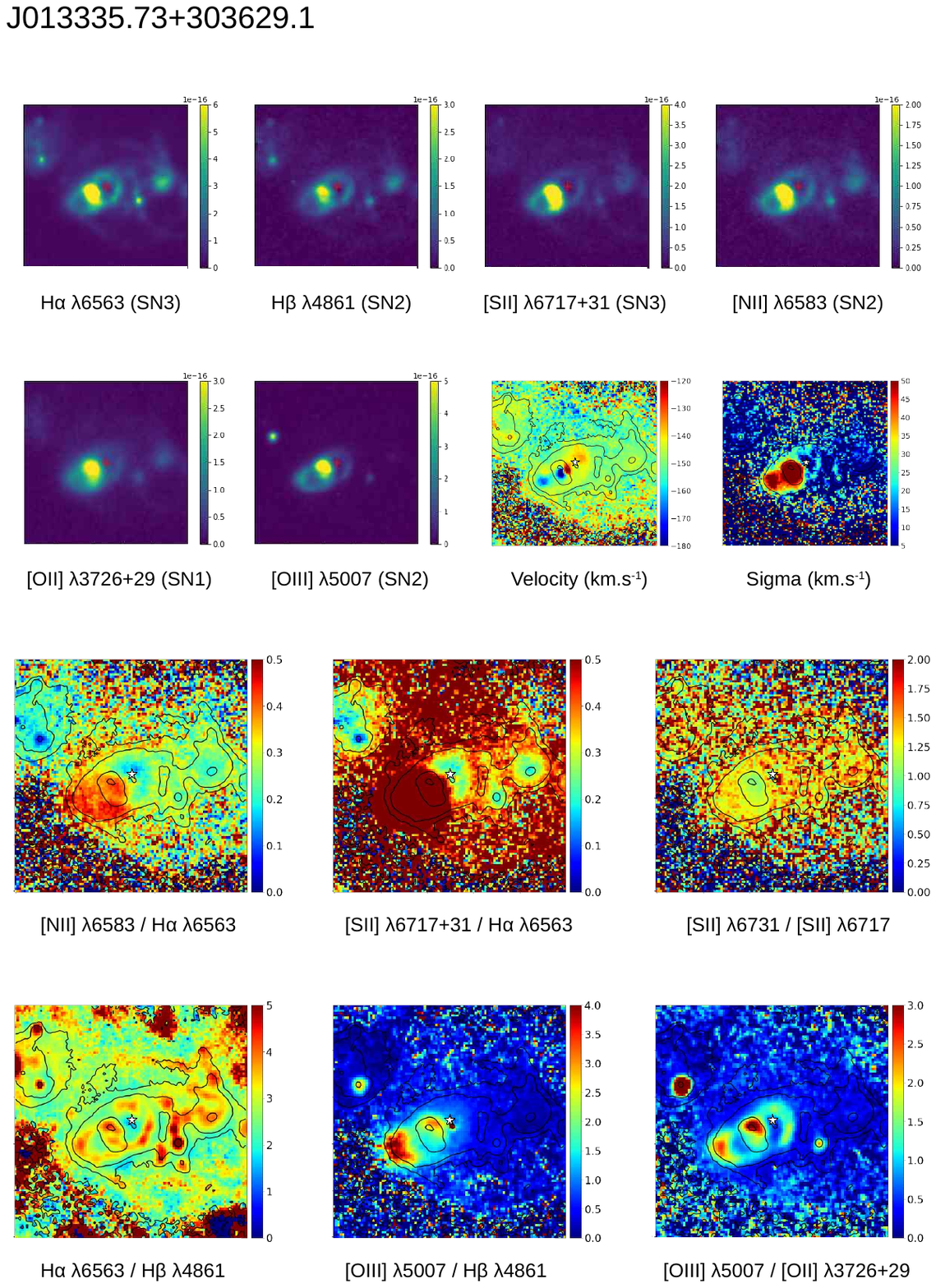}\par 
\label{fig:bubble7}
\caption{Same as Fig. \ref{fig:bubble1} for J013335.73+303629.1}
\end{figure*}

\begin{figure*}
    \includegraphics[page=8,width=0.9\linewidth]{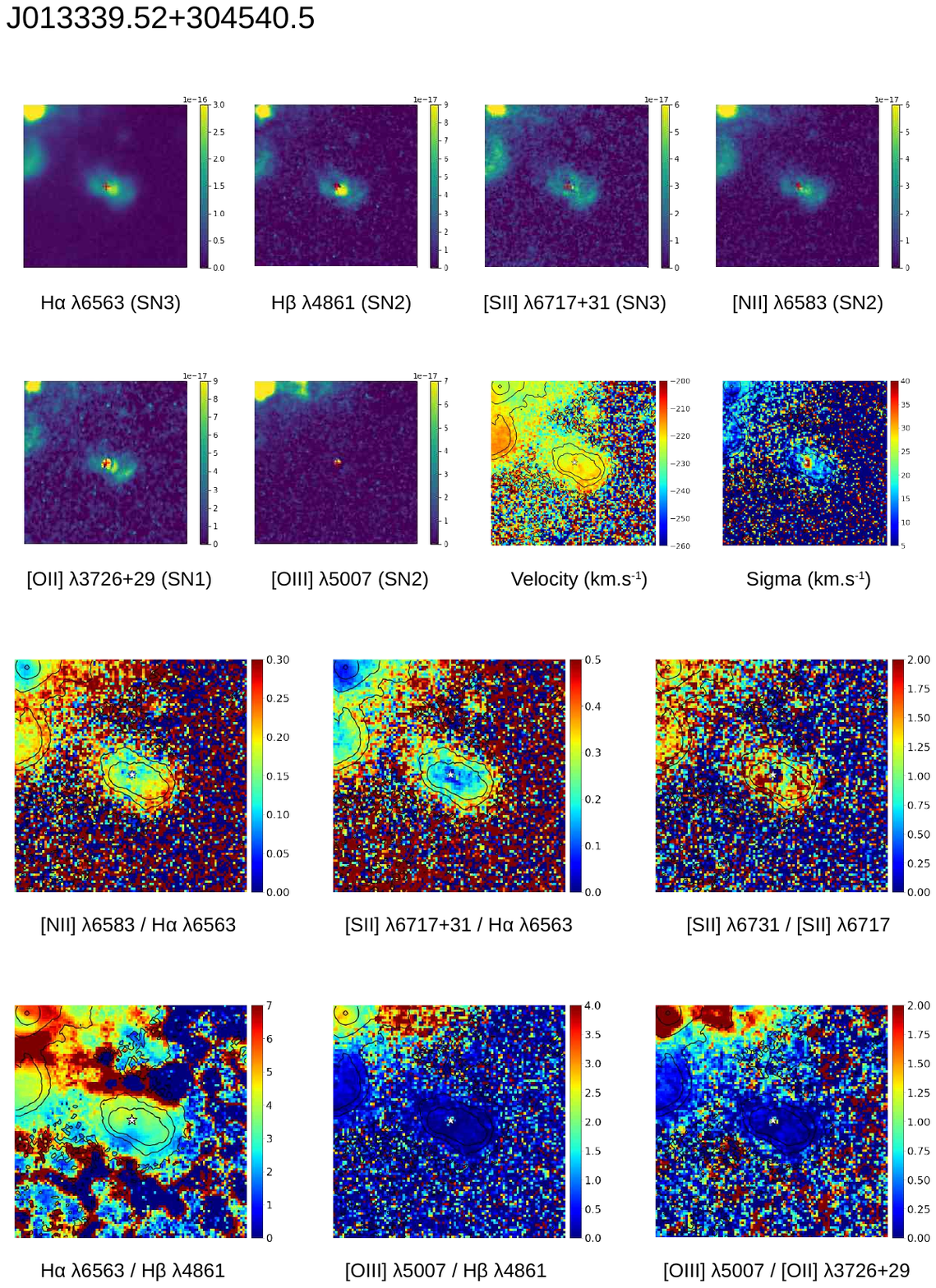}\par 
\label{fig:bubble8}
\caption{Same as Fig. \ref{fig:bubble1} for J013339.52+304540.5}
\end{figure*}

\begin{figure*}
    \includegraphics[page=9,width=0.9\linewidth]{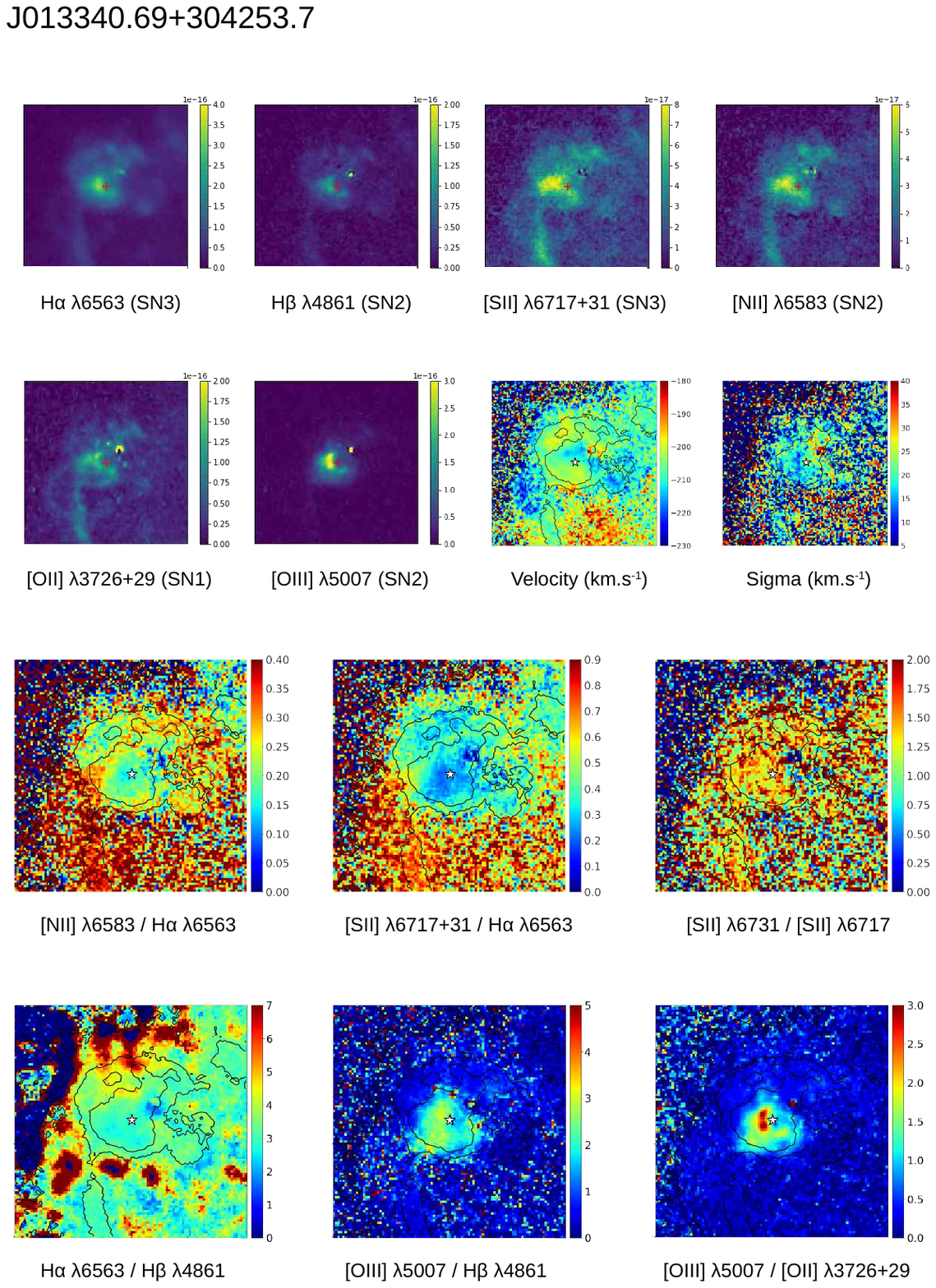}\par 
\label{fig:bubble9}
\caption{Same as Fig. \ref{fig:bubble1} for J013340.69+304253.7}
\end{figure*}

\begin{figure*}
    \includegraphics[page=10,width=0.9\linewidth]{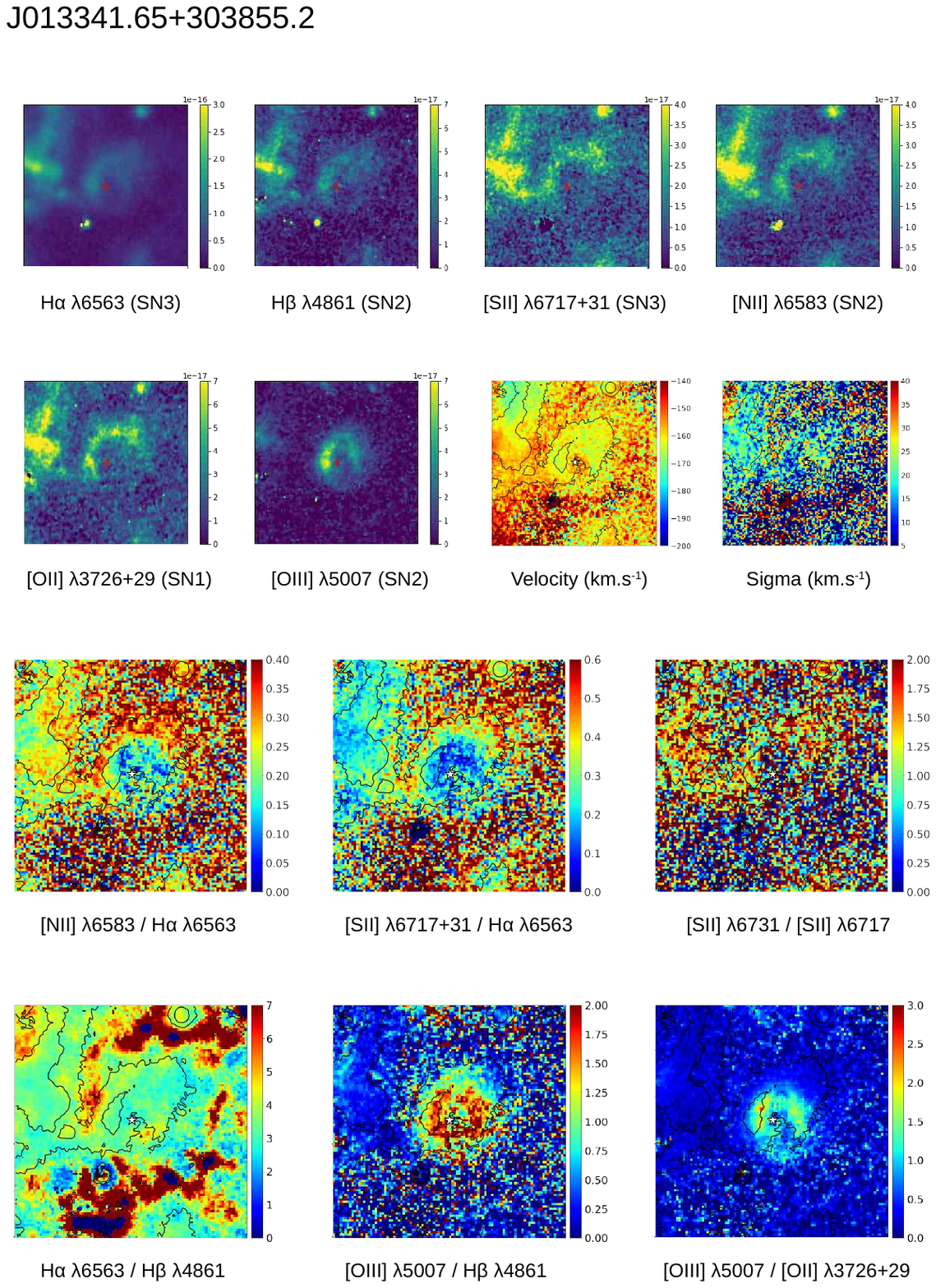}\par 
\label{fig:bubble10}
\caption{Same as Fig. \ref{fig:bubble1} for J013341.65+303855.2}
\end{figure*}

\begin{figure*}
    \includegraphics[page=11,width=0.9\linewidth]{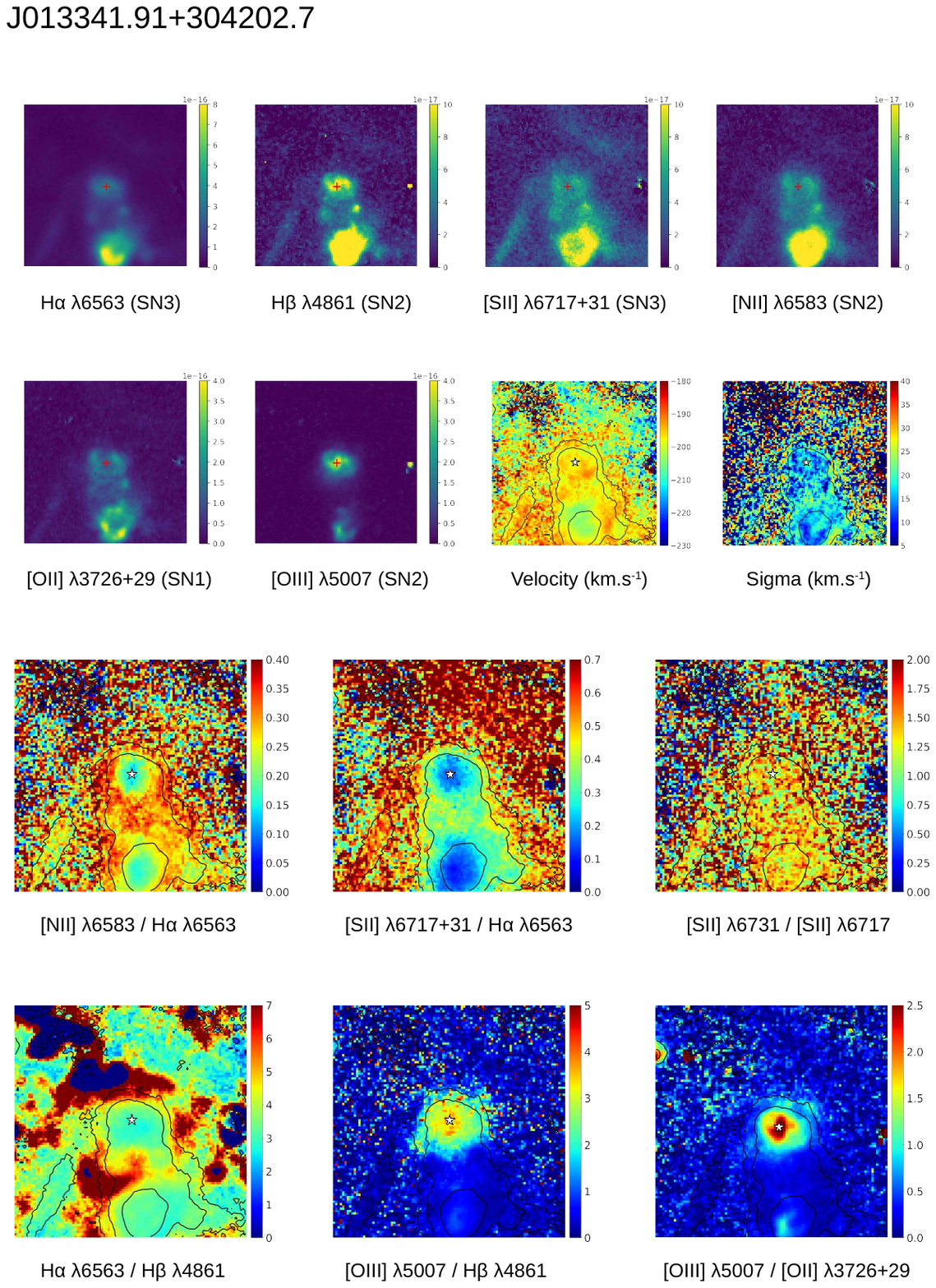}\par 
\label{fig:bubble11}
\caption{Same as Fig. \ref{fig:bubble1} for J013341.91+304202.7}
\end{figure*}

\begin{figure*}
    \includegraphics[page=12,width=0.9\linewidth]{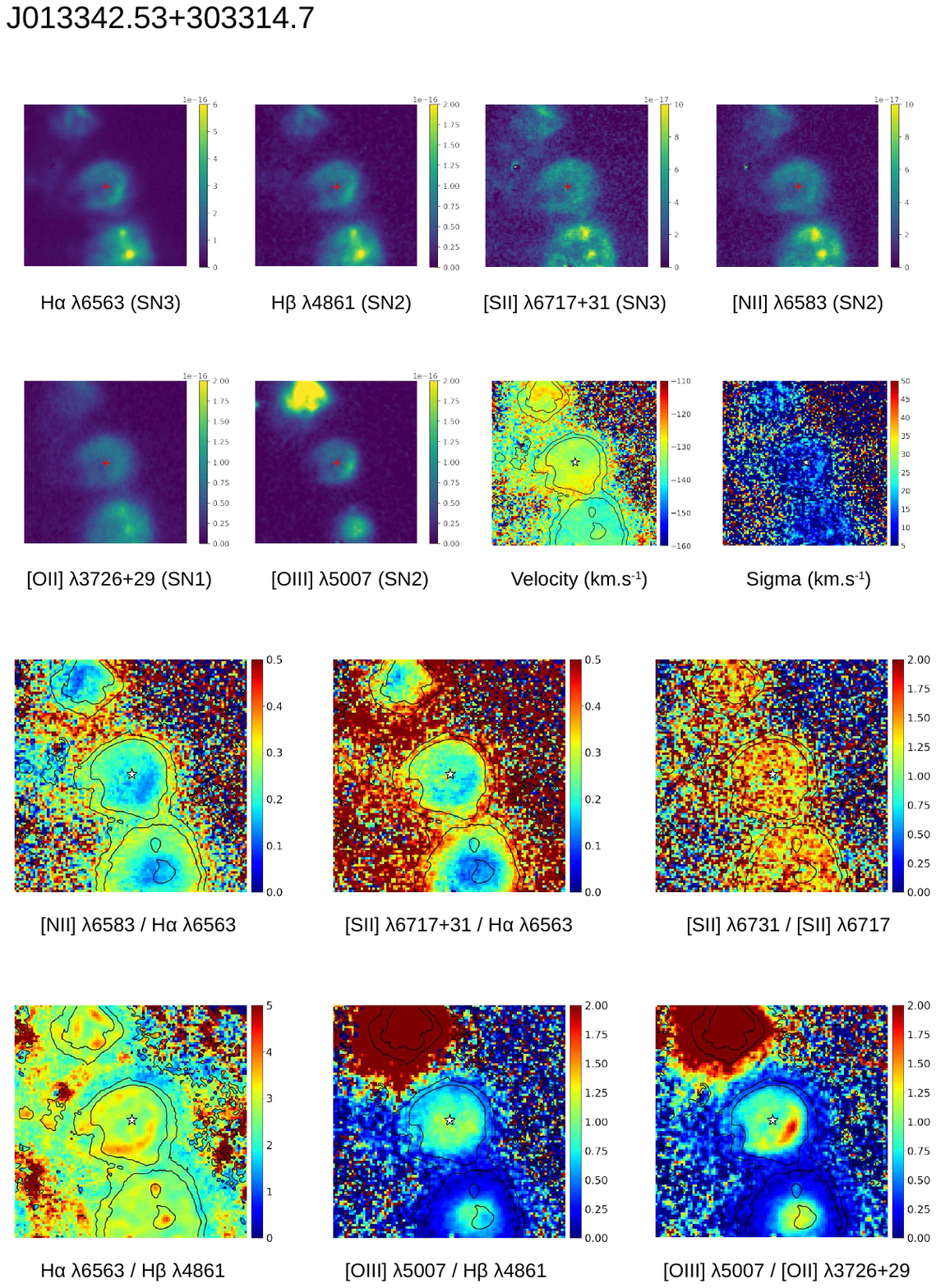}\par 
\label{fig:bubble12}
\caption{Same as Fig. \ref{fig:bubble1} for J013342.53+303314.7}
\end{figure*}

\begin{figure*}
    \includegraphics[page=13,width=0.9\linewidth]{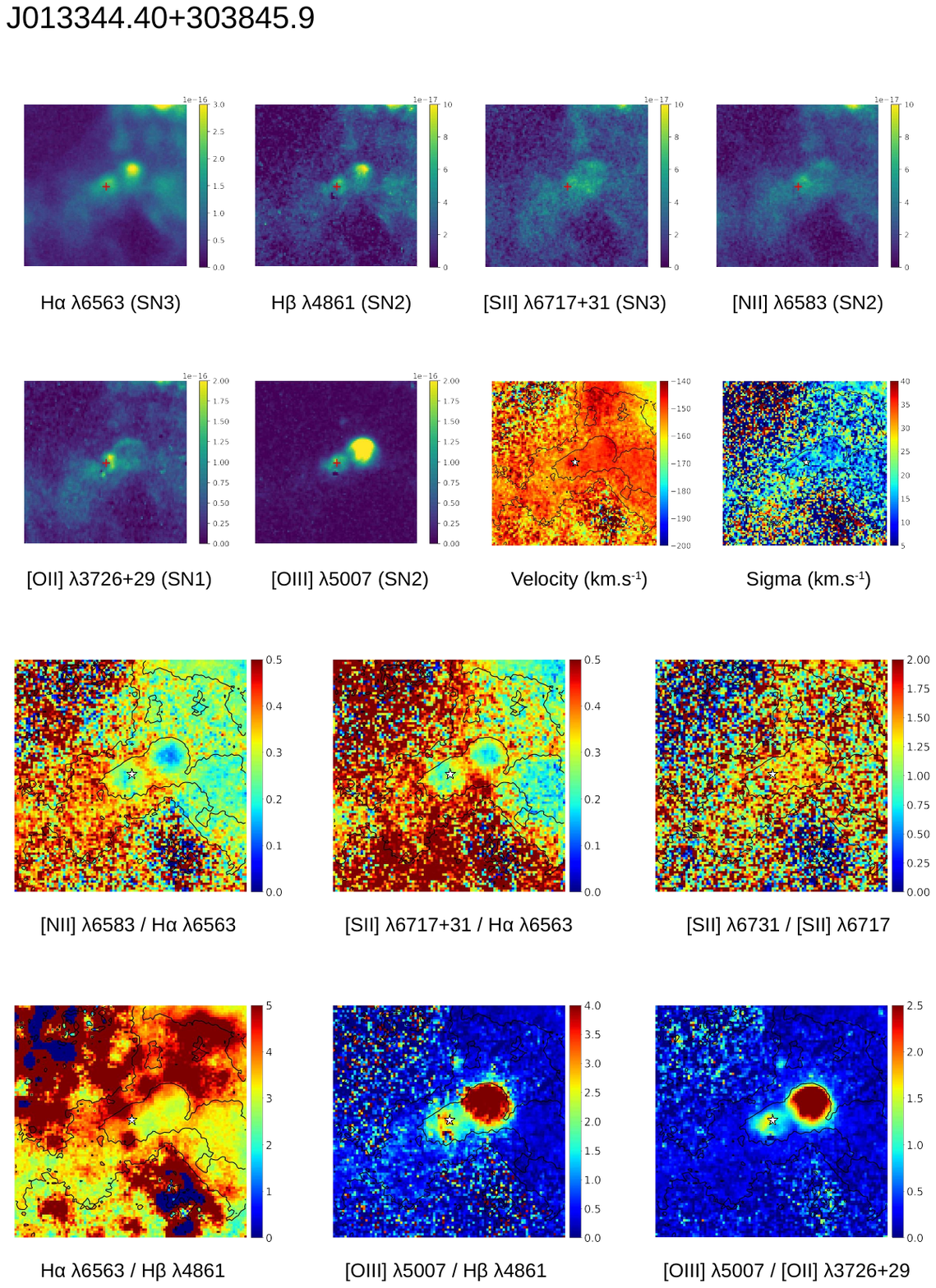}\par 
\label{fig:bubble13}
\caption{Same as Fig. \ref{fig:bubble1} for J013344.40+303845.9}
\end{figure*}

\begin{figure*}
    \includegraphics[page=14,width=0.9\linewidth]{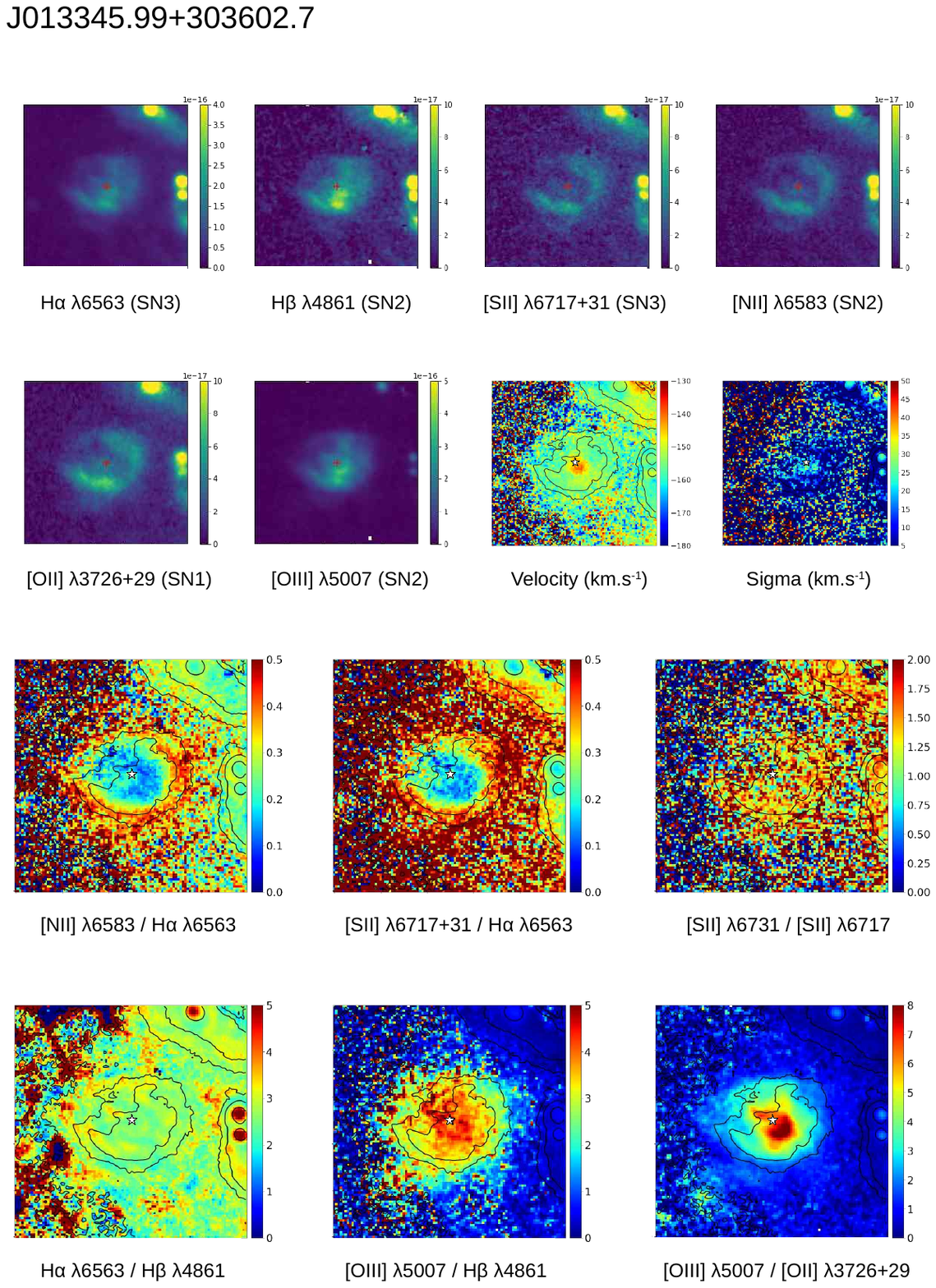}\par 
\label{fig:bubble14}
\caption{Same as Fig. \ref{fig:bubble1} for J013345.99+303602.7}
\end{figure*}

\begin{figure*}
    \includegraphics[page=15,width=0.9\linewidth]{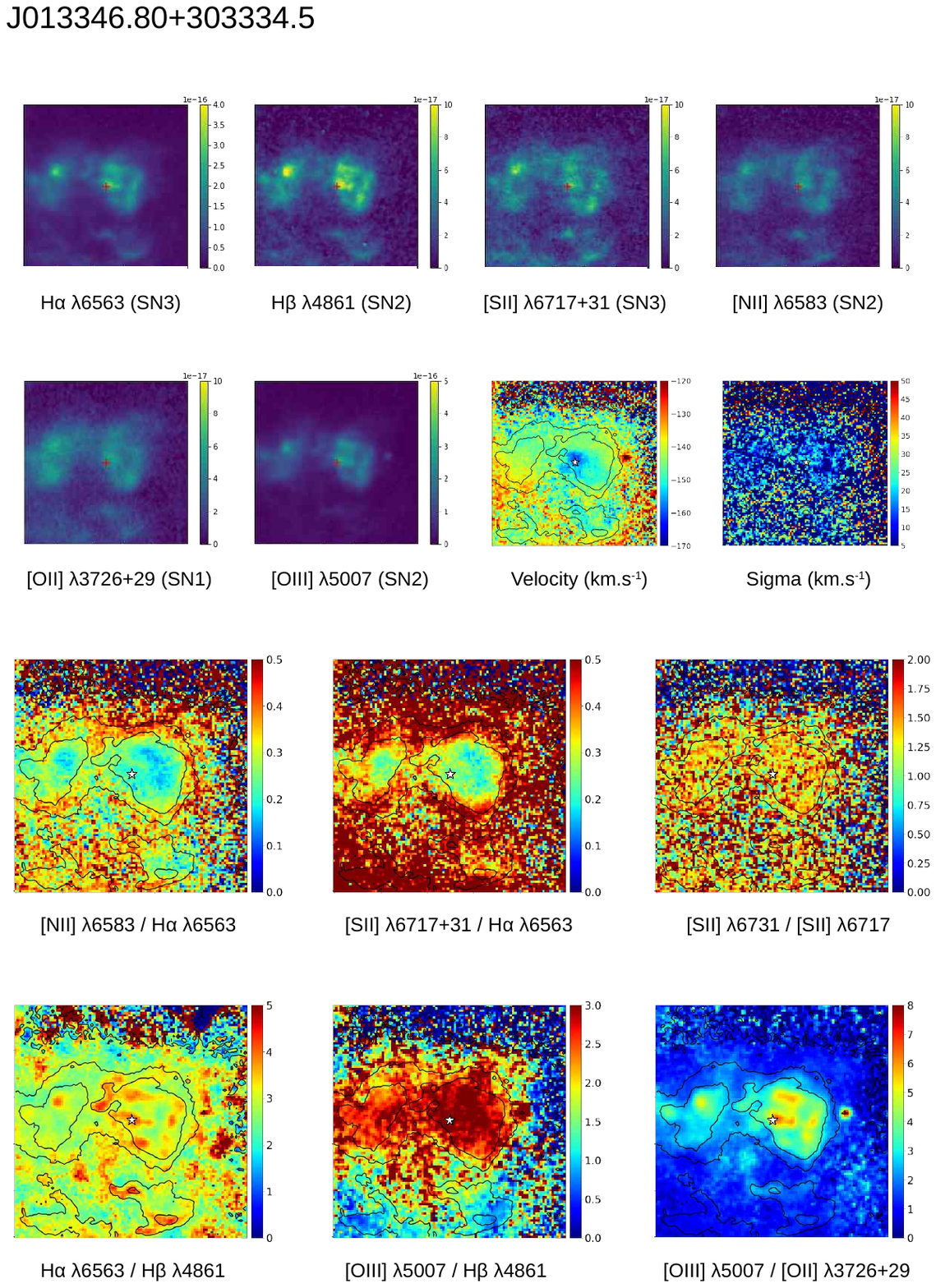}\par 
\label{fig:bubble15}
\caption{Same as Fig. \ref{fig:bubble1} for J013346.80+303334.5}
\end{figure*}

\begin{figure*}
    \includegraphics[page=16,width=0.9\linewidth]{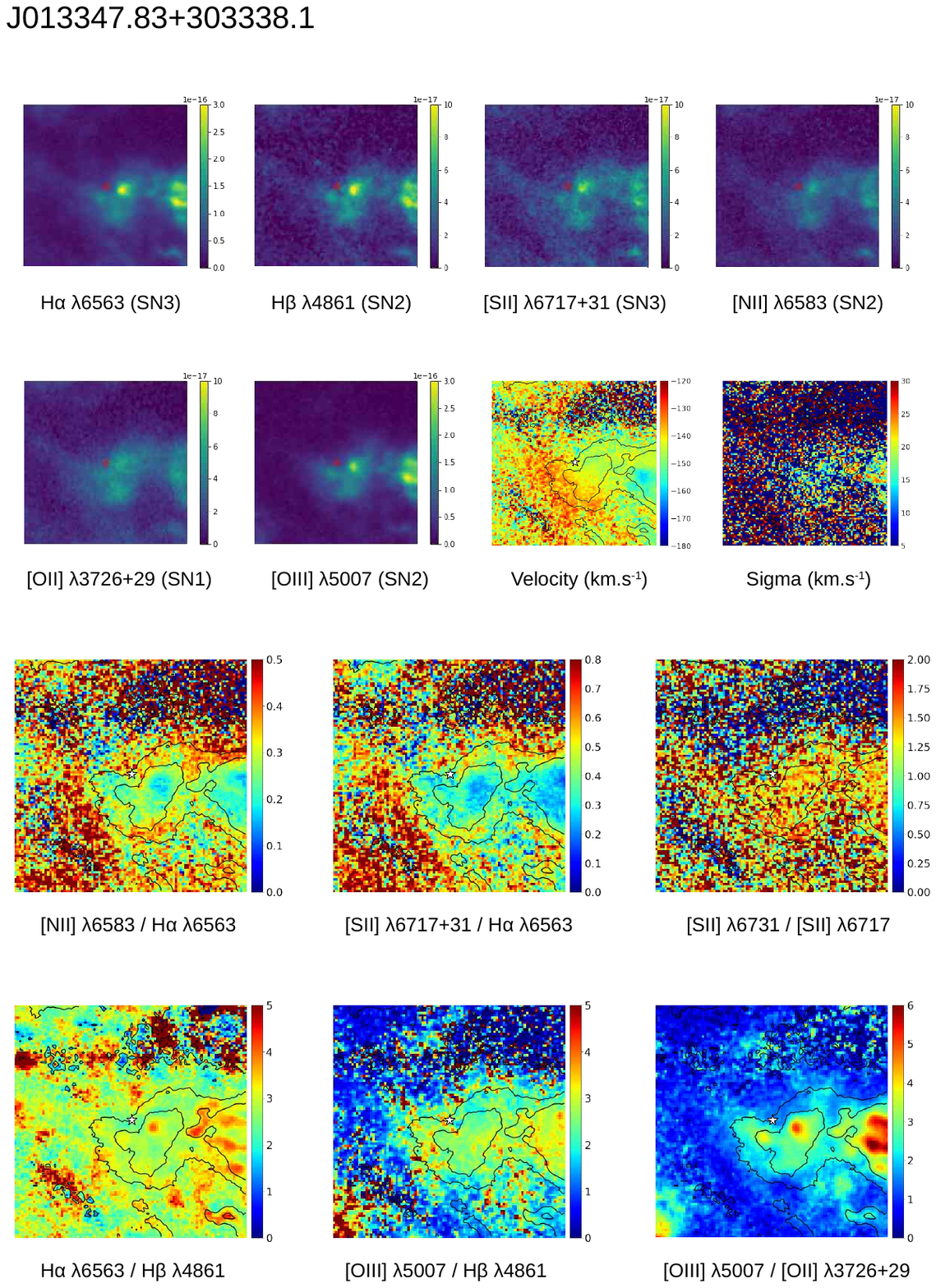}\par 
\label{fig:bubble16}
\caption{Same as Fig. \ref{fig:bubble1} for J013347.83+303338.1}
\end{figure*}

\begin{figure*}
    \includegraphics[page=17,width=0.9\linewidth]{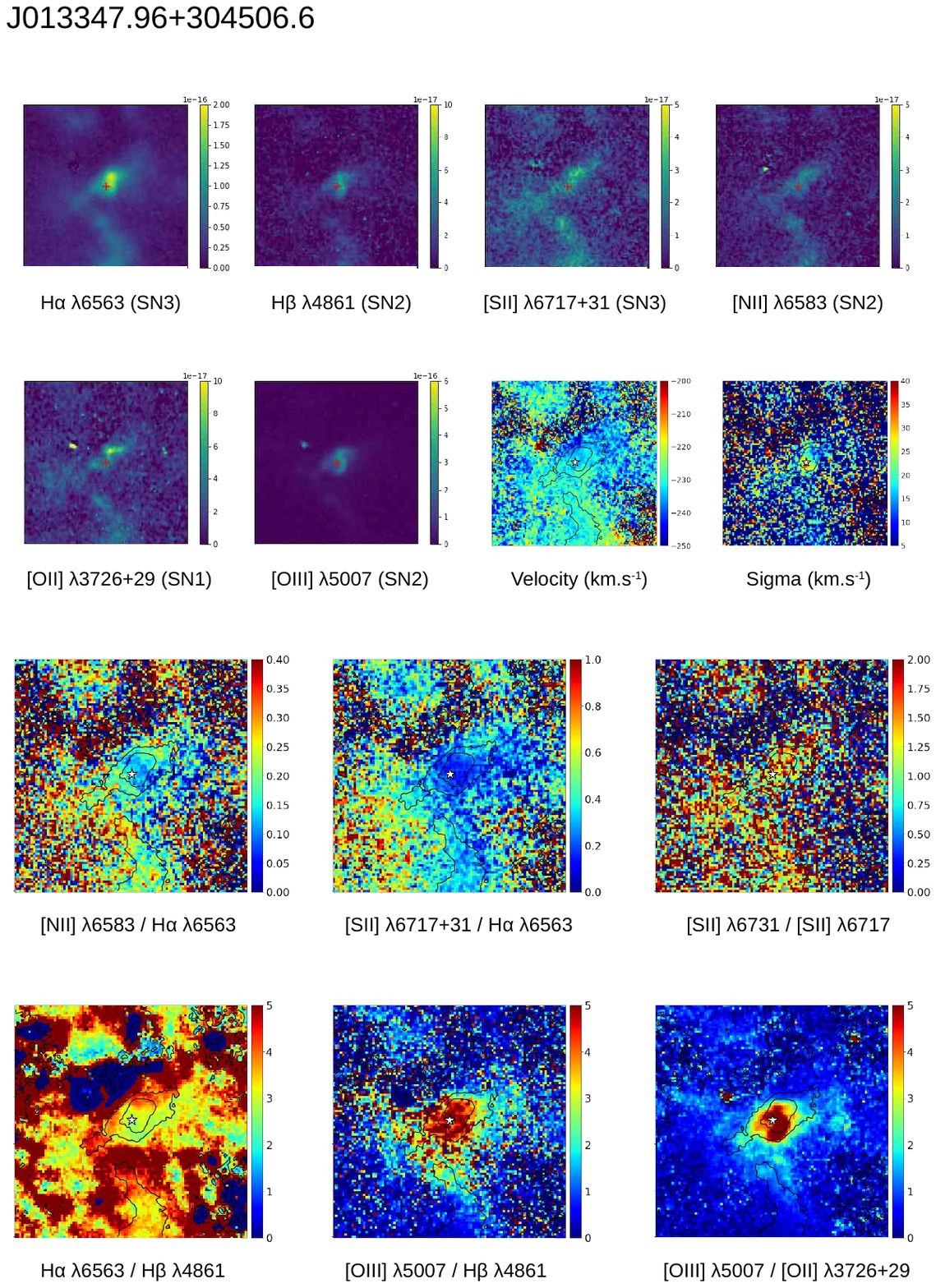}\par 
\label{fig:bubble17}
\caption{Same as Fig. \ref{fig:bubble1} for J013347.96+304506.6}
\end{figure*}

\begin{figure*}
    \includegraphics[page=18,width=0.9\linewidth]{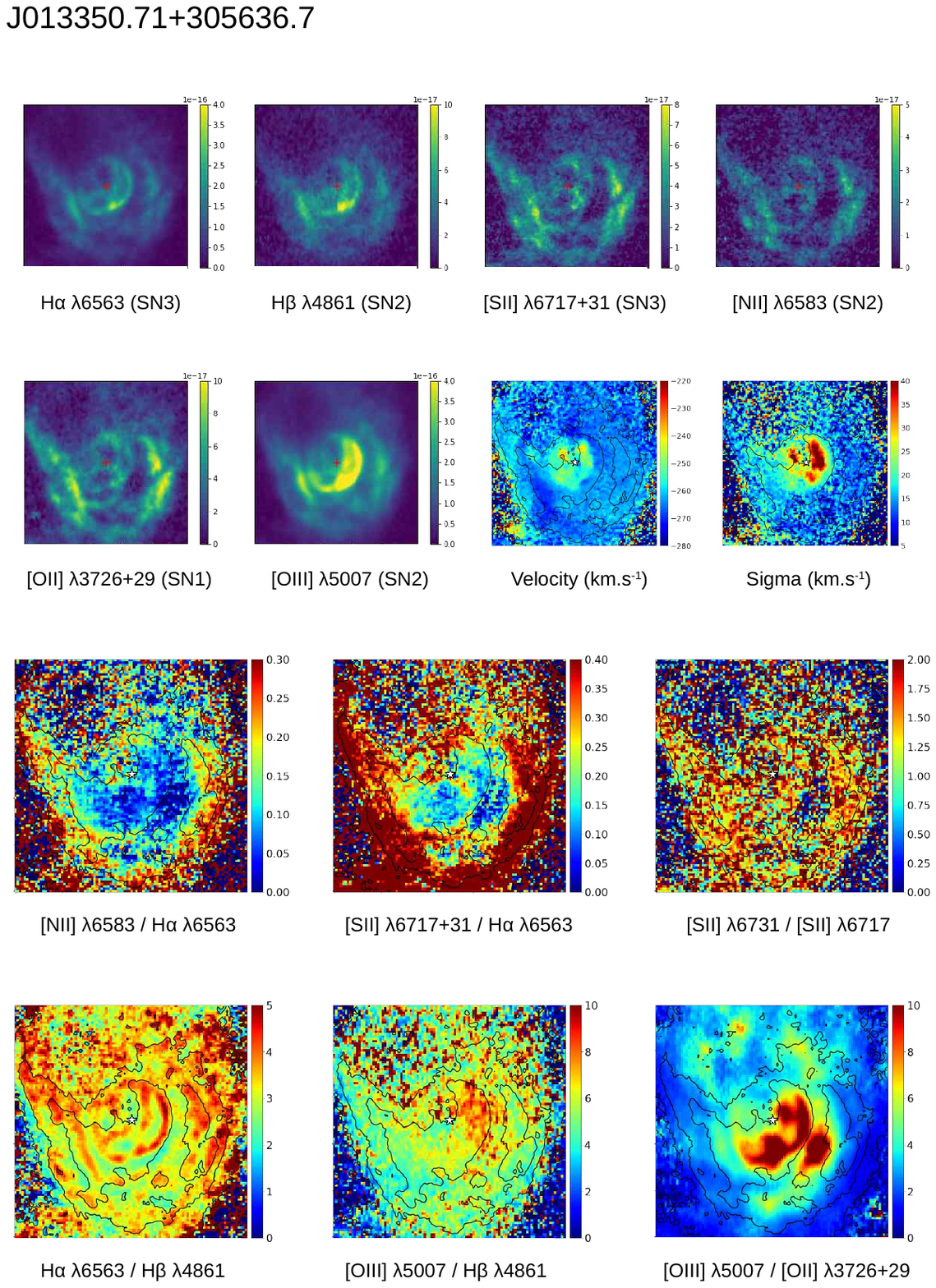}\par 
\label{fig:bubble18}
\caption{Same as Fig. \ref{fig:bubble1} for J013350.71+305636.7}
\end{figure*}

\begin{figure*}
    \includegraphics[page=19,width=0.9\linewidth]{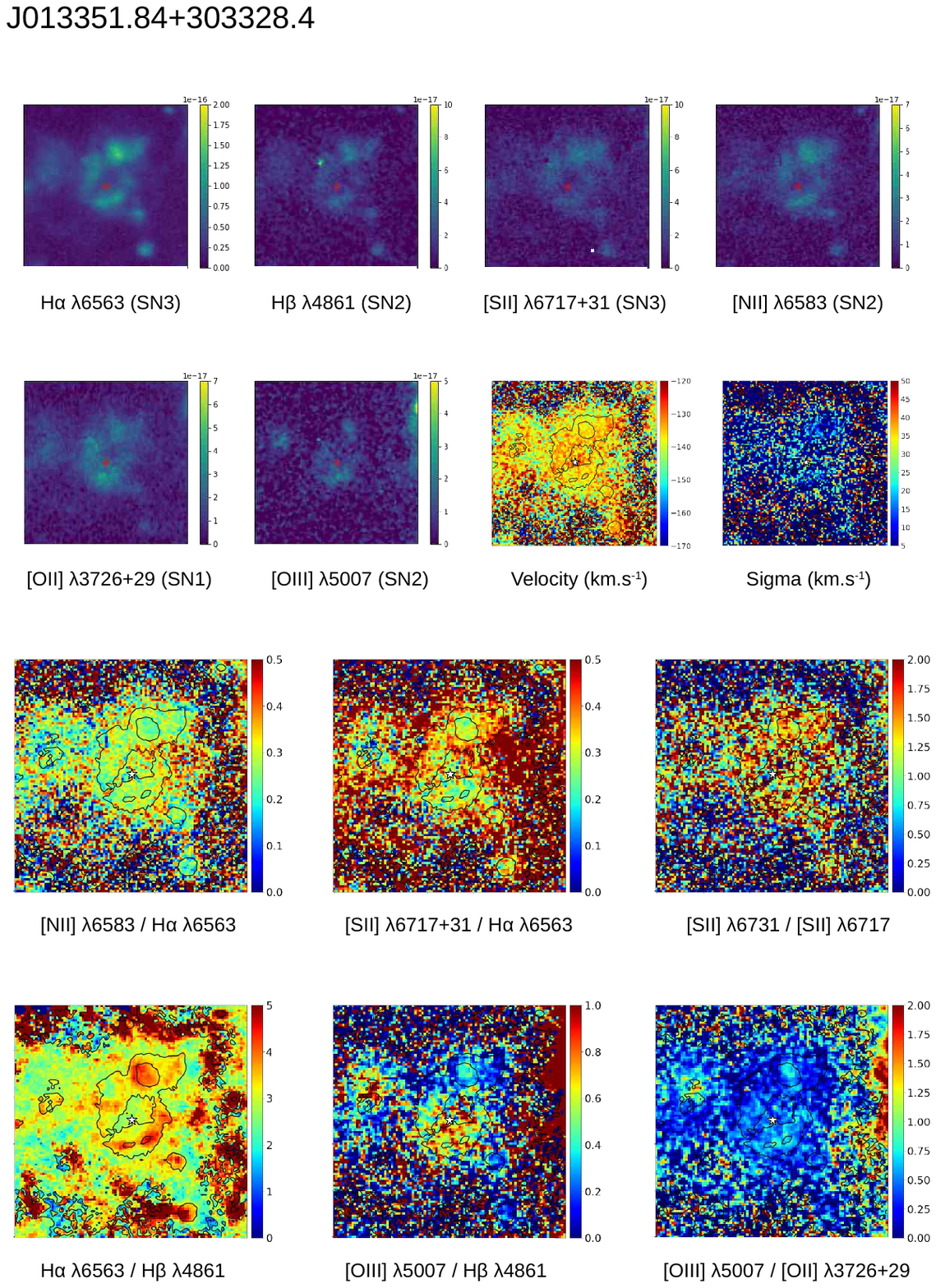}\par 
\label{fig:bubble19}
\caption{Same as Fig. \ref{fig:bubble1} for J013351.84+303328.4}
\end{figure*}

\begin{figure*}
    \includegraphics[page=20,width=0.9\linewidth]{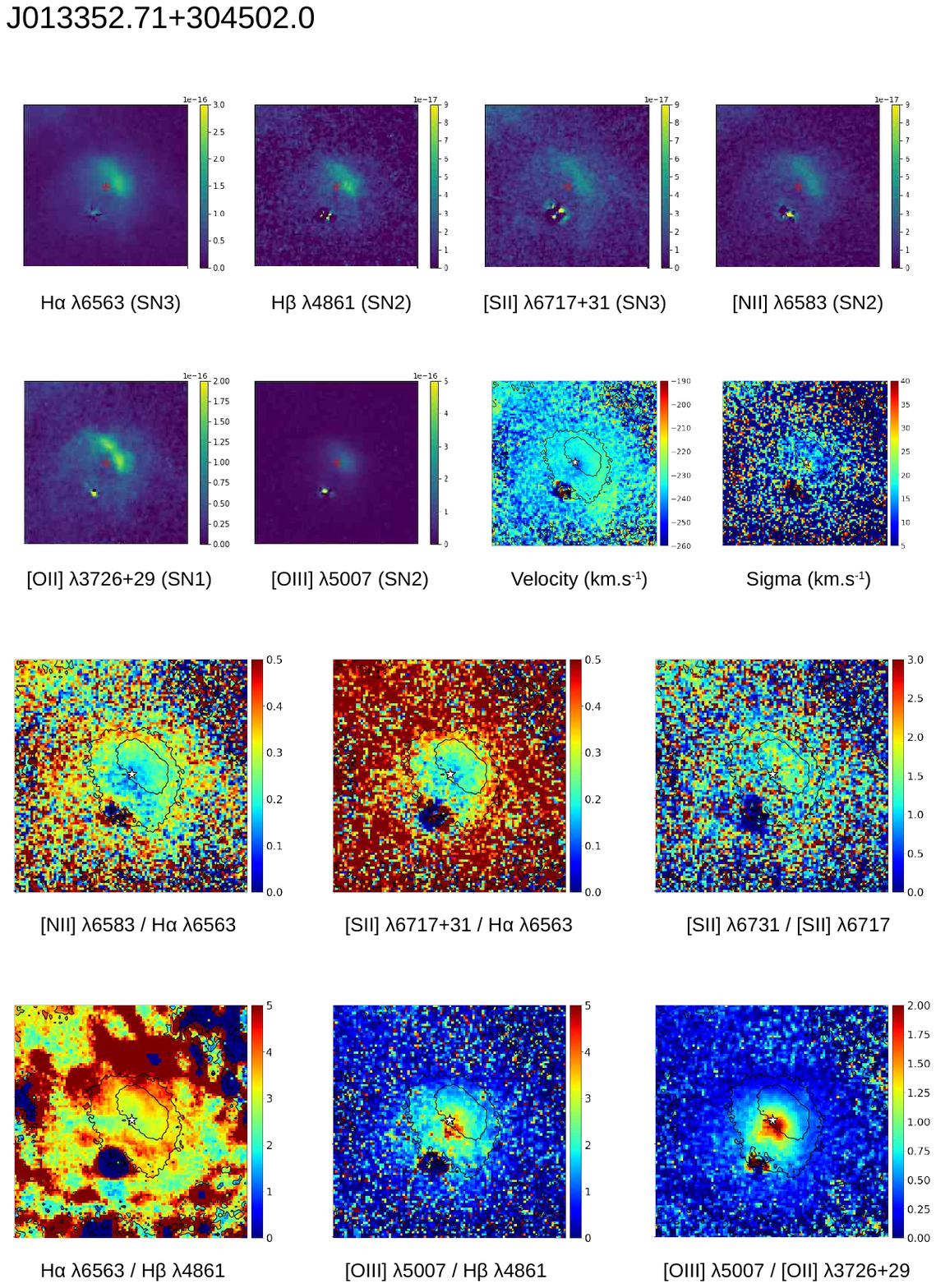}\par 
\label{fig:bubble20}
\caption{Same as Fig. \ref{fig:bubble1} for J013352.71+304502.0}
\end{figure*}

\begin{figure*}
    \includegraphics[page=21,width=0.9\linewidth]{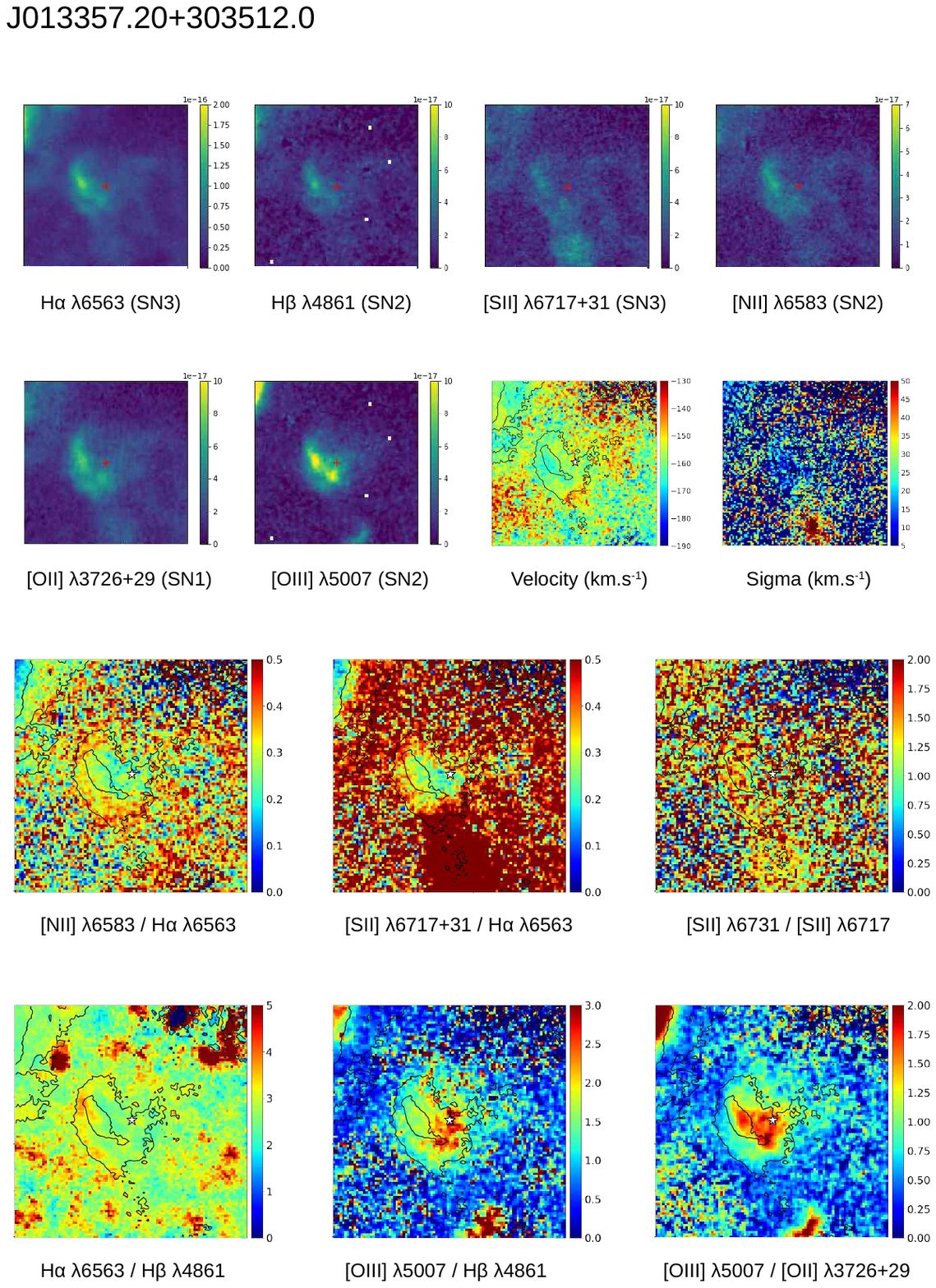}\par 
\label{fig:bubble21}
\caption{Same as Fig. \ref{fig:bubble1} for J013357.20+303512.0}
\end{figure*}

\begin{figure*}
    \includegraphics[page=22,width=0.9\linewidth]{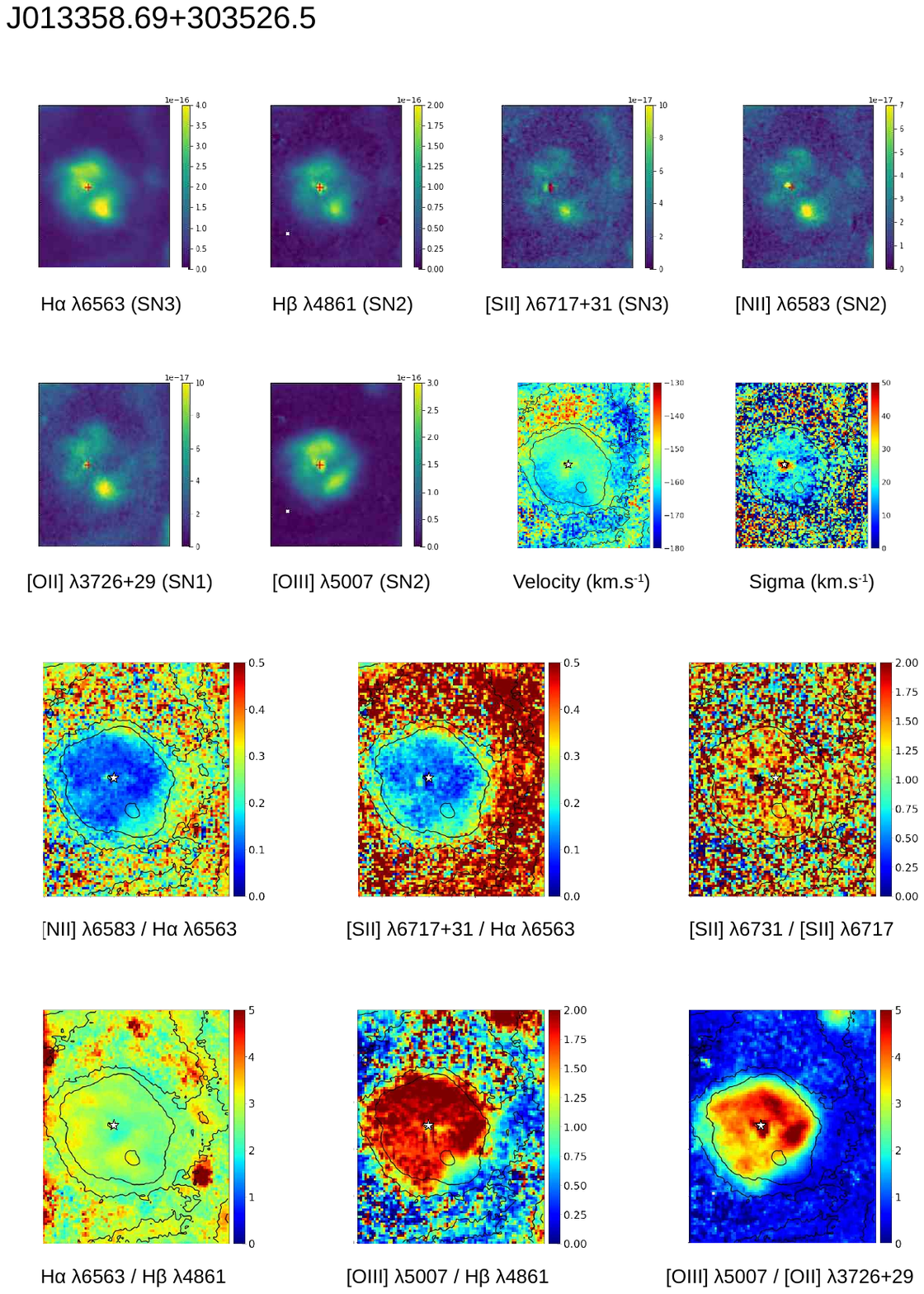}\par 
\label{fig:bubble22}
\caption{Same as Fig. \ref{fig:bubble1} for J013358.69+303526.5}
\end{figure*}

\begin{figure*}
    \includegraphics[page=23,width=0.9\linewidth]{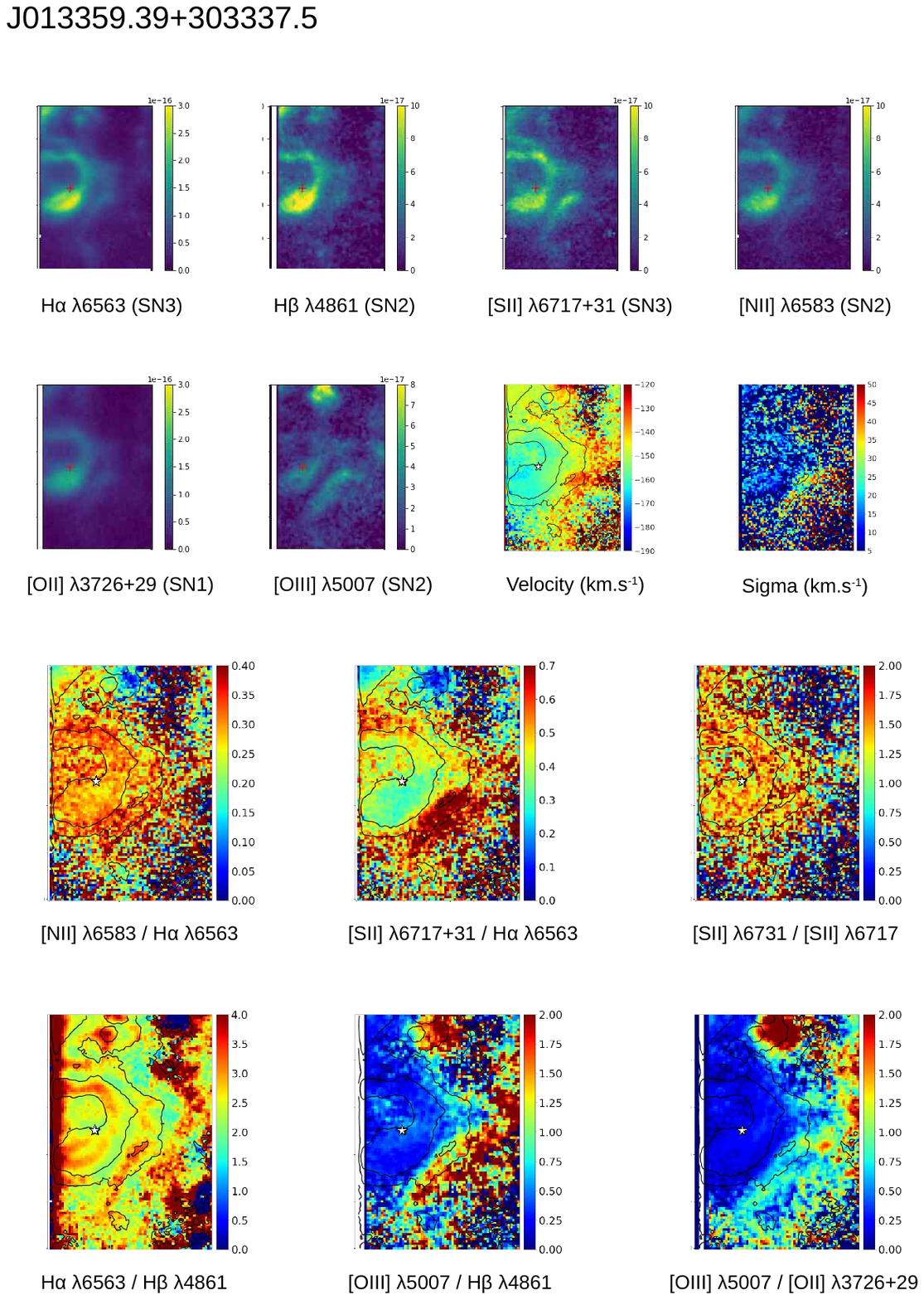}\par 
\label{fig:bubble23}
\caption{Same as Fig. \ref{fig:bubble1} for J013359.39+303337.5}
\end{figure*}

\begin{figure*}
    \includegraphics[page=24,width=0.9\linewidth]{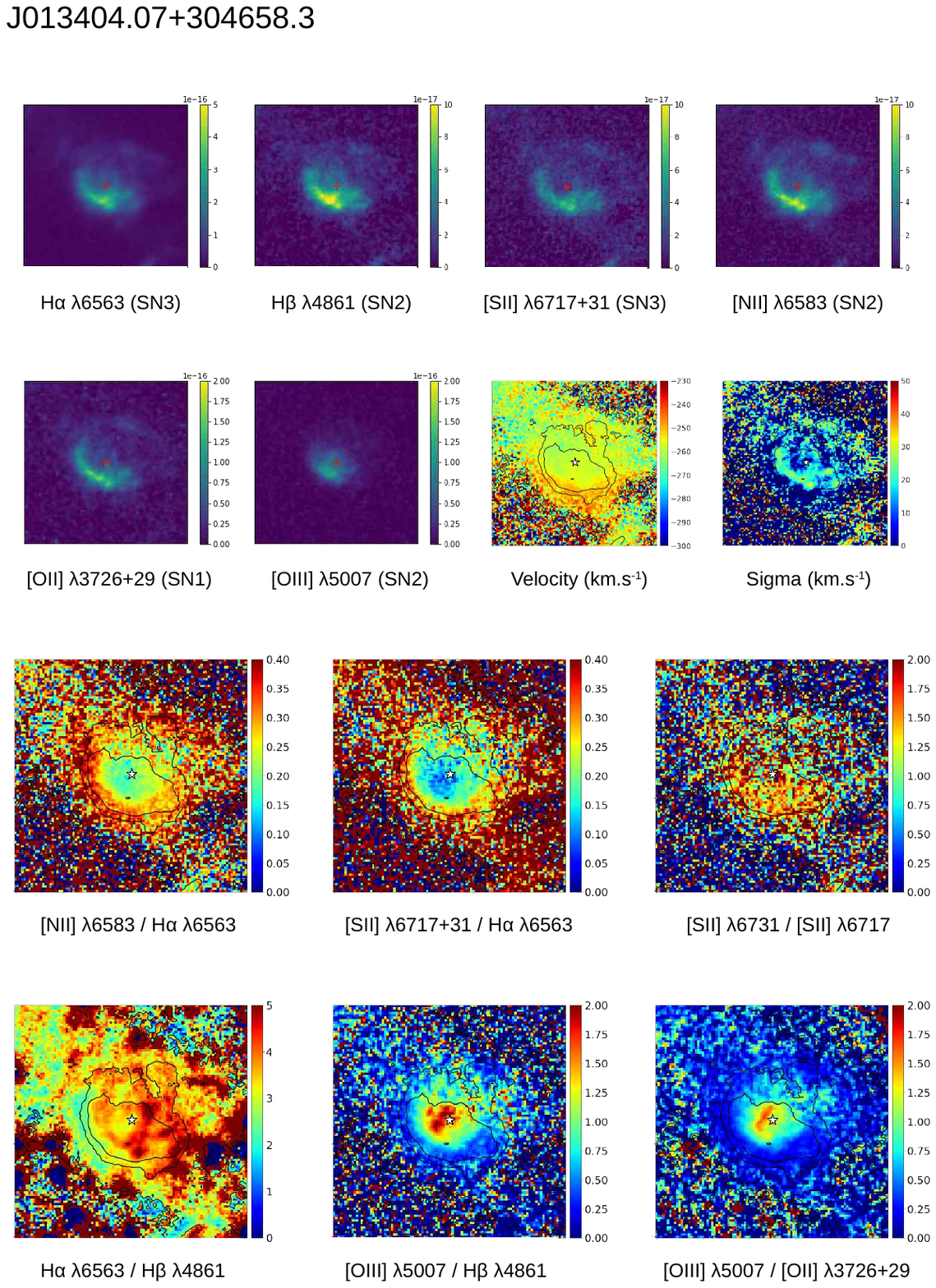}\par 
\label{fig:bubble24}
\caption{Same as Fig. \ref{fig:bubble1} for J013404.07+304658.3}
\end{figure*}

\begin{figure*}
    \includegraphics[page=25,width=0.9\linewidth]{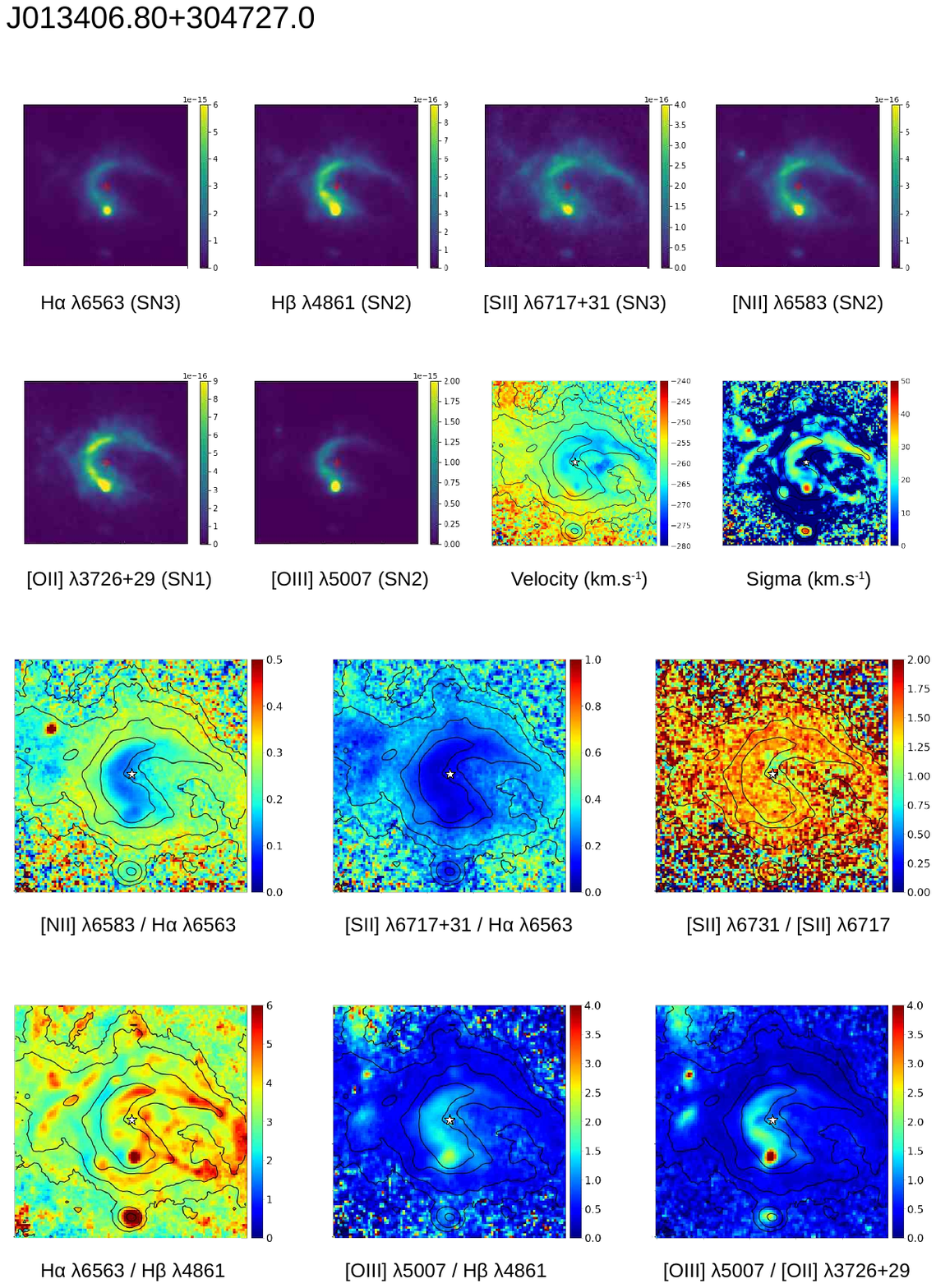}\par 
\label{fig:bubble25}
\caption{Same as Fig. \ref{fig:bubble1} for J013406.80+304727.0}
\end{figure*}

\begin{figure*}
    \includegraphics[page=26,width=0.9\linewidth]{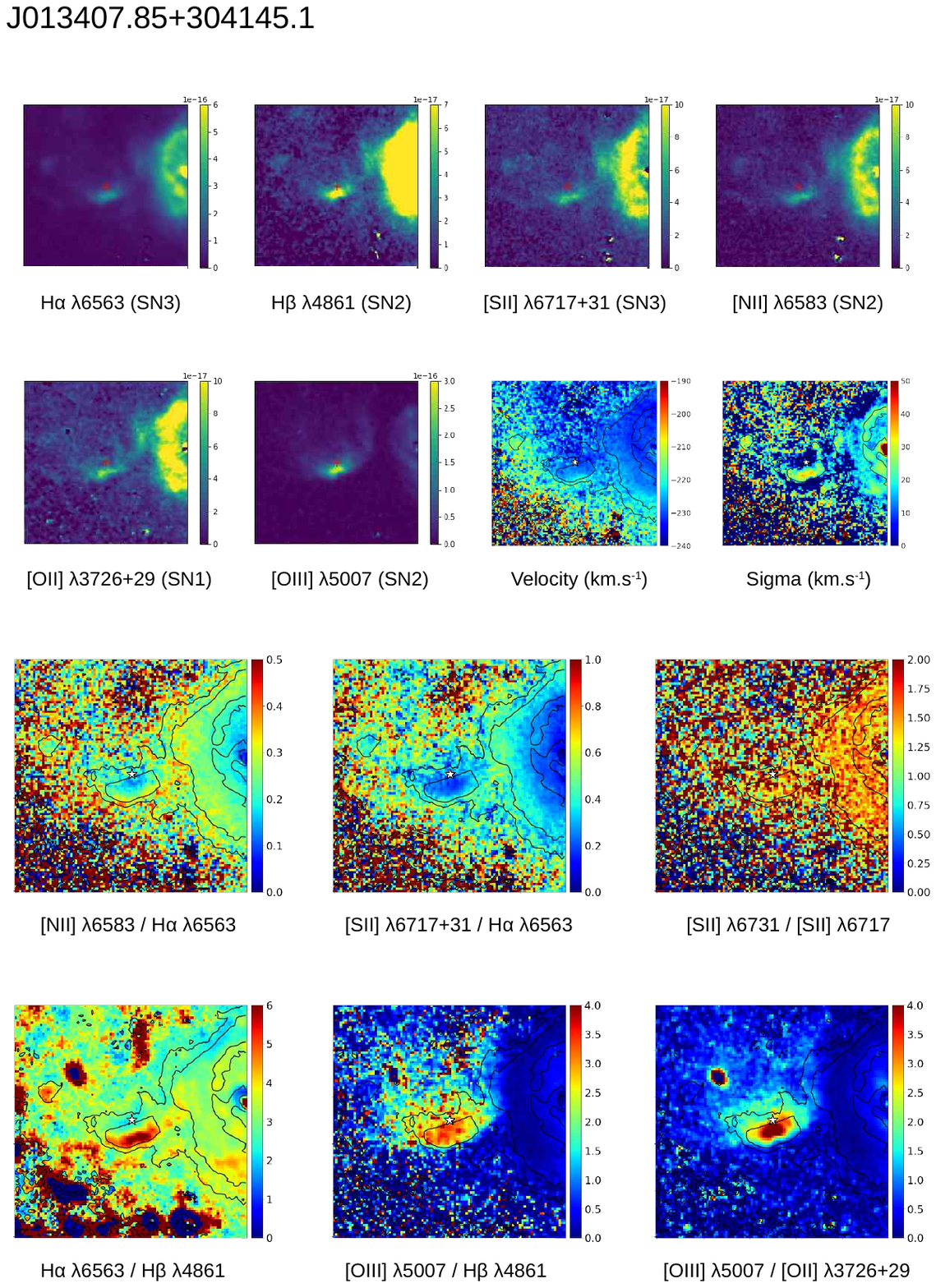}\par 
\label{fig:bubble26}
\caption{Same as Fig. \ref{fig:bubble1} for J013404.07+304145.1}
\end{figure*}

\begin{figure*}
    \includegraphics[page=27,width=0.9\linewidth]{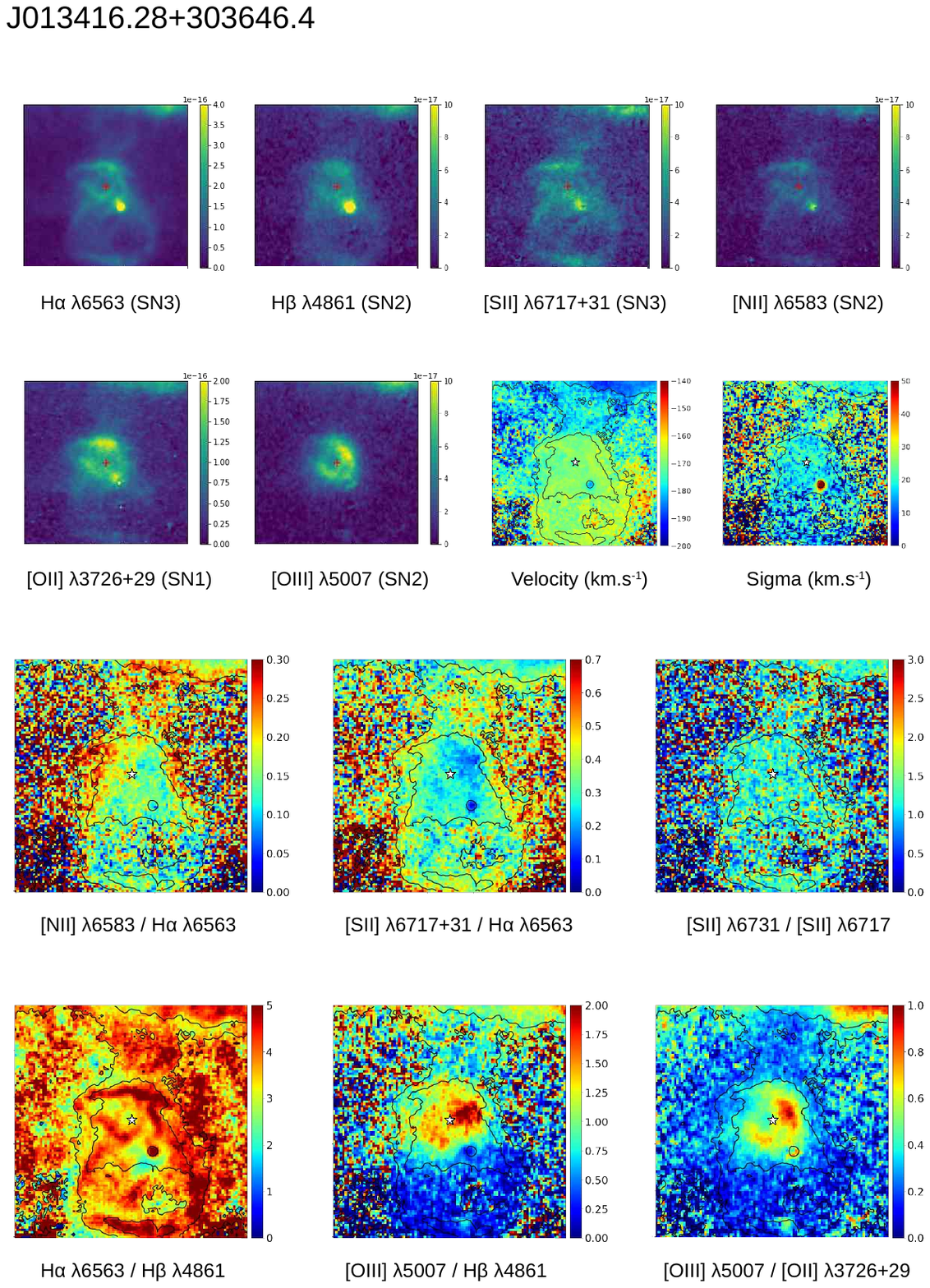}\par 
\label{fig:bubble27}
\caption{Same as Fig. \ref{fig:bubble1} for J013416.28+303646.4}
\end{figure*}

\begin{figure*}
    \includegraphics[page=28,width=0.9\linewidth]{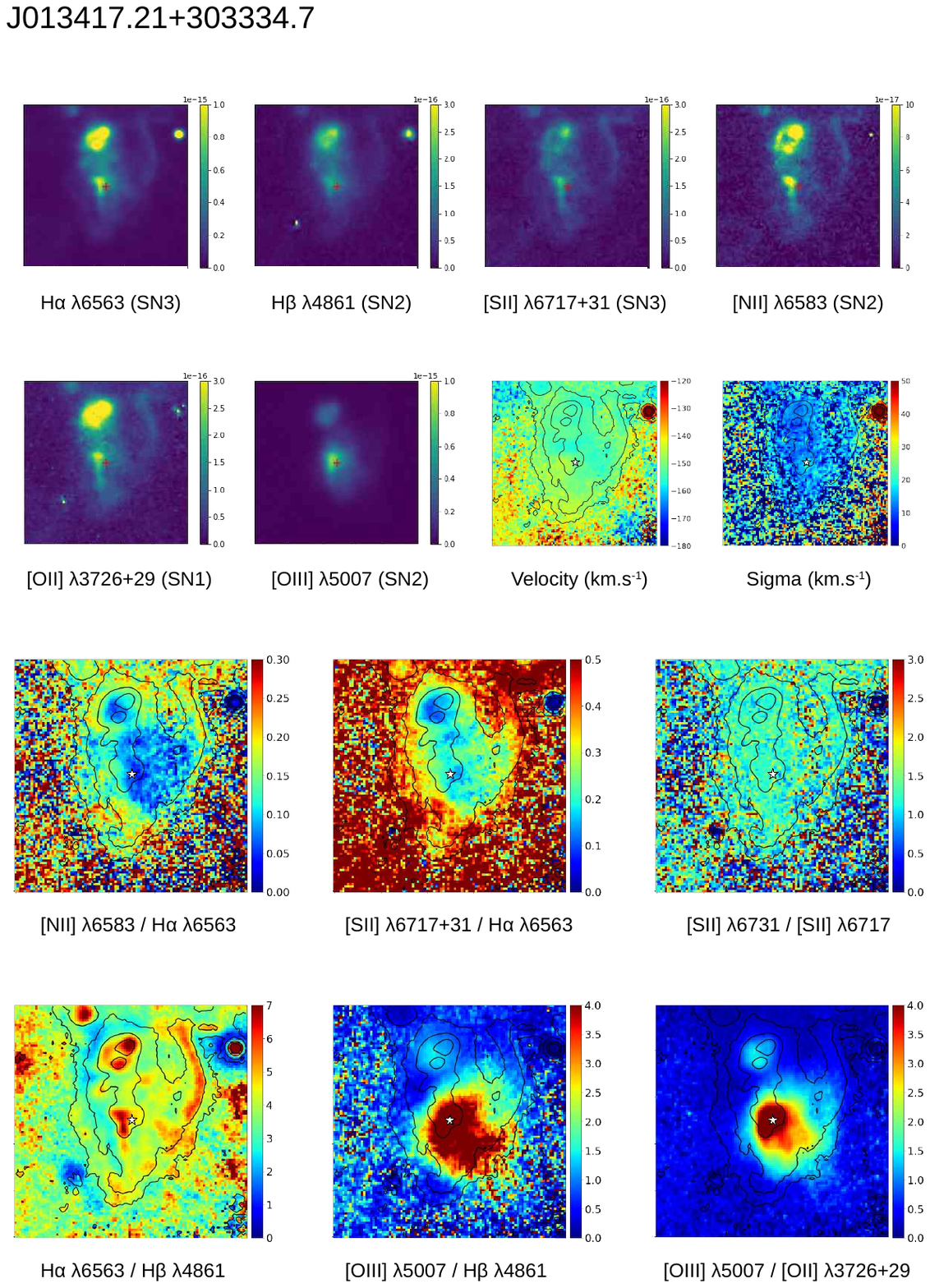}\par 
\label{fig:bubble28}
\caption{Same as Fig. \ref{fig:bubble1} for J013417.21+303334.7}
\end{figure*}

\begin{figure*}
    \includegraphics[page=29,width=0.9\linewidth]{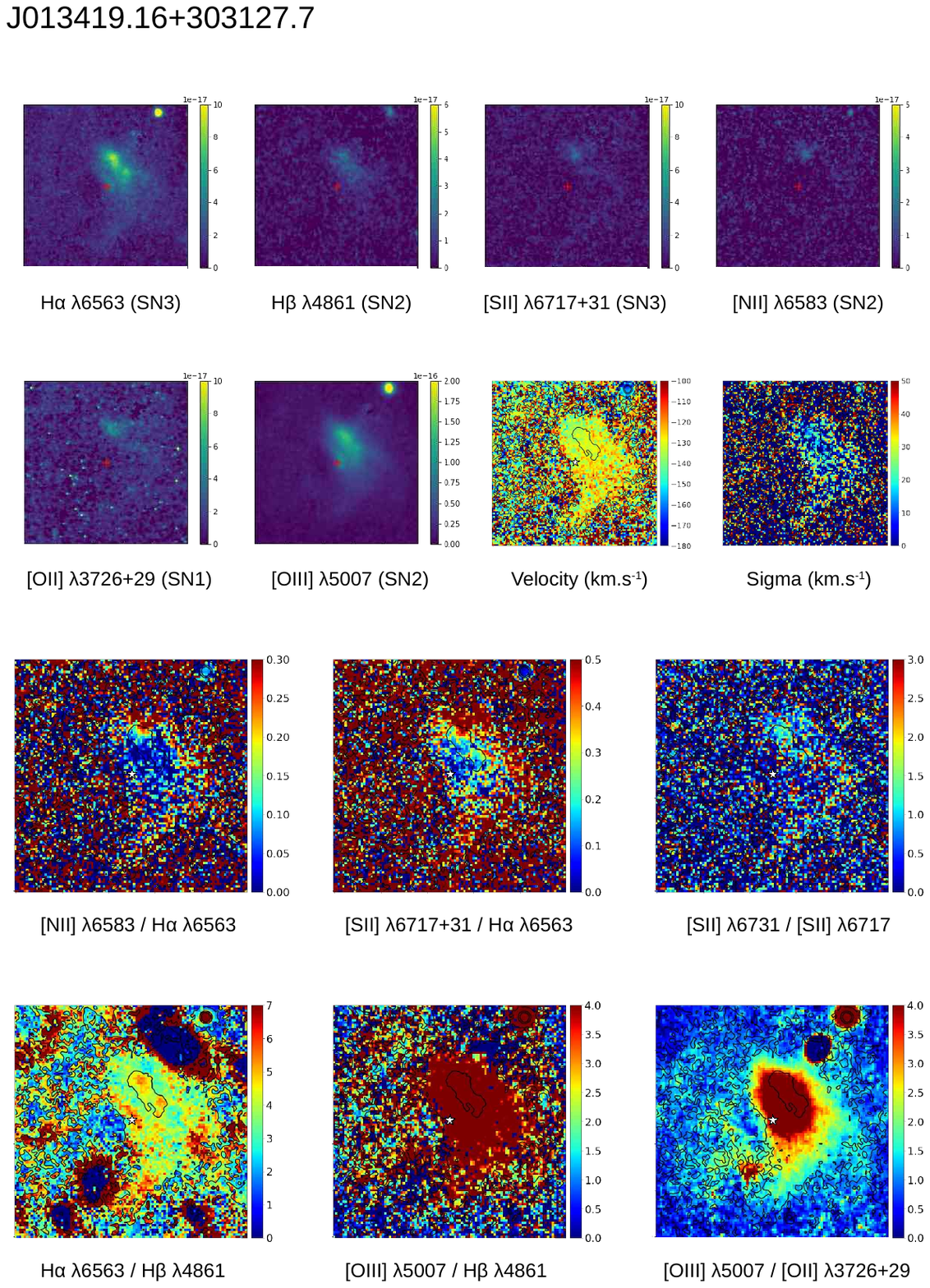}\par 
\label{fig:bubble29}
\caption{Same as Fig. \ref{fig:bubble1} for J013419.16+303127.7}
\end{figure*}

\begin{figure*}
    \includegraphics[page=30,width=0.9\linewidth]{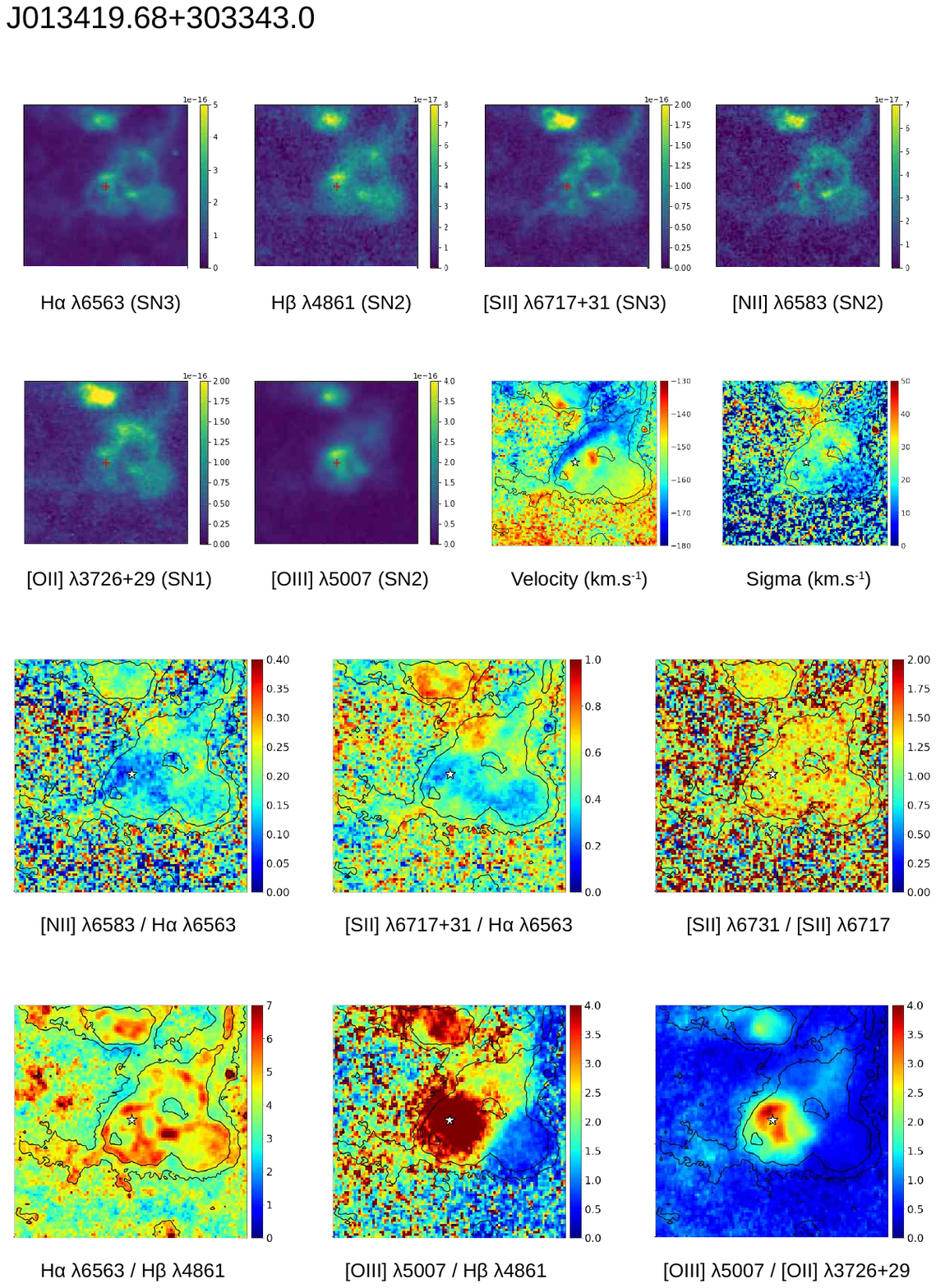}\par 
\label{fig:bubble30}
\caption{Same as Fig. \ref{fig:bubble1} for J013419.68+303343.0}
\end{figure*}

\begin{figure*}
    \includegraphics[page=31,width=0.9\linewidth]{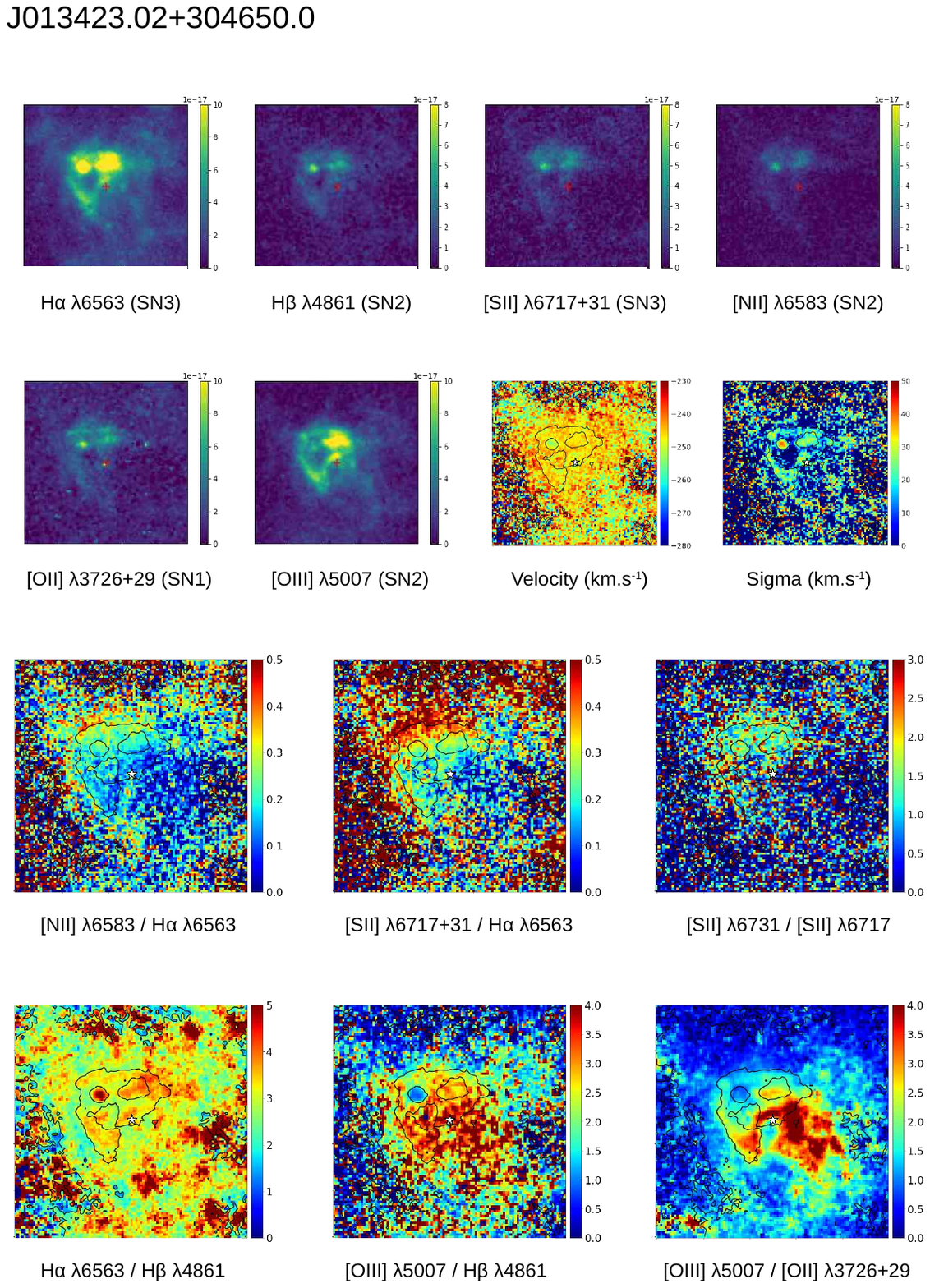}\par 
\label{fig:bubble31}
\caption{Same as Fig. \ref{fig:bubble1} for J013423.02+304650.0}
\end{figure*}

\begin{figure*}
    \includegraphics[page=32,width=0.9\linewidth]{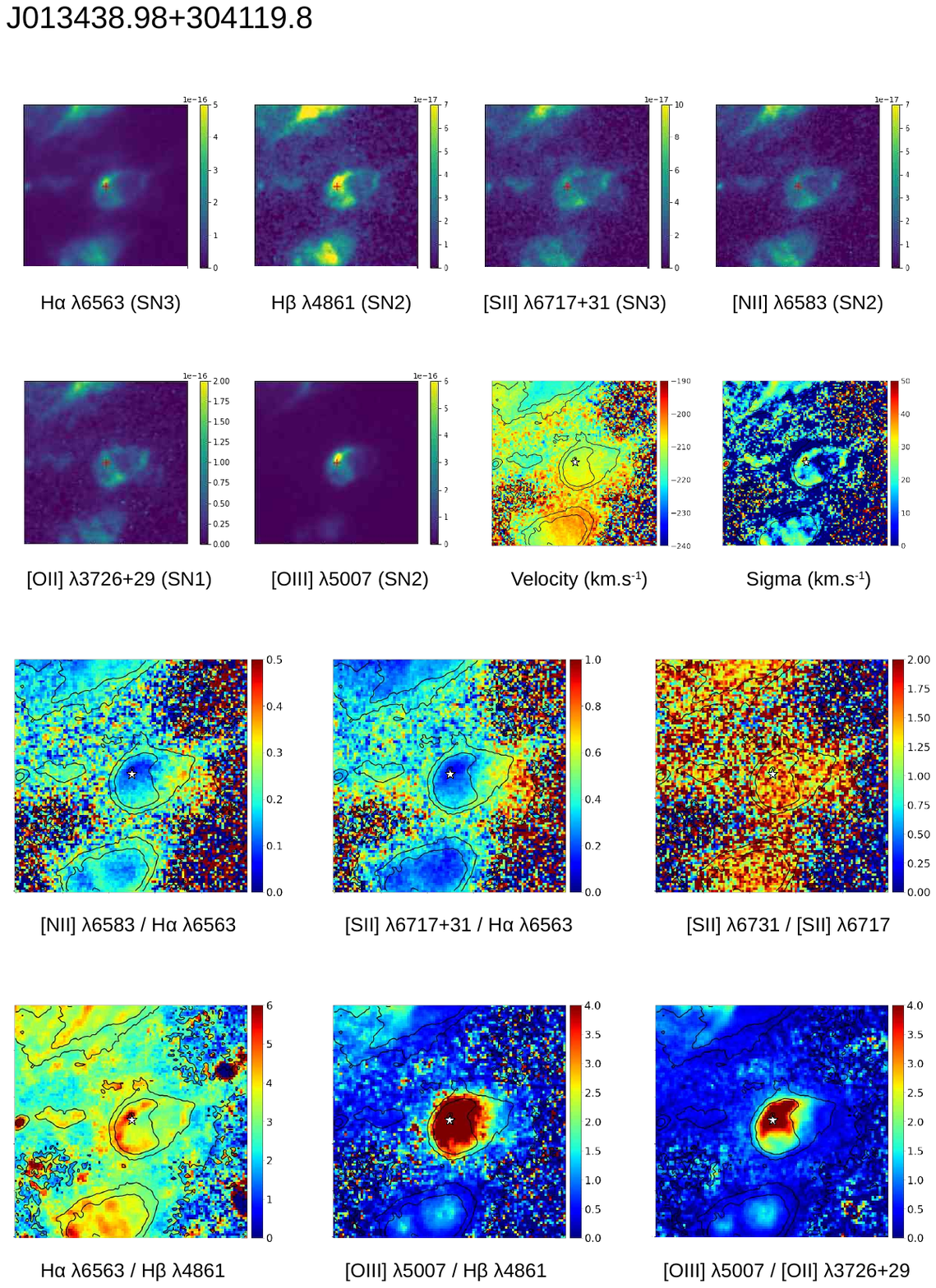}\par 
\label{fig:bubble32}
\caption{Same as Fig. \ref{fig:bubble1} for J013438.98+304119.8}
\end{figure*}

\begin{figure*}
    \includegraphics[page=33,width=0.9\linewidth]{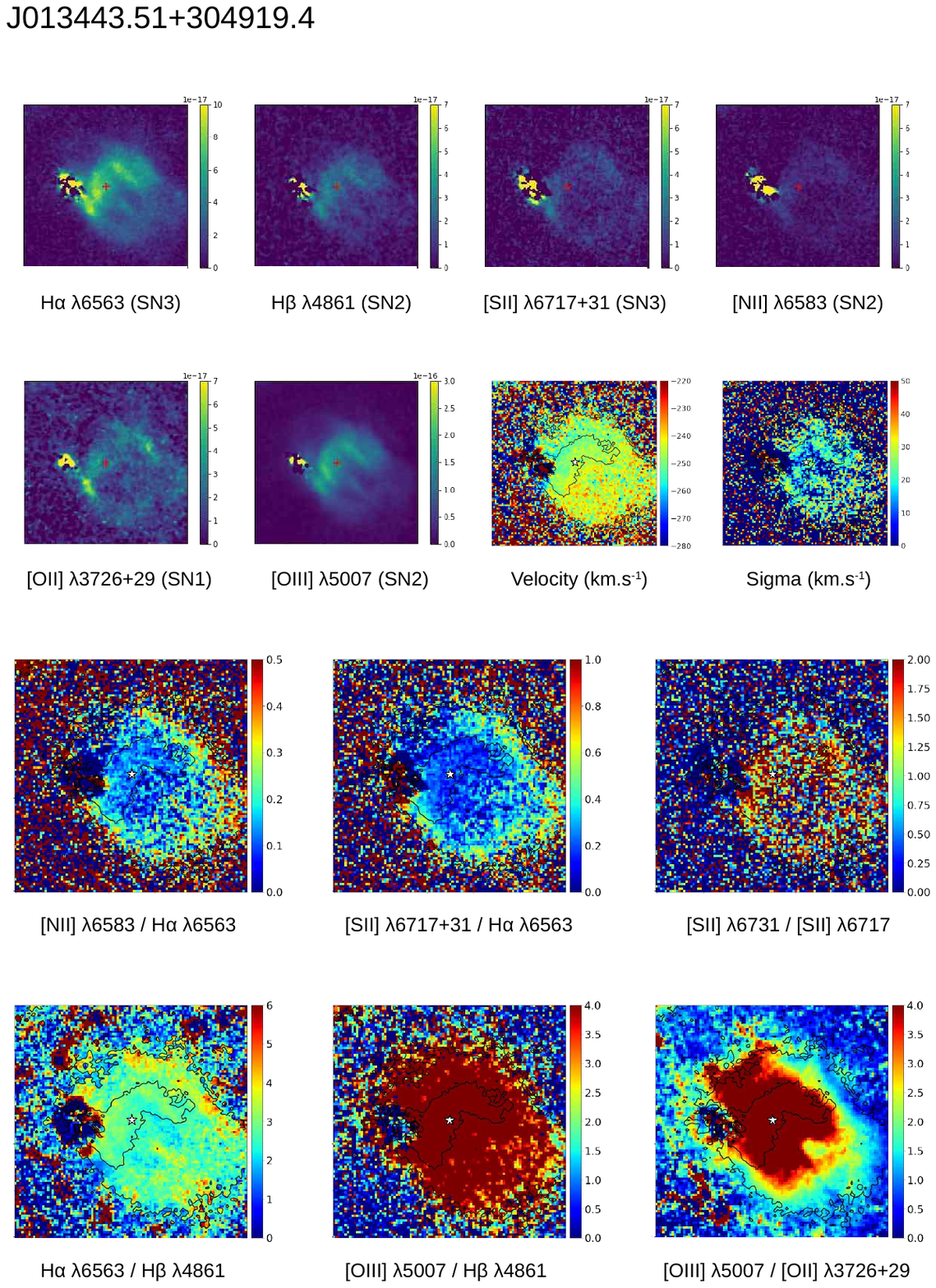}\par 
\caption{Same as Fig. \ref{fig:bubble1} for J013443.51+304919.4}
\label{fig:bubble33}
\end{figure*}


\bsp	
\label{lastpage}
\end{document}